\documentclass[12pt]{article} %***
\usepackage[sectionbib]{natbib}
\usepackage[table]{xcolor}
\usepackage{array,epsfig,fancyheadings,rotating}
\usepackage{sectsty, secdot}
\sectionfont{\fontsize{12}{14pt plus.8pt minus .6pt}\selectfont}
\renewcommand{\theequation}{\thesection\arabic{equation}}
\subsectionfont{\fontsize{12}{14pt plus.8pt minus .6pt}\selectfont}
\usepackage{lscape}
\usepackage{amsmath}
\usepackage{amssymb}
\usepackage{amsfonts}
\usepackage{multirow}
\usepackage{amsthm}
\usepackage{enumitem}
\usepackage{algorithm} 
\usepackage{algpseudocode}
\usepackage{mathrsfs}
\usepackage{booktabs} 
\usepackage{xcolor}
\newtheorem{theorem}{Theorem}
\newtheorem{lemma}{Lemma}

\theoremstyle{definition}

\newtheorem{remark}{Remark}

\newcommand\numberthis{\addtocounter{equation}{1}\tag{\theequation}}

\usepackage{xr}
\makeatletter
\newcommand*{\addFileDependency}[1]{
  \typeout{(#1)}
  \@addtofilelist{#1}
  \IfFileExists{#1}{}{\typeout{No file #1.}}
}
\makeatother

\newcommand{\blue}{\color{black}}

\newcommand{\bA}{\mathbf{A}}

\newcommand{\bC}{\mathbf{C}}
\newcommand{\bD}{\mathbf{D}}

\newcommand{\bG}{\mathbf{G}}

\newcommand{\bI}{\mathbf{I}}
\newcommand{\bJ}{\mathbf{J}}

\newcommand{\bQ}{\mathbf{Q}}

\newcommand{\bT}{\mathbf{T}}

\newcommand{\bW}{\mathbf{W}}
\newcommand{\bX}{\mathbf{X}}

\newcommand{\be}{\mathbf{e}}

\newcommand{\bg}{\mathbf{g}}
\newcommand{\bh}{\mathbf{h}}

\newcommand{\bss}{\mathbf{s}}

\newcommand{\bx}{\mathbf{x}}

\newcommand{\bz}{\mathbf{z}}
\newcommand{\zero}{\mathbf{0}}

\newcommand{\sA}{\mathcal{A}}
\newcommand{\sB}{\mathcal{B}}
\newcommand{\sC}{\mathcal{C}}

\newcommand{\sH}{\mathcal{H}}

\newcommand{\sJ}{\mathcal{J}}

\newcommand{\sP}{\mathcal{P}}

\newcommand{\sS}{\mathcal{S}}

\newcommand{\sU}{\mathcal{U}}
\newcommand{\sV}{\mathcal{V}}

\newcommand{\sX}{\mathcal{X}}
\newcommand{\sY}{\mathcal{Y}}

\newcommand{\fD}{\mathfrak{D}}

\newcommand{\fM}{\mathfrak{M}}

\newcommand{\fR}{\mathfrak{R}}

\newcommand{\E}{\mathbb{E}}
\newcommand{\Var}{\text{Var}}
\newcommand{\cov}{\text{cov}}
\newcommand{\tp}{\text{T}}

\newcommand{\diag}{\text{diag}}

\newcommand{\IMSE}{\text{IMSE}}
\newcommand{\IMAE}{\text{IMAE}}

\newcommand{\SNR}{\text{SNR}}

\newcommand{\bbeta}{\boldsymbol{\beta}}

\newcommand{\bphi}{\boldsymbol{\phi}}

\newcommand{\bgamma}{\boldsymbol{\gamma}}

\newcommand{\bSigma}{\boldsymbol{\Sigma}}

\newcommand{\bDelta}{\boldsymbol{\Delta}}

\newcommand{\bOmega}{\boldsymbol{\Omega}}

\newcommand{\abs}{\overrightarrow{\mathbf{s}}}

\begin{document}

%%%%%%%%%%%%%%%%%%%%%%%%%%%%%%%%%%%%%%%%%%%%%%%%%%%%%%%%%%%%%%%%%%%%%%%%%%%%%%%%%%%%%%%%%%%%%%%%%%%%%%%%%%%%%%%%%%%%%%%%%%%%
%%%%%%%%%%%%%%%%%%%%%%%%%%%%%%%%%%%%%%%%%%%%%%%%%%%%%%%%%%%%%%%%%%%%%%%%%%%%%%%%%%%%%%%%%%%%%%%%%%%%%%%%%%%%%%%%%%%%%%%%%%%%

\renewcommand{\baselinestretch}{2}

\markright{ \hbox{\footnotesize\rm Statistica Sinica
%{\footnotesize\bf 24} (201?), 000-000
}\hfill\\[-13pt]
\hbox{\footnotesize\rm
%\href{http://dx.doi.org/10.5705/ss.20??.???}{doi:http://dx.doi.org/10.5705/ss.20??.???}
}\hfill }

\markboth{\hfill{\footnotesize\rm FIRSTNAME1 LASTNAME1 AND FIRSTNAME2 LASTNAME2} \hfill}
{\hfill {\footnotesize\rm \uppercase{GMM for Varying-coefficient model}} \hfill}

\renewcommand{\thefootnote}{}
$\ $\par

%%%%%%%%%%%%%%%%%%%%%%%%%%%%%%%%%%%%%%%%%%%%%%%%%%%%%%%%%%%%%%%%%%%%%%%%%%%%%%%%%%%%%%%%%%%%%%%%%%%%%%%%%%%%%%%%%%%%%%%%%%%%

\fontsize{12}{14pt plus.8pt minus .6pt}\selectfont \vspace{0.8pc}
% \centerline{\large\bf \uppercase{Functional concurrent model under }} %in functional data analysis
\centerline{\large\bf \uppercase{Functional varying-coefficient model}} %in functional data analysis
\vspace{2pt} 
\centerline{\large\bf \uppercase{under heteroskedasticity}}
\vspace{2pt} 
\centerline{\large\bf \uppercase{with application to DTI data}}
\vspace{.4cm} 
\centerline{Pratim Guha Niyogi, Ping-Shou Zhong and Xiaohong Joe Zhou} 
\vspace{.4cm} 
\centerline{\it Johns Hopkins University and University of Illinois at Chicago}
\vspace{.55cm} \fontsize{9}{11.5pt plus.8pt minus.6pt}\selectfont

\begin{quotation}
\noindent {\it Abstract:}
In this paper, we develop a multi-step estimation procedure to simultaneously estimate the varying-coefficient functions using a local-linear generalized method of moments (GMM) based on continuous moment conditions. 
To incorporate spatial dependence, the continuous moment conditions are first projected onto eigen-functions and then combined by weighted eigen-values, thereby, solving the challenges of using an inverse covariance operator directly. 
We propose an optimal instrument variable that minimizes the asymptotic variance function among the class of all local-linear GMM estimators, and it outperforms the initial estimates which do not incorporate the spatial dependence. 
Our proposed method significantly improves the accuracy of the estimation under heteroskedasticity and its 
asymptotic properties have been investigated. 
Extensive simulation studies illustrate the finite sample performance, and the efficacy of the proposed method is confirmed by real data analysis.

\vspace{9pt}
\par
\end{quotation}\par

\def\thefigure{\arabic{figure}}
\def\thetable{\arabic{table}}

\renewcommand{\theequation}{\thesection.\arabic{equation}}

\fontsize{12}{14pt plus.8pt minus .6pt}\selectfont

%%%%%%%%%%%%%%%%%%%%%%%%%%%%%%%%%%%%%%%%%%%%%%%%%%%%%%%%%%%%%%%%%%%%%%%%%%%%%%%%%%%%%%%%%%%%%%%%%%%%%%%%%%%%%%%%%%%%%%%%%%%%
\section{Introduction}
Due to modern advancements in technology, varying-coefficient models in functional data have become popular to analyze data coming from several imaging technologies such as magnetic resonance imaging (MRI), diffusion tensor imaging (DTI), etc. We consider the problem of estimating non-parametric coefficient function $\bbeta(s)$ which is defined on the functional domain (for example, space) $\sS$ to understand the relationship between functional response $Y(s)$ and real-valued covariates denoted by $\bX =  (X_{1}, \cdots, X_{p})^{\tp}$, which takes the following form,
\begin{equation}
\label{Chapter3-gmm-Eq:model}
    Y(s) = \bX^{\tp}\bbeta(s) + U(s)
\end{equation}
where
$\bbeta(s) = (\beta_{1}(s), \cdots, \beta_{p}(s))^{\tp}$ is a $p$-dimensional vector of unknown smooth functions, and it is
assumed that $\bbeta(\cdot)$ is twice-differentiable with continuous second-order derivatives.
The random error $\{U_{i}(s): s\in \sS\}$ is assumed to be a stochastic process indexed by $s \in \sS$ and it characterizes the within-curve dependence
with mean zero and an unknown covariance function $\Sigma_{\bX}(s, s') = \cov\{U(s), U(s')|\bX\}$. To speed up theoretical exploration and facilitate fast computation, this paper mainly focuses on $\sS$ in a one-dimensional domain. The extension to a multivariate domain $\sS$ is provided in Section \ref{sec:extension} of the supplemental file.

The model (\ref{Chapter3-gmm-Eq:model}) allows  
heteroskedasticity in the covariance function so that 
$\Sigma_{\bx}(s,s')$ depends on $\bX$. There exists
limited research on heteroskedastic functional data.
For example, \citet{chiou2003functional,jiang2011functional, ding2021multivariate} considered covariates-dependent functional principal component analysis.  However, to the best
of our knowledge, no existing inference for the
varying coefficient model (VCM) with heteroskedastic functional data has been developed so far. The aim of this paper is to develop an efficient estimator for VCM with heteroskedastic functional data.
The model in Equation 
(\ref{Chapter3-gmm-Eq:model}) allows its 
regression coefficient to vary over some 
predictors of interest. It was introduced in the 
literature by \citet{hastie1993varying}. 
Because of the wide applicability of VCM, there exists abundant literature on the same.
For example, to name a few of them,
\citet{fan1999statistical,  wu2000kernel,  
fan2003adaptive, chiou2004functional, 
ramsay2005springer, wang2008variable, zhu2014spatially}. There is
a long list of literature on VCM and the aforementioned list is by no means to be exhaustive. A more comprehensive literature
review on VCM can be found in \citet{fan2008statistical}.
% \citet{hoover1998nonparametric, fan1999statistical,  wu2000kernel, huang2002varying, fan2003adaptive,huang2004polynomial, chiou2004functional, ramsay2005springer, zhang2007statistical, cardot2008varying,  fan2008statistical, wang2008variable, zhu2014spatially, kokoszka2017introduction}. 
The main difference between a standard VCM 
and functional VCM is in the error process
$U(s)$. The standard VCM typically assumes
that $U(s)$ are independent errors so that
$U(s)$ and $U(s')$ are independent for $s\neq s'$, while $U(s)$ is a dependent stochastic process in the functional VCM. 
One important challenge is to
consider dependence in the functional VCM. As noted by \citet{lin2001semiparametric}, commonly-used kernel methods are not able to make use of the dependence. Various progress has been made in incorporating dependence into estimation and statistical inference for sparse longitudinal data. For example, \citet{wang2003marginal} developed an innovative marginal kernel method to incorporate correlation and control bias. Further study in \citet{wang2005efficient} showed
that the method in \citet{wang2003marginal} achieves the semiparametric efficient bound. \citet{li2011efficient} further extended the method to include nonparametric covariance estimation. 
\cite{qu2006quadratic} developed an estimation method based on penalized spline and quadratic inference. However, these above-mentioned methods are mostly designed for sparse functional data with a small number of repeated measurements. The focus of the current paper is to develop a method to incorporate dependence for dense functional data with heteroskedastic dependence.
\par
%The notion of density is not well defined for functional responses (in general for any random function) \citep{delaigle2010defining}, as a result of which it is difficult to take advantage of likelihood-based inference, therefore we need to rely on the moment conditions. 
%Typically, we assume that the error term $U(s)$ satisfies the conditional mean zero assumption such as $\E\{U(s)|\bX\} = 0$. 
%By the iterated law of expectation, it is easy to see that for a given point $s \in \sS$, 
%we can define least square estimates as solution of the sample version of $\E\{\bX[Y(s) - \bX^{\tp}\bbeta(s)]\} = 0$. Equivalently, we can obtain these estimates 
%by a minimizer of the sample version of $\E\{ [Y(s) - \bX^{\tp}\bbeta(s)]^{2}\}$ which is termed as 
%non-parametric local-linear estimates \citep{fan1996local}.
%Since the above estimates rely only on the conditional mean-zero assumption, they become inefficient in the presence of heteroskedasticity. 
In this paper, we develop a functional generalized method of moments (GMM) estimation procedure for such VCM, which does not require distributional assumption and can accommodate heteroskedasticity of unknown form. There exist rich literature of applying GMM to varying coefficient models without functional data and heteroskedasticity. For example, \citet{cai2008nonparametric} proposed a one-step local-linear GMM estimator that corresponds to the local-linear GMM discussed in \citet{su2013local} with an identity weight matrix.
\citet{tran2009local} provided a local constant two-step GMM estimator with a specified weight matrix by minimizing the asymptotic variance. 
\citet{su2013local}  developed a local-linear GMM estimator procedure of functional-coefficient instrument variable (IV) models with a general weight matrix under exogenous conditions. 
\citet{cai2006functional} proposed a two-step local-linear estimation procedure to estimate the functional coefficient which includes the estimation of high-dimensional non-parametric model in first step and later estimates the functional coefficients using the first-step non-parametric estimates as generated regressor. As opposed to the classical GMM, for non-parametric local-linear GMM estimator, the integrated mean square error increases as the number of IVs increases for its arbitrary choice \citep{bravo2021second}.  

\par
\begin{figure}[h]
    \centering
    \includegraphics[height = 6cm, width = 10cm]{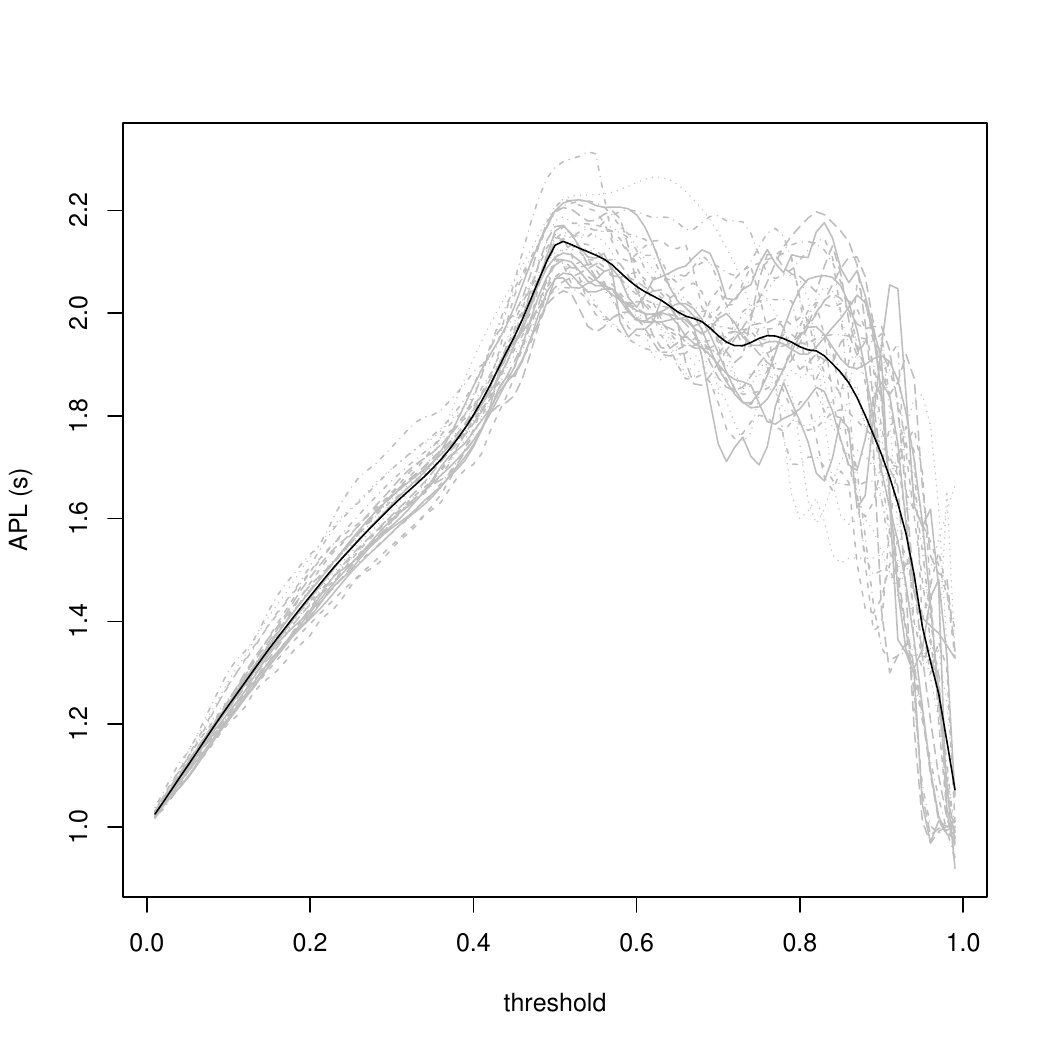}
    \caption{\texttt{Apnea-data}:  Smoothed average path length (APL) from 29 patients over different thresholds(s). Black solid line indicates the mean of APL over thresholds.}
    \label{Chapter3-GMM-Fig:FA-APL}
\end{figure}

The current work is motivated by the problem encountered in diffusion tensor imaging (DTI) where multiple diffusion properties are measured along common major white matter fiber tracts across multiple individuals to characterize the structure and orientation of white matter in the human brain.
Recently a study has been performed to understand white matter structural alternation using DTI for obstructive sleep apnea patients \citep{xiong2017brain}. 
As an illustration, we present smoothed functional data to analyze the efficiency properties of the network generated by diffusion properties of the water molecules. 
In Figure \ref{Chapter3-GMM-Fig:FA-APL}, we plot the graphical characteristics of one of the diffusion properties called fractional anisotropy (FA) over different significant levels to obtain the graphical connectivity
from 29 patients. 
Scientists are often interested to know the individual association of average path length (APL) of the network generated from FA with a set of covariates of interests such as age and lapses score. Moreover, in this data, there is sufficient evidence of heteroskedasticity in the covariates. Details about the data-set and associated variables are described in Section \ref{Chapter3-gmm-Section:data-analyis}.
We therefore need an estimation procedure which
(1) does not need knowledge of the distribution, (2) can handle the heteroskedasticity of covariates, 
(3) can estimate the non-parametric coefficient functions from VCM, and
(4) has a systematic technique for computing an efficient estimator. 

In this article, we develop a local-linear GMM estimation procedure for VCM. 
For given IVs, we propose an optimal local-linear GMM estimator motivated by \citet{lu2020gmm}.
However, the key difference in our approach from the later is that we model the variance of integrated squared error using a non-parametric function of covariates whereas they assume a parametric form in case of classical regression. Therefore, we can ensure that the proposed estimator is at least as efficient as local-linear estimates (initial estimator) and more efficient than that under the presence of heteroskedasticity. 
\par
This paper is organized as follows. In Section \ref{Chapter3-gmm-Section:model}, we introduce our varying-coefficient model and propose a local-linear GMM estimator. In Section \ref{Chapter3-gmm-Section:method}, we present a multi-step estimation procedure. We establish asymptotic results in Section \ref{Chapter3-gmm-Section:theory}. 
We perform a set of simulations studied to understand the finite sample performance of the proposed estimator and present those in Section \ref{Chapter3-gmm-Section:simulation}. 
In Section \ref{Chapter3-gmm-Section:data-analyis}, we apply the proposed method in a real imaging data-set on obstructive sleep apnea (OSA).
In Section \ref{Chapter3-gmm-Section:discussion}, we conclude this article with some discussion. The extension of the proposed method, additional simulation results, and all technical details are provided in the supplementary material.

% ----------------------------------------------- %
\section{Varying-coefficient functional model and moment conditions}
\label{Chapter3-gmm-Section:model}
In this section, we first introduce heteroskedastic conditions for  SVC model and thereafter, propose a mean-zero function for
constructing the GMM estimator.
\subsection{Model}
Let $\{Y_{i}(s), \bX_{i}\}$ for $i =1, \cdots , n$ be independent copies of $\{Y(s), \bX\}$. 
Instead of observing the entire functional trajectory, one can observe $Y(s)$ only on the discrete spatial grid $\{s_{1}, \cdots, s_{r}\}$ on the functional domain $\sS$.
Data can be Gaussian or non-Gaussian and homoskedastic or heteroskedastic depending upon the real applications. 
Therefore, the observed data for the $i$-th individual are $\{s_{j}, Y_{i}(s_{j}), \bX_{i}: j = 1, \cdots, r\}$. 
For simplifying the notation, define $Y_{ij} = Y_{i}(s_{j})$ and $U_{ij} =U_{i}(s_{j})$.
Considering the functional principal component analysis (FPCA) model for $U_{i}(s)$, we assume that $U_{i}(s)$ is square-integrable and admits the Karhunen-Lo\`eve expansion \citep{karhunen1946spektraltheorie, loeve1946functions}. 
Let $\omega_{1}(\bX) \geq \omega_{2}(\bX) \geq \cdots \geq 0$ be ordered eigen-values of the linear operator determined by $\Sigma_{\bx}$ with $\sum\limits_{k=1}^{\infty}\omega_{k}(\bX)$ being finite and $\psi_{k}(s)$'s being the corresponding orthonormal eigen-functions or principal components. 
Thus, the spectral decomposition \citep{j1909xvi} is given by $\Sigma_{\bx}(s, s') = \sum\limits_{k=1}^{\infty}\omega_{k}(\bX)\psi_{k}(s)\psi_{k}(s')$.
% \begin{equation}
%     \Sigma(s, s') = \sum_{k=1}^{\infty}\omega_{k}(\bX)\psi_{k}(s)\psi_{k}(s')
% \end{equation}
Therefore, $U_{i}(s)$ admits the Karhunen-Lo\`eve expansion as follows. 
\begin{equation}
U_{i}(s) = \sum_{k=1}^{\infty}\xi_{k}(\bX_{i})\psi_{k}(s),
\end{equation}
where $\xi_{k}(\bX_{i}) = \int_{\sS} U_{i}(s)\psi_{i}(s)ds$, which is termed as the $k$-th functional principal score for $i$-th individual. The $\xi_{k}(\bX_{i})$ are uncorrelated over $k$ with
$\E\{\xi_{k}(\bX_{i})|\bX_{i}\} = 0$ and $\Var\{\xi_{k}(\bX_{i})|\bX_{i}\} = \omega_{k}(\bX_{i})$, $k \geq 1$. 
Furthermore, assume that the eigen-values vary with $\bX_{i}$ such that $\omega_{k}(\bX_{i}) = \theta_{k}\sigma^{2}(\bX_{i})$ for some unknown function $\sigma(\bX) \geq 0$ and $\theta_{1} \geq \theta_{2} \geq \cdots \geq 0$.
For identifiability, we need some restrictions on $\theta_{k}$s, such as known or fixed $\theta_{1}$. 
Therefore, the above assumption on eigen-values for spectral decomposition allow us to incorporate heteroskedasticity into the model. To best of our knowledge, this is the first attempt to model SVC with unknown heteroskedasticity. 
\subsection{Local-linear mean-zero function}
Let us reiterate our main objective: we want to efficiently estimate the varying-coefficient functions based on GMM for the case of continuum moment conditions together with infinite-dimensional parameters.  Therefore, we need to construct a mean-zero function which will be described in this sub-section.
\par
Since $\bbeta(\cdot)$ in model (\ref{Chapter3-gmm-Eq:model}) is twice continuously differentiable, %(due to assumption \ref{Chapter3-gmm-Cond:beta})
we can apply the Taylor series expansion to $\bbeta(s_{j})$ around an interior point $s_{0}$ and get $\bbeta(s_{j}) = \bbeta(s) + \dot{\bbeta}(s_{0})(s_{j}-s_{0}) + \ddot{\bbeta}(s^{*})(s_{j}-s_{0})^{2}/2$, 
% \begin{equation}
%     \bbeta(s_{j}) = \bbeta(s) + \dot{\bbeta}(s_{0})(s_{j}-s_{0}) + \ddot{\bbeta}(s^{*})(s_{j}-s_{0})^{2}/2
% \end{equation}
where $s^{*}$ lies between $s_{j}$ and $s_{0}$ for all $j = 1, \cdots, r$ and $\dot{\bbeta}$ and $\ddot{\bbeta}$ denote the gradients of $\bbeta$ and $\dot{\bbeta}$ with respect to $s$. 
Thus, $\bbeta(s_{j})$ can be approximated as $\beta_{k}(s_{j}) \approx \beta(s_{0}) + {\partial \beta_{k}(s_{0})}/{\partial s}\times (s_{j} - s_{0})$. So 
in matrix notation, the first-order Taylor series expansion of the coefficient functions becomes 
\begin{equation}
\label{Chapter3-gmm-Eq:beta-approx}
    \bbeta(s_{j}) \approx \bA(s_{0})\bz_{h}(s_{j} - s_{0}),
\end{equation}
where $\bz_{h}(s_{j} - s_{0}) = \left( 1, {(s_{j} - s_{0})}/{h} \right)^{\tp}$ and $\bA(s_{{0}}) = [\bbeta(s_{0}), h\dot{\bbeta}(s_{0})]$ 
which is a $p\times 2$ matrix.
Hence, applying the approximation procedure in Equation (\ref{Chapter3-gmm-Eq:beta-approx}), we can rewrite model (\ref{Chapter3-gmm-Eq:model}) as
\begin{align*}
    \label{Chapter3-gmm-Eq:model-approx}
    Y_{ij} &\approx \bX_{i}^{\tp}\left\{\bA(s_{{0}})\bz_{h}(s_{j}-s_{0}) \right\} + U_{ir}
%    & = \left\{\bz_{h}(s_{j}-s_{0}) \otimes \bX_{i} \right\}^{\tp}\vect\{\bA(s_{0})\} + U_{ir}\\
    = \bW_{ij}(s_{0})^{\tp}\bgamma(s_{0}) + U_{ij},
    \numberthis
\end{align*}
such that $s_{j}$ are sufficiently close to $s_{0}$, where 
$\bW_{ij}(s_{0}) = [\bz_{h}(s_{j}-s_{0})\otimes \bX_{i}]$
and $\bgamma(s_{0}) = (\bbeta(s_{0})^{\tp}, h\dot{\bbeta}(s_{0})^{\tp})^{\tp}$, both of which are vectors of length $2p\times 1$.
\par
Let $K(\cdot)$ be a symmetric probability density function
which is used as kernel and $h > 0$ be the bandwidth; thus, the re-scaled kernel function is defined as $K_{h}(\cdot) = h^{-1}K(\cdot)$.
It is easy to see that for a given location $s_{0} \in \sS$, we can construct a least squares estimator of $\bgamma(s)$ defined in Equation (\ref{Chapter3-gmm-Eq:model-approx})
by minimizing the sample version of the mean squared error 
$\E\{ [Y_{ij} - \bW_{ij}^{\tp}(s_{0})\bgamma(s_{0})]^{2}| \bX_{i} \}$. 
Let $\fM(\bX)$ be a $q$-dimensional IV with $q \geq p$;
the moment condition can be written as $\E\left\{r^{-1}\sum_{j=1}^{r}K_{h}(s_{j}-s_{0})\bDelta_{ij}(s_{0})\right\} = {\zero_{q}}$ where $\bDelta_{ij}(s_{0}) = \fM(\bX_{i})\left\{Y_{ij}-\bW_{ij}(s_{0})^{\tp}\bgamma(s_{0}) \right\}$ is a zero mean stochastic process {with dimension $q$}. 
There exists abundance of literature on constructing 
IVs for optimizing parameter estimations in semi-parametric 
models with homoskedastic or heteroscedastic (known or unknown) error distributions (e.g., \citet{Newey1994,Amemiya1977,Ai1997,Ma2006,Ghoshetal2023}) or
parameters defined by moment conditions with or without nuisance unknown nonparametric functions (e.g.,  \citet{newey1990efficient,ai2003efficient,ChenPouzo2009}).
However, due to the focus of the paper is on estimating nonparametric functions, and the existence of functional dependence and heteroskedasticity of an unknown form, these existing approaches can not be directly undertaken for the model we considered. 
There are also some papers discussing choosing IVs for optimizing nonparametric function estimators for 
independent errors (e.g., \citet{cai2008nonparametric,su2013local}).  
We take this opportunity to investigate the choice of IVs in our framework. For details, please refer to the Remark below Theorem \ref{Chapter3-gmm-Theorem:final}.
\par
Motivated by the idea of local-linear estimator {and local GMM methods in \citet{cai2008nonparametric} and
\citet{su2013local}, 
we consider the local-linear IVs $\bQ_{i{j}}(s_{0}) = (\fM(\bX_{i}), \fM(\bX_{i})(s_{j}-s_{0})/h)^{\tp}$. 
Therefore, 
consider the following non-parametrically localizing augmented orthogonal moment conditions for estimating $\bbeta(s)$.
\begin{align}
    \bg_{i}\{ \bgamma(s_{0}) \} 
    &=r^{-1}\sum_{j=1}^{r}K_{h}(s_{j}-s_{0})\bz_{h}(s_{j}-s_{0})\otimes\bDelta_{ij}(s_{0}),
    \label{ggamma}
\end{align}
and note that $\{\bg_{i}(\bgamma(s))\}: i=1, \cdots, n\}$ are independent and 
$\E\{\bg_{i}(\bgamma(s))\} = \zero_{2q\times 1}$ for $s \in \sS$.
\par
Most of the VCMs that exist in the literature assume homoskedasticity in covariates and are limited to weakly dependent non-parametric models \citep{su2013local, sun2016functional}, which differs significantly in our model assumptions. In contrast, we assume a spatially VCM under heteroskedasticity of unknown form. 

\section{Multi-step estimation procedure}
\label{Chapter3-gmm-Section:method}
This section develops a multi-step estimation procedure to estimate $\bbeta(s)$ simultaneously across all $s \in \sS$.
Essentially, the multi-step procedure can be broken down as, Step-I: an initial estimation; Step-II:  estimation of the variance function, mean zero function, and eigen-components and Step-III: GMM estimation. 
The key ideas of each step are described below. 
\begin{enumerate}[label=Step-\Roman*.]
    \item Calculate the least squares estimates of $\bbeta(s)$ as initial estimates, denoted by $\breve{\bbeta}(s)$ across all $s \in \sS$. 
	\item Estimate the conditional variance of integrated square residuals non-parametrically and subsequently estimate the covariance of mean-zero function. Estimate the eigen-components using multivariate FPCA.  
	\item Project the continuous moment conditions  onto eigen-functions and then combine them by weighted eigenvalues to incorporate spatial dependence and thus obtain the 
	updated estimate of $\bbeta(s)$, denoted by $\widehat{\bbeta}(s)$ across all $s \in \sS$.
\end{enumerate}
	
\subsection{Step-I: Initial least squares estimates}
We consider a local-linear smoother \citep{fan1996local} to obtain an initial estimator of $\bbeta(\cdot)$ ignoring functional dependencies. 
In this case, the non-linear least squares function of the model \ref{Chapter3-gmm-Eq:model} can be defined as an objective function given by $\sJ_{init}\{\bbeta(\cdot)\} = ({nr})^{-1}\sum\limits_{i=1}^{n}\sum\limits_{j=1}^{r}\{ Y_{ij} - \bX_{i}^{\tp}\bbeta(s_j)\}^{2}$. By the local-linear smoothing method we estimate $\bgamma$ at functional point $s_{0}$, by minimizing 
\begin{align*}
    \label{Chapter3-gmm-Eq:Model1InitEq}
    \sJ_{init}\{\bgamma(s_{0})\} &= 
            ({nr})^{-1}\sum_{i = 1}^{n}
            \sum_{j = 1}^{r}
            K_{h}(s_{j} -s_{0})\left\{
                Y_{ij} - \bW_{ij}(s_{0})^{\tp}\bgamma(s_{0})
            \right\}^{2}.
    \numberthis
\end{align*}
The solution of the above least-squares problem can be expressed as 
\begin{align*}
    \breve{\bgamma}(s_{0}) &= \left\{
        ({nr})^{-1}\sum_{i=1}^{n}\sum_{j=1}^{r}K_{h}(s_{j}-s_{0})
        \bW_{ij}(s_{0})\bW_{ij}(s_{0})^{\tp}
    \right\}^{-1}\\
    &\qquad \times \left\{
        ({nr})^{-1}\sum_{i=1}^{n}\sum_{j=1}^{r}K_{h}(s_{j}-s_{0})\bW_{ij}(s_{0})Y_{ij}
    \right\}.
    \numberthis
\end{align*}
Consequently, the estimator of the coefficient function vector $\bbeta(s)$ at $s_{0}$ is $\breve{\bbeta}(s_{0}) = [(1,0) \otimes \bI_{p}]\breve{\bgamma}(s_{0})$.
We determine the tuning parameter $h$ by using some data-driven techniques such as cross-validation and generalized cross-validation.

\subsection{Step-II: Intermediate steps}
Step-II consists of two important steps in determining the class of GMM estimator. 
First in Step-II.A, we propose a method to obtain optimal IVs and therefore estimate the eigen-components which are used in local-linear GMM objective function in Step-III. 
To estimate eigen-components, we essentially need to use a multivariate version of FPCA which is quite uncommon in the literature. We borrow the method proposed by \citet{wang2008karhunen}.
%some of which can be found in \citet{viviani2005functional, wang2008karhunen, berrendero2011principal, chiou2014multivariate, happ2018multivariate}.

\subsubsection*{Step-II.A: Choice of instrument variables (IVs)}
\label{sec:IV}
Choosing IVs is critical, and the required identification condition is $q \geq p$, which ensures that the dimension of $\bQ_{ij}(s_{0})$ is at least equal to the dimension of $\bgamma(s_{0})$.
In our model as discussed in Section \ref{Chapter3-gmm-Section:model}, the error term has a potential heteroskedasticity of unknown form. 
We define a set of independent and identically distributed random variables $R_{1}, R_{2}, \cdots, R_{n}$ for $n$ individuals where $R_{i} = \int U_{i}^{2}(s)ds$ for each $i$, termed as integrated square of residuals, and $\E\{R_{i}|\bX_{i}\} = \sigma^{2}(\bX_{i})\sum_{k = 1}^{\infty}\theta_{k}$. 
Therefore, consider the following non-parametric regression problem. 
\begin{equation}
\label{Chapter3-gmm-Eq:Het}
    \log{R_{i}} = \log \sigma^{2}(\bX_{i}) + \epsilon_{i},
\end{equation}
where $\epsilon_{i}$ is the mean zero random variable with constant variance. 
The above model in Equation (\ref{Chapter3-gmm-Eq:Het}) boils down to the problem of 
estimation of $\log \sigma^{2}(\bX_{i})$ by regressing the logarithmic value of the integrated squared residuals variable on the covariates $\bX_{i}$. 
This approach is along the lines of \citet{yu2004likelihood, wasserman2006all}, although used in a different context. 
Since $U_{i}$s are not observable, we replace $U_{i}$ by an efficient estimate that is obtained from Step-I, viz.,
$\breve{U}_{i}(s) = Y_{i}(s) - \bX_{i}^{\tp}\breve{\bbeta}(s)$ for all $s\in \sS$.
For application, this step can easily be implemented using 
``\texttt{gam}" function available in \texttt{mgcv} package in R to get an estimate of the non-parametric mean function, denoted by $\widehat{\mu}(\bX)$ and therefore $\widehat{\sigma}^{2}(\bX) = \exp\{\widehat{\mu}(\bX)\}$. Given the estimate of $\sigma(\cdot)$, we can, therefore, choose IVs as $\fM(\bX_{i}) = \left(\bX_{i}, \bX_{i}/\widehat{\sigma}^{2}(\bX_{i})\right)^{\tp}$.

\subsubsection*{Step-II.B: Estimation of eigen-components}
Without loss of generality, assume for simplicity that the spectrum of functional domain $\sS = [0,1]$ 
and the dimension of mean-zero function $\bg\{\bgamma(s)\} = (g_{1}\{\bgamma(s)\}, \cdots, g_{2q}\{\bgamma(s)\})^{\tp}$ is $2q$.  
Note that $\bg\{\bgamma(s)\}$ in (\ref{ggamma}) is defined on an interval $[0,1]$ such that $\sum\limits_{l = 1}^{2q}\int\E\{g_{l}^{2}\{\gamma(s)\}\}ds$ is finite and the covariance function 
$\bC(s,s') = \E\left[\bg\{{\bgamma}(s)\}\bg\{{\bgamma}(s)\}^{\tp}\right]$.
Under condition \ref{Chapter3-gmm-Cond:g} mentioned in Section \ref{Chapter3-gmm-Section:theory}, 
using the lining-up method in {\cite[Chapter~5]{wang2008karhunen} and \cite{ramsay2005springer}}, 
define a new stochastic process $e(s_*)$ on the interval $[0, 2q]$ with eigen-function $\phi_{e}$ such that, 
$e(s_*) = g_{l}\{\bgamma(s_* - (l-1))\}$ and $\phi_{e}(s_*) = \phi_{l}\{\bgamma(s_* - (l-1))\}$ for $l-1 \leq s_* < l$, $l=1, \cdots 2q$, where we define the eigen-function for each $g_{l}$ as $\phi_{l}$ for $l =1, \cdots, 2q$.
Therefore, the covariance function between $e(s_*)$ and $e(s'_*)$ can be expressed as $C_{l, l'}(s_*, s'_*) = \cov\{ e(s_*), e(s'_*)\}$ for $l-1 \leq s_* < l$ and $l' - 1 \leq s'_* < l'$; $l, l' = 1, \cdots, 2q$. 
Note that, for $2q$-dimensional processes, the Fredholm integral equation %\citep{porter1990integral} 
is equivalent to $2q$-simultaneous integral equations where
each of them corresponds to a specific functional interval of $e(s_*)$.
For $l-1 \leq s_* < l$; $l=1, \cdots, 2q$, the Fredholm integral equation yields $\int_{0}^{2q}\cov\{e(s_*), e(s'_*)\}\phi_{e}(s_*)ds_* = \lambda\phi_{e}(s_*)$.
Now for $0< s < 1$, the above relation is equivalent to the following.
\begin{align*}
\sum_{l'=1}^{2q}\int_{0}^{1}
     \cov\{g_{l}\{\bgamma(s)\}, g_{l'}\{\bgamma(s')\}\}\phi_{l'}(s')ds' = \lambda\phi_{l}(s).
    \numberthis
\end{align*}
In multivariate setting, the orthogonality condition becomes 
\begin{align*}
\label{Chapter3-gmm-Eq:orthogonal}
\textbf{1}(l = l') &= \int_{0}^{2q}\phi_{e,l}(s_*)\phi_{e,l'}(s_*)ds_*
= \sum_{k=1}^{2q}\int_{0}^{1}\phi_{k, l}(s)
    \phi_{k, l'}(s)ds.
    \numberthis
\end{align*}
Using the generalized Mercer's theorem \citep{j1909xvi}, the results for the covariance function can be briefly shown using the lining-up method.
Assume that the covariance function is continuous after the lining up processes, so for $(l-1) \leq s_* < l$ and $(l'-1) \leq s_*' < l'$;
$l, l' = 1, \cdots, 2q$, the covariance function between $g_{l}(s)$ and $g_{l'}(s')$ can be expressed as 
\begin{align*}
    C_{l, l'}(s, s') %&= \cov\{g_{l}(s), g_{l'}(s)\}
    = \sum_{k=1}^{\infty}\lambda_{k}
    \phi_{k, l}\{s_*-(l-1)\}
    \phi_{k, l'}\{s_*'-(l'-1)\}.%\\
   % &= \sum_{k = 1}^{\infty}\lambda_{k}
  %  \phi_{k, l}(s)\phi_{k, l'}(s')
    \numberthis
\end{align*}
Therefore, using the above argument, 
we can define the multivariate
spectral decomposition $\bC(s, s') = \sum\limits_{k = 1}^{\infty}\lambda_{k}\bphi_{k}(s)\bphi_{k}(s')^{\tp}$ for $s,s'\in [0,1]$
with the orthogonality condition (\ref{Chapter3-gmm-Eq:orthogonal}).
After the lining-up process, data are univariate and hence we can adopt the existing techniques of estimating functional eigen-values and eigen-function in the literature \citep{yao2005functional,muller2010empirical, li2010uniform} to estimate $\lambda$ and $\phi_{e}(s)$, and hence can estimate $\bphi(s)$ by stacking all components for aliened eigen-functions $\phi_{e}(s)$.

\subsection{Step-III: Final estimates}
Finally, we demonstrate our proposed estimator based on local-linear GMM where the proposed mean-zero function can be projected onto eigen-function and then combined by the weighted eigen-values. 
For any positive $\alpha$, the objective function is given by 
\begingroup
\allowdisplaybreaks
\begin{align*}
    &\sJ\{\bgamma(s_{0})\} = \sum_{k=1}^{\infty}\frac{\widehat{\lambda}_{k}}{\widehat{\lambda}_{k}^{2} + \alpha}
        \left\{ 
             \overline{\bg}(\bgamma(s_{0}))^{\tp}\widehat{\bphi}_{k}(s_{0})
        \right\}^{2}\\
        &= \sum_{k=1}^{\infty}\frac{\widehat{\lambda}_{k}}{\widehat{\lambda}_{k}^{2} + \alpha}
        \left\{
        ({nr})^{-1}\sum_{i=1}^{n}\sum_{j=1}^{r}
        K_{h}(s_{j}-s_{0})\widehat{\bphi}_{k}(s_{0})^{\tp}\bQ_{ij}(s_{0})\left[
            Y_{ij} - \bW_{ij}(s_{0})^{\tp}\bgamma(s_{0})
        \right]
        \right\}^{2}.
        \numberthis
\end{align*}
\endgroup
By minimizing the above objective function, we obtain
\begingroup
\allowdisplaybreaks
\begin{align*}
        &\sum_{k=1}^{\infty}\frac{\widehat{\lambda}_{k}}{\widehat{\lambda}_{k}^{2} + \alpha}
        \left\{
            ({nr})^{-1}\sum_{i=1}^{n}\sum_{j=1}^{r}
            K_{h}(s_{j}-s_{0})\widehat{\bphi}_{k}(s_{0})^{\tp}\bQ_{ij}(s_{0})\bW_{ij}(s_{0})
        \right\}\\
        &\qquad \times
        \left\{
            ({nr})^{-1}\sum_{i=1}^{n}\sum_{j=1}^{r}
            K_{h}(s_{j}-s_{0})\widehat{\bphi}_{k}(s_{0})^{\tp}\bQ_{ij}(s_{0})
            \left[
                Y_{ij} - \bW_{ij}(s_{0})^{\tp}\bgamma(s_{0})
            \right]
        \right\}\\
        &:= \sum_{k=1}^{\infty}\frac{\widehat{\lambda}_{k}}{\widehat{\lambda}_{k}^{2} + \alpha}
        \sX_{k}(s_{0})\left\{
            \sY_{k}(s_{0}) - \sX_{k}(s_{0})^{\tp}\bgamma(s_{0})
        \right\},
    \numberthis
\end{align*}
\endgroup
where $\sX_{k}(s_{0}) = ({nr})^{-1}\sum\limits_{i=1}^{n}\sum\limits_{j=1}^{r}K_{h}(s_{j}-s_{0})\bW_{ij}(s_{0})\bQ_{ij}(s_{0})^{\tp}\widehat{\bphi}_{k}(s_{0})$ and $\sY_{k}(s_{0}) = ({nr})^{-1}\sum\limits_{i=1}^{n}\sum\limits_{j=1}^{r}K_{h}(s_{j}-s_{0})\bQ_{ij}(s_{0})^{\tp}\widehat{\bphi}_{k}(s_{0})Y_{ij}$. Therefore, the final estimate of the coherent function is $\widehat{\bbeta}(s_{0}) = [(1, 0)\otimes\bI_{p}]\widehat{\bgamma}(s_{0})$ where
\begin{equation}
    \widehat{\bgamma}(s_{0}) = \left\{
    \sum_{k=1}^{\infty}\frac{\widehat{\lambda}_{k}}{\widehat{\lambda}_{k}^{2} + \alpha}
    \sX_{k}(s_{0})\sX_{k}(s_{0})^{\tp}
    \right\}^{-1}
    \left\{
    \sum_{k=1}^{\infty}\frac{\widehat{\lambda}_{k}}{\widehat{\lambda}_{k}^{2} + \alpha}
    \sX_{k}(s_{0})\sY_{k}(s_{0})
    \right\}.
\end{equation}
\par
The Algorithm \ref{Chapter3-gmm-Algo:estimation} in the supplementary material summarizes the proposed method. For demonstration purposes, we choose the tuning parameters using cross-validation as discussed in the algorithm. In the proposed algorithm, $\alpha$ controls the number of eigen-values, and can be chosen so that condition \ref{Chapter3-gmm-Cond:eigen} defined in Section \ref{Chapter3-gmm-Section:theory} is valid. 
Furthermore, it is essential to establish a continuity criterion for alignment to provide theoretical validation. Even when there is a lack of continuity in $\phi_{e}$, empirical studies suggest that the final outcomes remain suitable for practical application.
We also discuss the extension of the proposed method to the multivariate 
domain with $s\in [0,1]^d$ in Section \ref{sec:extension} of the supplementary material.

\section{Asymptotic results}
\label{Chapter3-gmm-Section:theory}
In this section, we provide some assumptions and then study the asymptotic properties of the local-linear GMM estimator. Here, we allow the sample size $n$ and the number of functional domains $r$ to grow to infinity. Detailed technical proofs are provided in the online Supplementary Material.
\par
Let $\bbeta_{0}(s_{0})$ be the true value of $\bbeta(s_{0})$ at location $s_{0}$. 
Consider the following conditions that will be useful in asymptotic results. 
\begin{enumerate}[label=(C\arabic*)]
    \item\label{Chapter3-gmm-Cond:kernel} Kernel function $K(\cdot)$ is a symmetric density function defined on the bounded support $[-1, 1]$ and is Lipschitz continuous.
    \item\label{Chapter3-gmm-Cond:density} Density function $f_{T}$ of $T$ is bounded above and away from infinity, and also below and away from zero. Moreover, $f$ is differentiable and the derivative is continuous. 
    \item\label{Chapter3-gmm-Cond:X} $\E\{\|X\|^{a}\} < \infty$ and $\E\{\sup_{s\in \sS} |U(s)|^{a}\} < \infty$ for some positive $a>1$. Define, $\E\{\fM(\bX)\bX^{\tp}\} = \bOmega$ with rank $p$.
    \item\label{Chapter3-gmm-Cond:beta} The true coefficient function $\bbeta_{0}(s)$ is three-times continuously differentiable and $\Sigma_{\bx}(s, s')$ are twice continuously differentiable. 
    \item\label{Chapter3-gmm-Cond:donsker} $\{U(s): s\in [0,1]\}$ and $\{\fM(\bX) U(s): s\in [0,1]\}$ are Donsker class, where $\bX \subset \fM(\bX)$.
    \item\label{Chapter3-gmm-Cond:g} 
        \begin{enumerate}
            \item $\lim_{s \searrow1} \E\{ |g_{l}\{\bgamma(s-1)\} - g_{l}\{\bgamma(0)\}|^{2}\} = 0$ for $l = 1, \cdots, 2q$
            \item $\lim_{s\nearrow 1} \E\{ |g_{l-1}\{\bgamma(s)\} - g_{l}\{\bgamma(0)\}|^{2}\} = 0$ for $l = 2, \cdots, 2q$.
        \end{enumerate}
    \item\label{Chapter3-gmm-Cond:covg} All second-order partial derivatives of $\bC(s, s')$ exist and are bounded on the support of the functional domain.
    \item\label{Chapter3-gmm-Cond:eigen} For some $\kappa_{0} \geq 1$ and $\alpha^{-1} = o\left(\sum\limits_{k=1}^{\kappa_{0}}\lambda_{k}^{-1}/\sum\limits_{k=\kappa_{0}+1}^{\infty}\lambda_{k} \right)$. 
    \item\label{Chapter3-gmm-Cond:limit} The numbers of individuals and functional grid-points are growing to infinity such that $h\rightarrow 0$ and $rh\rightarrow\infty$ as $n\to\infty$. 
    For $a > 2$,  $h^{-4}(\log n/n)^{1-2/a} \rightarrow 0$ and
    $|\log h|^{1-2/a}/h \leq r^{1-2/a}$ for $a \in (2,4)$.
    %$(h^{4} + h^{3}/r + h^{2}/r^{2})^{-1}(\log n/n)^{1-2/a} \rightarrow 0$ 
    %as $n \rightarrow \infty$.
\end{enumerate}
\begin{remark}
Conditions \ref{Chapter3-gmm-Cond:kernel} and \ref{Chapter3-gmm-Cond:density} are commonly used in the literature of non-parametric regression. 
The bound condition for the density function in \ref{Chapter3-gmm-Cond:density} of functional points is standard for random design. 
Similar results can be obtained for fixed design where the grid-points are pre-fixed according to the design density $\int_{0}^{s_{j}}f(s)dt = j/r$ for $j = 1, \cdots, r$, for $r \geq 1$.
The condition \ref{Chapter3-gmm-Cond:X} is similar to that in \cite{li2010uniform} which requires the bound on the higher order moment of $\bX$.  Moreover the condition on rank of $\bOmega$ is required for identification of the functional coefficient and its first order derivatives \citep{su2013local}. 

To obtain the asymptotic expression of $\widehat{\bbeta}(s)$, observed for fixed sample size, there exists $\kappa_{0}$ such that $k \leq \kappa_{0}$, $\lambda_{k}^{2}$ is much larger than $\alpha$, thus, the ratio $\lambda_{k}/(\lambda_{k}^{2} + \alpha) \approx \lambda_{k}^{-1}$. 
    On the other hand, if $k > \kappa_{0}$, $\lambda_{k}^{2} << \alpha$, as a result, the fraction $\lambda_{k}/(\lambda_{k}^{2} + \alpha)$ can be approximately written as $\lambda_{k}/\alpha$. 
    Therefore, 
    by the assumption \ref{Chapter3-gmm-Cond:eigen}, we can write, for $s\in \sS$,
    $
        \sum\limits_{k = 1}^{\kappa_{0}}\lambda_{k}^{-1}\bphi_{k}(s)\bphi_{k}(s')^{\tp}
        + \alpha^{-1}\sum\limits_{k = \kappa_{0}+1}^{\infty}\lambda_{k}\bphi_{k}(s)\bphi_{k}(s')^{\tp} = \sum\limits_{k=1}^{\kappa_{0}}\lambda_{k}^{-1}\bphi_{k}(s)\bphi_{k}(s')^{\tp}\left\{ 1+ o(1)\right\}.
    $
Condition \ref{Chapter3-gmm-Cond:limit} provide the range of bandwidth. Under the fixed sampling design, this condition can be weakened, see \cite{zhu2012multivariate}. Due to the limited space, detailed remarks on conditions (C4) and (C6) are included in Section S3 of the supplemental file.
\end{remark}
\par
The following Theorem provides weak convergence of the initial estimates in the above Step-I. Define $\bSigma_{\bx}^{*}(s_{0}, s_{0}) = \lim\limits_{n\rightarrow\infty}\frac{1}{n}\sum\limits_{i=1}^{n}\E\{\bX_{i}\bX_{i}^{\tp}\Sigma_{\bx_{i}}(s_{0}, s_{0})\}$.

\begin{theorem}
\label{Chapter3-gmm-Theorem:init}
Let $\nu_{a, b} = \int t^{a}K^{b}(t)dt$. Under conditions  \ref{Chapter3-gmm-Cond:kernel}-\ref{Chapter3-gmm-Cond:donsker}, and \ref{Chapter3-gmm-Cond:limit},
% \ref{Chapter3-gmm-Cond:kernel}, \ref{Chapter3-gmm-Cond:density}, 
% \ref{Chapter3-gmm-Cond:X}, 
% \ref{Chapter3-gmm-Cond:beta}, \ref{Chapter3-gmm-Cond:donsker}, and \ref{Chapter3-gmm-Cond:limit} 
$$\Big\{
    \sqrt{n}\left(
        \breve{\bbeta}(s_{0}) - \bbeta_{0}(s_{0}) - 0.5h^{2}\nu_{21}\ddot{\bbeta}_{0}(s_{0})
    \right)\Big.: s_{0} \in \sS\Big\}$$ weakly converges to a mean zero Gaussian process with a covariance matrix 
    $\bOmega_{\bx}^{-1}\bSigma_{\bx}^{*}(s_{0}, s_{0})\bOmega_{\bx}^{-1}$
    %$\Sigma(s_{0}, s_{0})\bOmega_{\bx}^{-1}$,
where $\bOmega_{\bx} = \E\{\bX\bX^{\tp}\}$. 
% and $\Sigma(s_0,s_0)$ is the variance of $U(s_0)$.
%Moreover, $\sup_{s_{0}\in \sS}|\breve{\bbeta}(s_{0}) - \bbeta_{0}(s_{0})| = O(\delta_{n1}(h) + h)$ almost surely. 
\end{theorem}
Next, we study the convergence rates of the estimated eigen-components based on the proposed lining-up method. For simplicity, define, $\delta_{n1}(h) = \left\{(1+(hr)^{-1})\log n/n \right\}^{1/2}$ and 
$\delta_{n2}(h) = \left\{(1+(hr)^{-1} + (hr)^{-2})\log n/n \right\}^{1/2}$.
The following lemma is output of the asymptotic expansion of eigen-components of an estimated covariance function.
\begin{lemma}
\label{lemma1}
Under assumptions \ref{Chapter3-gmm-Cond:kernel}-\ref{Chapter3-gmm-Cond:X}, \ref{Chapter3-gmm-Cond:g}-\ref{Chapter3-gmm-Cond:limit},
% \ref{Chapter3-gmm-Cond:kernel}, \ref{Chapter3-gmm-Cond:density}, 
% \ref{Chapter3-gmm-Cond:X},
% \ref{Chapter3-gmm-Cond:g}, 
% \ref{Chapter3-gmm-Cond:covg}, 
% \ref{Chapter3-gmm-Cond:eigen}, and \ref{Chapter3-gmm-Cond:limit}
the following convergence holds almost surely for any finite-dimensional mean-zero function $\bg(s)$.
\begin{enumerate}
    \item $\left|\widehat{\lambda}_{k}  - \lambda_{k} \right| = O\{h^{2} + \delta_{n1}(h) + \delta_{n2}(h)\}$
    \item $\sup_{s_{0}\in \sS}\left| \widehat{\bphi}_{k}(s_{0}) - \bphi_{k}(s_{0}) \right| = O\{h^{2} + \delta_{n1}(h) + \delta_{n2}^{2}(h)\}$
\end{enumerate}
for all $k = 1, \cdots, \kappa_{0}$.
\end{lemma}
We skip the proof of the above lemma as it is well-developed in the literature of functional data analysis including \citet{hall2004generalized, hall2006properties, li2010uniform}. If $r^{-1}=O(\{n/\log n\}^{1/4})$
and $h=O(\{n/\log n\}^{-1/4})$, then Lemma \ref{lemma1} implies that both eigenvalues $\widehat{\lambda}_{k}$
and eigenfunctions $\widehat{\bphi}_{k}(s_{0})$ converge at the order of $O\{(\log n/n)^{1/2}\}$.
Next, we show the asymptotic results of the proposed estimation. 
\begin{theorem}
\label{Chapter3-gmm-Theorem:final}
Let $\nu_{a, b} = \int t^{a}K^{b}(t)dt$. Under the conditions \ref{Chapter3-gmm-Cond:kernel}-\ref{Chapter3-gmm-Cond:limit},
% \ref{Chapter3-gmm-Cond:kernel},
% \ref{Chapter3-gmm-Cond:density}, 
% \ref{Chapter3-gmm-Cond:X}, 
% \ref{Chapter3-gmm-Cond:beta}, 
% \ref{Chapter3-gmm-Cond:donsker}, 
% \ref{Chapter3-gmm-Cond:g}, 
% \ref{Chapter3-gmm-Cond:covg},
% \ref{Chapter3-gmm-Cond:eigen}, and
% \ref{Chapter3-gmm-Cond:limit} 
for the proposed local-linear GMM estimator $\widehat{\bbeta}(s)$, the following results hold.
$$\left\{
    \sqrt{n}\left(
        \widehat{\bbeta}(s) - \bbeta_{0}(s) - 0.5h^{2}\nu_{21}\ddot{\bbeta}(s))
    \right)
    : s \in \sS
\right\}$$ weakly converges to a mean zero Gaussian process with a covariance function 
$\sA(s_{0}, s_{0}) = \sB^{-1}(s_{0}, s_{0})\bOmega^{\tp}\bC_{\kappa_{0}, 11}^{-1}(s_{0}, s_{0})\bSigma_{\fM}^{*}(s_{0}, s_{0})\bC^{-1}_{\kappa_{0}, 11}(s_{0}, s_{0})\bOmega\sB^{-1}(s_{0}, s_{0})$, where
$\sB(s_{0}, s_{0}) = \bOmega^{\tp}\bC_{\kappa_{0}, 11}^{-1}(s_{0}, s_{0})\bOmega$,
% $\sA(s, s) = 
% \left(\bOmega\bC_{\kappa_{0}, 11}^{-1}(s, s)
% \bOmega^{\tp}\right)^{-1}
% \bOmega\bC_{\kappa_{0}, 11}^{-1}(s, s)\bSigma(s, s)
% \bC_{\kappa_{0}, 11}^{-1}(s, s)\bOmega
% \left(\bOmega\bC_{\kappa_{0}, 11}(s, s)^{-1}\bOmega^{\tp}\right)^{-1}.$
$\bOmega$ is defined in condition (C3), $\bC_{\kappa_{0}, 11}^{-1}$ is given by
\begin{equation}
 \bC_{\kappa_{0}}^{-1}(s, s') = \sum_{k=1}^{\kappa_{0}}\lambda_{k}^{-1}\bphi_{k}(s)\bphi_{k}(s')^{\tp} = \begin{pmatrix}
    \bC_{\kappa_{0}, 1, 1}^{-1}(s, s') & 0\\
    0 & \bC_{\kappa_{0}, 2, 2}^{-1}(s, s')
    \end{pmatrix}
\end{equation}
and 
\begin{equation}
    \bSigma_{\fM}^{*}(s_{0}, s_{0}) = \lim_{n \rightarrow \infty}\frac{1}{n}\sum_{i=1}^{n}\E\{\fM(\bX_{i})\fM(\bX_{i})^{\tp}\Sigma_{\bx_{i}}(s_{0}, s_{0})\}.
\end{equation}
\end{theorem}

\begin{remark}
The asymptotic variance-covariance of $\widehat{\bbeta}(s)$  depends on the choices of IVs in $\bOmega$. The suggested
choice of IVs in Step-II.A of section 3.2 maybe optimal in the sense that it minimizing the variance-covariance matrix of $\hat{\bbeta}(s)$ among the class of all local-linear GMM estimators. Please refer to the supplemental file for a detailed discussion.
\end{remark}

\section{Simulation studies}
\label{Chapter3-gmm-Section:simulation}
We conduct numerical studies to compare finite sample performance under different correlation structures and heterogeneity conditions. 
% Based on the number of unknown coefficient functions, we consider the following two situations. 
%\subsection*{Scenario A (Single covariate):}
Data are generated from the model $Y_{i}(s) = X_{i}\bbeta(s) + U_{i}(s)$,
where we generate trajectories observed at $r$ spatial locations for the $i$-th curve, $i = 1, \cdots, n$. 
Assume that the functional fixed effect is $\beta(s) = \cos(2\pi s)$ and the corresponding fixed effect covariate is generated from a normal distribution with unit mean and variance. 
The error process is generated as $U_{i}(s) = \xi_{1}(X_{i})\psi_{1}(s) + \xi_{2}(X_{i})\psi_{2}(s)$
where $\xi_{1}(X_{i})$ and $\xi_{2}(X_{i})$ are independent central normal random variables with variance $3\sigma^{2}(X_{i})\theta_{0}^{2}$ and $1.5\sigma^{2}(X_{i})\theta_{0}^{2}$ 
where $\theta_{0}$ is determined by the relative importance of random error signal-to-noise ratio, denoted as $\SNR_{\theta}$ which is interpreted as the ratio of the standard deviation of the additive prediction without noise divided by the standard error of the random noise function. For example, $\SNR_{\theta} = 0.5$  means that the contribution of each functional random noise to the variability in $Y(s)$ is about twice of that of the fixed effect \citep{scheipl2015functional}.
Here, we use scaled orthonormal functions $\psi_{1}(s) \propto (1.5 - \sin(2\pi s) - \cos(2\pi s) )$ and $\psi_{2}(s) \propto \sin(4\pi s)$; due to orthonormality, the proportionality constants can easily be determined. Contributions to the conditional variances in $\xi_{k}(X)$ are specified as (S.0) $\sigma^{2}(x) = 1$ (homoskedastic), (S.1) $\sigma^{2}(x) = (1+x^{2}/2)^{2}$, (S.2) $\sigma^{2}(x) = \exp(1+x^{2}/2)$, (S.3) $\sigma^{2}(x) = \exp(1+ |{x}| + x^{2})$ and (S.4) $\sigma^{2}(x) = (1+|x|/2)^{2}$.
We sample the trajectories at $r$ equidistant spatial points $\{ s_{1}, \cdots, s_{r}\}$ on $[0, 1]$. Let $s_{i} = (j-0.5)/r$ for $j=1, \cdots, r$ for $i$-th curve. The number of spatial points is assumed to be $r=200$ for each case. 
We set number of trajectories $n \in \{30, 50, 100, 200, 500\}$ and the controlling parameter $\theta_{0}$ is determined using signal-to-noise ratio, $\SNR_{\theta}$ which is assumed to be either 0.5 or 1. Here, we perform 500 simulation replicates. To make it consistent with theoretical results and numerical examples, we use ``\texttt{FPCA}" function in R which is available in \texttt{fdapace} package \citep{fdapace} to estimate the eigen-functions. 
Bandwidths are selected using five-fold generalized cross-validation in all situations  
and for estimation, the Epanechnikov kernel $K(x) = 0.75(1-x^{2})_{+}$ is used; where $(a)_{+} = \max(a, 0)$. 
Accuracy of the parameter estimation is assessed using integrated mean square error and integrated mean absolute error,
which for the $b$-th replication is defined as $\IMSE_{b} =  \left[\sum\limits_{j=1}^{r}\left(\widehat{\bbeta}_{b}(s_{j}) - \bbeta(s_{j})\right)^{2}\Delta(s_{r})\right]$ and
 $\IMAE_{b} =  \left[\sum\limits_{j=1}^{r}|\widehat{\bbeta}_{b}(s_{j}) - \bbeta(s_{j})|\Delta(s_{r})\right]$ respectively,
with $\Delta(s_{j}) = s_{j} - s_{j-1}$ where $s_{0} = 0$ and $s_{1} < \cdots < s_{r}$ are the observed points over the support set of observational points. Let $h^{*}$ be the optimal bandwidth obtained from five-fold cross-validation, which when multiplied by a constant within a certain range provides improved results. According to the comments under Lemma \ref{lemma1}, undersmoothing is needed, so we use $\widehat{\bbeta}$ corresponding to bandwidth $0.75h^{*}$ for our numerical studies. Similar strategies have also been applied in \cite{cai2006functional}, \cite{wei2017heteroskedasticity} and \cite{Wangetal2017}. The constant 0.75 is not an optimal choice but it was recommended based on our numerical experiments.

We present Tables \ref{Chapter3-gmm-Table:univariate-0.5} {here} and \ref{Chapter3-gmm-Table:univariate-1} in the supplementary document where IMSEs and IMAEs are averaged over 500 replications for each situation. In parentheses, the corresponding standard deviations are reported. We denote by \texttt{LLE} and \texttt{LLGMM} the local-linear smoothing estimator described in Step-I and local-linear GMM with weight matrix proposed in Step-III in Section \ref{Chapter3-gmm-Section:method} respectively. 
In addition, we have compared the proposed method with that of \citet{wei2017heteroskedasticity}, setting the spatial autoregressive parameter to zero. This approach is referred to as \texttt{LLWS} in Tables \ref{Chapter3-gmm-Table:univariate-0.5} and \ref{Chapter3-gmm-Table:univariate-1}.
As expected, for all situations, the IMSE and IMAE are significantly reduced if we increase sample size and/or $\SNR_{\theta}$. For the homoskedastic case, the error rates of \texttt{LLE} are similar for \texttt{LLGMM} but under the presence of heteroskedasticity of unknown form, our proposed method outperforms. More simulation results with multiple covariates are included in Section \ref{sec:trisim} of the supplementary material.

%\begin{landscape}

\begin{table}
\vspace{-1.5cm}
\tabcolsep 3pt
    \centering
    \caption{Comparison among the proposed LLGMM with the local linear estimator (LLE) and \citet{wei2017heteroskedasticity}'s approach (LLWS) for $\SNR_{\theta} = 0.5$.}
    \label{Chapter3-gmm-Table:univariate-0.5}
    \begin{tabular}{rrrrrrrrrrr}
    \toprule
     & \multicolumn{2}{c}{n = 30} &
        \multicolumn{2}{c}{n = 50} &
        \multicolumn{2}{c}{n = 100} &
        \multicolumn{2}{c}{n = 200} &
        \multicolumn{2}{c}{n = 500}\\
    \midrule    
     Method & IMSE & IMAE &  
                    IMSE & IMAE & 
                    IMSE & IMAE &
                    IMSE & IMAE &
                    IMSE & IMAE \\
    \midrule            
    %homoskedastic
    &\multicolumn{10}{c}{Case: S.0}\\
    \texttt{LLE} & 0.0626 & 0.1864 & 0.0372 & 0.1429 & 0.0189 & 0.1016 & 0.0099 & 0.0737 & 0.0041 & 0.0472 \\ 
    & \scriptsize (0.0619) & \scriptsize (0.0951) & \scriptsize (0.0398) & \scriptsize (0.0757) & \scriptsize (0.0200) & \scriptsize (0.0540) & \scriptsize (0.0108) & \scriptsize (0.0391) & \scriptsize (0.0044) & \scriptsize (0.0256)\\

    {\blue\texttt{LLWS}}& 0.0630 & 0.1865 & {\blue 0.0375} & {\blue 0.1435} & {\blue 0.0191} & {\blue 0.1022} & {\blue 0.0099} & {\blue 0.0737} & {\blue 0.0041}  & {\blue 0.0471}\\ 
    & \scriptsize (0.0621) & \scriptsize (0.1168) & {\blue \scriptsize (0.0399)} & {\blue \scriptsize (0.1435)} & {\blue \scriptsize (0.0205)} & {\blue \scriptsize (0.1022)} & {\blue \scriptsize (0.0109)} & {\blue \scriptsize (0.0737)} & {\blue \scriptsize (0.0044)} & {\blue \scriptsize (0.0471)}\\
    
    \texttt{LLGMM} &  0.0630 & 0.1865 & 0.0388 & 0.1460 & 0.0198 & 0.1038 & 0.0100 & 0.0740 & 0.0042 & 0.0474 \\ 
    & \scriptsize (0.0627) & \scriptsize (0.1865) & \scriptsize (0.0408) & \scriptsize (0.1460) & \scriptsize (0.0208) & \scriptsize (0.1038) & \scriptsize (0.0109) & \scriptsize (0.0740) & \scriptsize (0.0046) & \scriptsize (0.0474)\\
    \midrule

    &\multicolumn{10}{c}{Case: S.1}\\
    \texttt{LLE} & 0.1513 & 0.2909 & 0.0939 & 0.2271 & 0.0516 & 0.1679 & 0.0261 & 0.1189 & 0.0106 & 0.0766 \\
    & \scriptsize (0.1509) & \scriptsize (0.1528) & \scriptsize (0.1019) & \scriptsize (0.1225) & \scriptsize (0.0541) & \scriptsize (0.0906) & \scriptsize (0.0288) & \scriptsize (0.0647) & \scriptsize (0.0109) & \scriptsize (0.0402)\\
    
    {\texttt{LLWS}} & {\blue 0.1366} & {\blue 0.2754} & {\blue 0.0816} & {\blue 0.2123} & {\blue 0.0443} & {\blue 0.1556} & {\blue 0.0227} & {\blue 0.1109} & {\blue 0.0091} & {\blue 0.0708} \\ 
    & \scriptsize (0.1223) & \scriptsize (0.2033) & {\blue \scriptsize (0.0848)} & {\blue \scriptsize (0.2123)} & {\blue \scriptsize (0.0461)} & {\blue \scriptsize (0.1556)} & {\blue \scriptsize (0.0249)} & {\blue \scriptsize (0.1109)} & {\blue \scriptsize (0.0094)} & {\blue \scriptsize (0.0708)}\\
    
    \texttt{LLGMM} & 0.1187 & 0.2579 & 0.0585 & 0.1820 & 0.0292 & 0.1262 & 0.0135 & 0.0867 & 0.0050 & 0.0517 \\ 
    & \scriptsize (0.1107) & \scriptsize (0.2579) & \scriptsize (0.0574) & \scriptsize (0.1820) & \scriptsize (0.0308) & \scriptsize (0.1262) & \scriptsize (0.0143) & \scriptsize (0.0867) & \scriptsize (0.0053) & \scriptsize (0.0517)\\
    \midrule
    &\multicolumn{10}{c}{Case: S.2}\\
    \texttt{LLE} & 0.2026 & 0.3407 & 0.1381 & 0.2810 & 0.0812 & 0.2169 & 0.0468 & 0.1632 & 0.0209 & 0.1094 \\ 
    & \scriptsize (0.1854) & \scriptsize (0.1732) & \scriptsize (0.1308) & \scriptsize (0.1423) & \scriptsize (0.0727) & \scriptsize (0.1047) & \scriptsize (0.0420) & \scriptsize (0.0787) & \scriptsize (0.0174) & \scriptsize (0.0516)\\

    {\texttt{LLWS}} & 0.3817 & 0.3520 & {\blue 0.0804} & {\blue 0.2105} & {\blue 0.0372} & {\blue 0.1409} & {\blue 0.0164} & {\blue 0.0902} & {\blue 0.0048} & {\blue 0.0486} \\ 
    & \scriptsize \blue (0.1551) & \scriptsize (0.2373) & \scriptsize\blue (0.0792) & \scriptsize\blue  (0.2105) & \scriptsize\blue (0.0397) & \scriptsize\blue (0.1409) & \scriptsize\blue (0.0202) & \scriptsize\blue (0.0902) & \scriptsize\blue (0.0059) & \scriptsize\blue (0.0486)\\
  
    \texttt{LLGMM} & 0.1427 & 0.2812 & 0.0557 & 0.1462 & 0.0134 & 0.0817 & 0.0045 & 0.0471 & 0.0015 & 0.0262 \\
    & \scriptsize (0.1407) & \scriptsize (0.2812)  & \scriptsize (0.3642) & \scriptsize (0.1462) & \scriptsize (0.0169) & \scriptsize (0.0817) & \scriptsize (0.0061) & \scriptsize (0.0471) & \scriptsize (0.0031) & \scriptsize (0.0262)\\
    \midrule
    &\multicolumn{10}{c}{Case: S.3}\\
    \texttt{LLE} & 0.2569 & 0.3996 & 0.1762 & 0.3330 & 0.1018 & 0.2538 & 0.0581 & 0.1913 & 0.0265 & 0.1291 \\ 
    & \scriptsize (0.1933) & \scriptsize (0.1679) & \scriptsize (0.1240) & \scriptsize (0.1299) & \scriptsize (0.0688) & \scriptsize (0.0950) & \scriptsize (0.0365) & \scriptsize (0.0644) & \scriptsize (0.0163) & \scriptsize (0.0458)\\
    
    {\texttt{LLWS}} & 0.0781 & 0.1763 & \blue 0.0328 & \blue 0.1069 & \blue 0.0126 & \blue 0.0619 & \blue 0.0055 & \blue 0.0376 & \blue 0.0023 & \blue 0.0243 \\ 
    & \scriptsize (0.0738) & \scriptsize (0.1860) & \scriptsize\blue (0.0506) & \scriptsize\blue (0.1069) & \scriptsize\blue (0.0211) & \scriptsize\blue (0.0619) & \scriptsize\blue (0.0100) & \scriptsize\blue (0.0376) & \scriptsize\blue (0.0039) & \scriptsize\blue (0.0243)\\
    
    \texttt{LLGMM} & 0.0746 & 0.1275 & 0.1067 & 0.0600 & 0.0021 & 0.0201 & 0.0004 & 0.0093 & 0.0003 & 0.0064 \\ 
    & \scriptsize (0.1178) & \scriptsize (0.1275) & \scriptsize (0.2031) & \scriptsize (0.0600) & \scriptsize (0.0125) & \scriptsize (0.0201) & \scriptsize (0.0019) & \scriptsize (0.0093) & \scriptsize (0.0018) & \scriptsize (0.0064)\\
    \midrule
    &\multicolumn{10}{c}{Case: S.4}\\
    \texttt{LLE} & 0.0971 & 0.2330 & 0.0588 & 0.1798 & 0.0309 & 0.1298 & 0.0158 & 0.0928 & 0.0065 & 0.0596 \\ 
    & \scriptsize (0.0962) & \scriptsize (0.1198) & \scriptsize (0.0633) & \scriptsize (0.0953) & \scriptsize (0.0332) & \scriptsize (0.0694) & \scriptsize (0.0176) & \scriptsize (0.0500) & \scriptsize (0.0068) & \scriptsize (0.0317)\\
    
    {\texttt{LLWS}} & 0.0958 & 0.2306 & \blue 0.0577 & \blue 0.1782 & \blue 0.0303 & \blue 0.1285 & \blue 0.0155 & \blue 0.0920 & \blue 0.0063 & \blue 0.0589 \\
    & \scriptsize\blue (0.0951) & \scriptsize\blue (0.2231) & \scriptsize\blue (0.0614) & \scriptsize\blue (0.1782) & \scriptsize\blue (0.0327) & \scriptsize\blue (0.1285) & \scriptsize\blue (0.0172) & \scriptsize\blue (0.0920) & \scriptsize\blue (0.0067) & \scriptsize\blue (0.0589)\\
    
    \texttt{LLGMM} & 0.0958 & 0.2306 & 0.0576 & 0.1792 & 0.0303 & 0.1287 & 0.0153 & 0.0914 & 0.0063 & 0.0585 \\ 
    & \scriptsize (0.0960) & \scriptsize (0.2306) & \scriptsize (0.0584) & \scriptsize (0.1792) & \scriptsize (0.0319) & \scriptsize (0.1287) & \scriptsize (0.0170) & \scriptsize (0.0914) & \scriptsize (0.0066) & \scriptsize (0.0585)\\
    \bottomrule
    \end{tabular}
\end{table}
%\end{landscape}

\section{Real data analysis}
\label{Chapter3-gmm-Section:data-analyis}
For illustrating the application of our proposed method and the estimation procedure therein, we use \texttt{Apnea-data} to understand white matter structural alterations using diffusion tensor imaging (DTI) in obstructive sleep apnea (OSA) patients \citep{xiong2017brain}.
The data consist of 29 male patients (age range: 30-55 years) who underwent the study for the diagnosis of continuous positive airway pressure (CPAP) therapy.
DTI was performed at 3T, followed by the analysis using tract-based spatial statistics to investigate the difference in fractional anisotropy (FA) and other diffusion properties between the groups based on lapses. FA measures the degree of anisotropy of a diffusion process. The image acquisition is as follows: Images are recorded on a 3T MRI scanner using a commercial 32-channel head coil. 
An axial T1-weighted image of the brain (3D-BRAVO) is collected with repetition time (TR) = 12ms, echo time (TE) = 5.2ms, flip angle = $13^{\circ}$, inversion time = 450 ms, matrix = $384\times 256$, voxel size = $1.2\times 0.57\times 0.69$mm and scan time = 2 min 54 sec. DTI are obtained in the axial plane using a spin-echo echo planner imaging sequence with TR = 4500ms, TE = 89.4ms, field of view = $20\times 20$cm$^{2}$, matrix size = $160\times 132$, slice thickness = 3mm, slice spacing = 1mm, b-values = 0, 1000 s/mm$^{2}$.
\par
FA varies systematically along the trajectory of each white matter fascicle. 
Several pre- and post-processing steps were performed by the FSL software. The brain was extracted using brain segmentation tools. After generating FA maps using FMRIB diffusion toolbox, images from all individuals were aligned to an FA standard template through non-linear co-registration. 
The Johns Hopkins University (JHU) white matter tractography atlas was used as a standard template for white matter parcellation with 50 regions of interest (ROIs). All imaging parameters were calculated by averaging the voxel values in each ROI. See \citet{xiong2017brain} for more details on the data and the preprocessing steps.
\par
For each subject, we calculate the similarity matrix $\bC$ with dimension $50 \times 50$. The $(k,l)$-th element of the matrix $\bC$ is defined as $c_{kl} = |y_{k} - y_{l}|$ where $y_{k}$ is the measure of FA associated with $k$-th ROI. 
For simplicity, we scale the similarity matrix such that the range of the elements of the matrix is $[0,1]$. 
To create the network, we threshold each similarity matrix to build an adjacency matrix $\bG$ with elements $\{1,0\}$ depending on whether the similarity values exceed the threshold or not.
Since this threshold controls the topology of the data, we contract the adjacent matrix over a set of threshold parameters from 0.01 to 0.99, and this set is denoted as $\sS$ with cardinality $99$. The role of the threshold is to investigate the graph networks formed by the ROIs with different degrees of anisotropy. When the 
threshold value is small, most ROIs are connected with edges in 
the graph. In this case, the focus is more on global and entire 
brain regions. When the threshold value is high, there will be 
fewer ROIs connected with edges. In this case, the focus is more on a few brain regions whose degrees of anisotropy are more dissimilar from each other. In summary, the threshold allows us to tune our focus on different collections of connected ROIs, and it allows one to zoom in and zoom out. 
\par
A popular measure of connectivity in a given graph is average path length (APL) which is defined as the average number of steps along the shortest path for all possible pairs of the network nodes. Therefore it measures the efficacy of the information on a network \citep{albert2002statistical}. 
For a series of threshold parameters $(s)$, we observe APL for FA as shown in Figure \ref{Chapter3-GMM-Fig:FA-APL}. 
Scientists are often interested to know the association of APL of the network generated from FA with a set of covariates of interests such as age and lapses score. \par
We fit the model (\ref{Chapter3-gmm-Eq:model})
to APL that is collected over continuous spatial domains (viz, thresholds) from all individuals in which $\bX_{i}$ included the clinical variables such as lapses, and age. We discard the subjects from the analysis with missing clinical variables and therefore the sample size $n = 27$. Here we used three-fold cross-validation to obtain the tuning parameters and the fraction of variance explained (FVE) is set at 0.99. 
In Figure \ref{Chapter3-GMM-Fig:FA-beta}, we present the estimated coefficient functions corresponding to age and number of lapses associated with APL where it can be observed that the coefficient of network property is negative with age but positive with lapses counts. Moreover, the effect for the APL is found to be increasing when the threshold is small to moderate and decreasing at moderate to large threshold; whereas, the effect of APL is more-or-less similar up to the larger values of threshold, and after that, it significantly decreases. 

\begin{figure}[bhtp!]
    \centering
    \includegraphics[width = 0.3\textwidth]{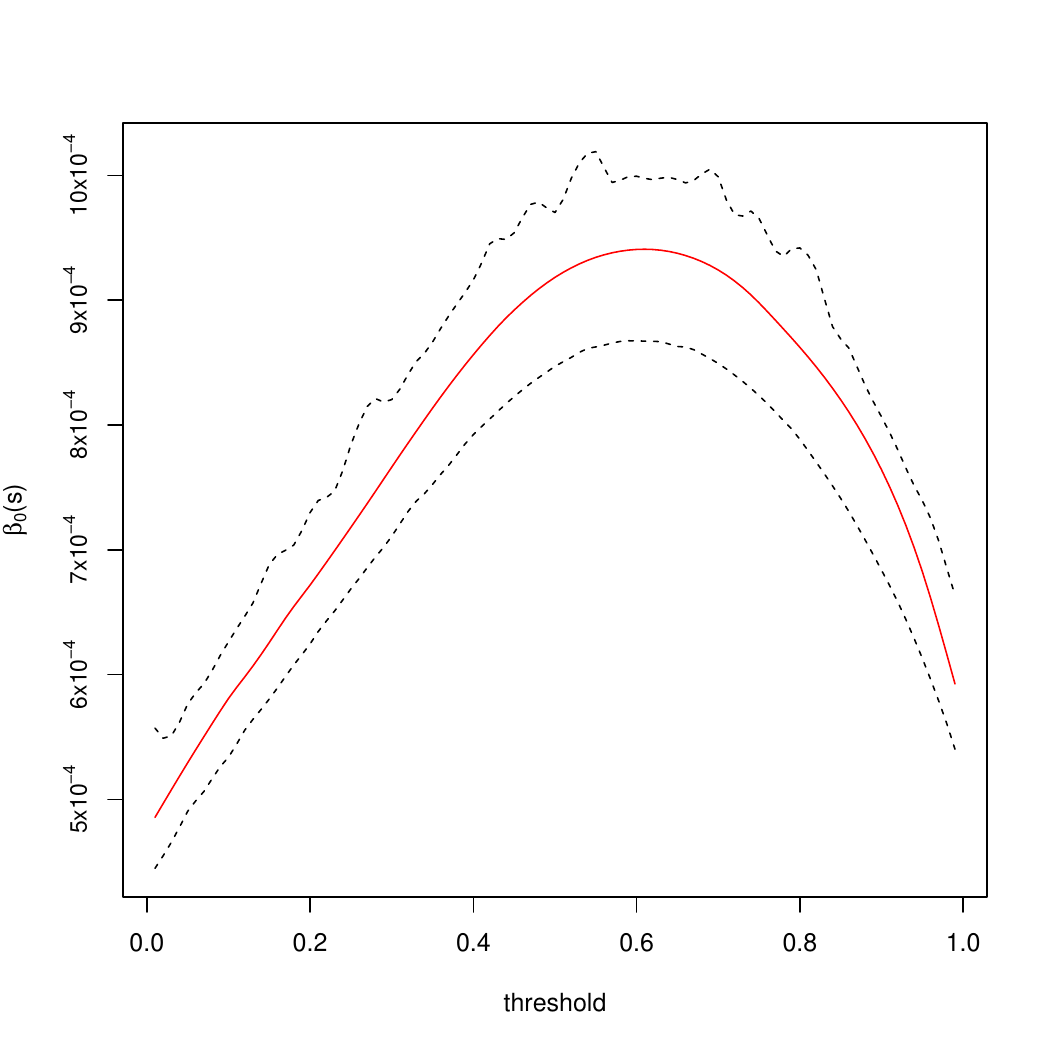}
    \includegraphics[width = 0.3\textwidth]{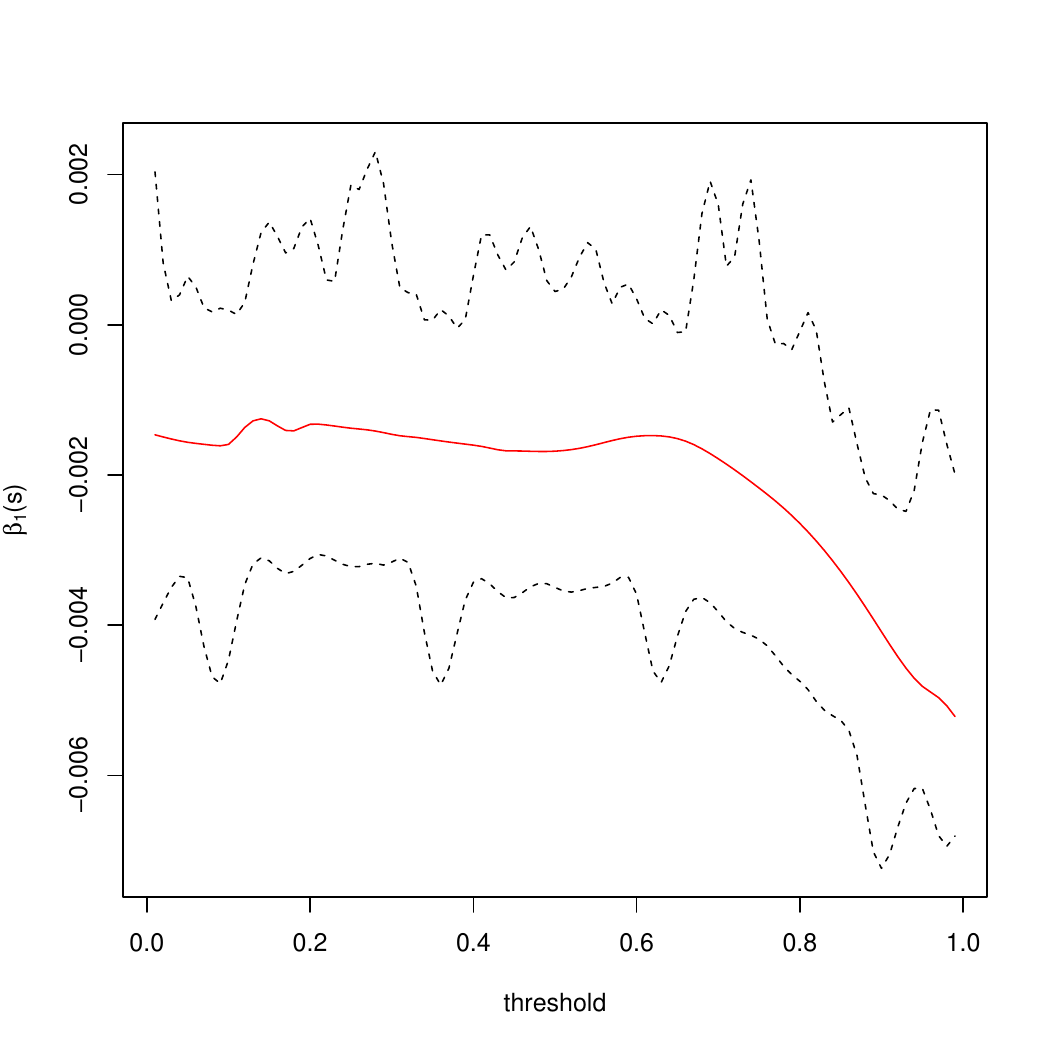}
    \includegraphics[width = 0.3\textwidth]{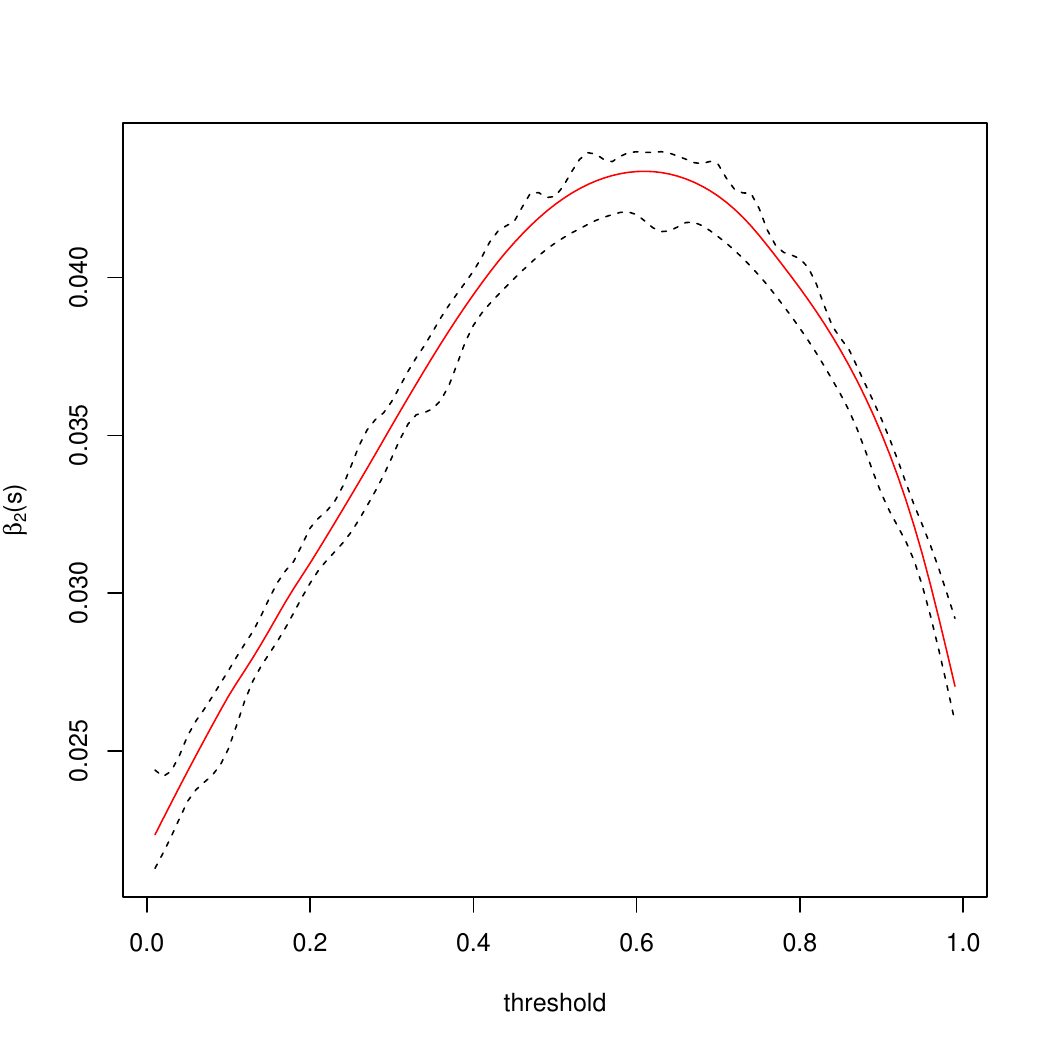}
    \caption{\texttt{Apnea-data} analysis: Plots of estimated intercept (left), coefficient functions of age (middle) and number of lapses (right) for average path length associated with Fractional Anisotropy (FA) in DTI analysis. The red line is the estimates of the functions and the dotted black lines are point-wise 95\% confidence intervals.}
    \label{Chapter3-GMM-Fig:FA-beta}
\end{figure}

In \cite{xiong2017brain}, the authors firstly divided patients into a nonsleepy group (lapses $\leq 5$) and a sleepy group (lapses $>$ 5), and then compared FA values between the sleepy and nonsleepy group using two sample tests for all the ROIs. The authors found that ``the alterations in FA of individual fiber tracts occurred mainly in the internal/external capsule, corona radiata, corpus callosum, and sagittal stratum regions''.
Our finding in this paper is consistent with that in \cite{xiong2017brain}. First, the coefficient function for the number of lapses is significant for all the threshold values. This indicates the association between the number of lapses and FA values. Second, we observe that the coefficient function for the number of lapses achieves its maximum when the threshold is around 0.65. The brain regions contributing most to the APL for the brain network when the threshold is 0.65, including corpus callosum, cerebral peduncle, internal/external capsule, corona radiata, cingulum  hippocampus and tapetum, have the largest correlation with the number of lapses. These brain regions found by the proposed method include those found in \cite{xiong2017brain}.

\section{Discussion}
\label{Chapter3-gmm-Section:discussion}
 In this article, we propose an efficient estimation procedure for varying-coefficient models which is commonly used in neuroimaging and econometrics. 
 Our procedure stands out for its efficiency in handling the integration of heteroskedasticity within the realm of functional data analysis. To the best of our knowledge, this is the first attempt to incorporate such conditions into the model. Such a model is therefore equipped with a more complex relationship between the functional response and real-valued covariates. 
Additionally, our method is easy to implement in a wide range of applications due to the multi-stage structure of the algorithm. The applicability of the proposed method is illustrated by numerical studies.

%%%%%%%%%%%%%%%%%%%%%%%%%%%%%%%%%%%%%%%%%%%%%%%%%%%%%%%%%%%%%%%%%%%%%%%%%%%%%%%%%%%%%%%%%%%%%%%%%%%%%%%%%%%%%%%%%%%%%%%%%%%%
\section*{Supplementary material}
The online supplementary material contains the proposed algorithm, the extension of the proposed method for multivariate functional domain, comments on assumptions, additional simulation results, proofs of the theorems presented in Section \ref{Chapter3-gmm-Section:theory}, and a discussion on the choice of IVs.

\section*{Acknowledgements}
The research was partially supported by an NIH grant R03NS128450 and an NSF grant FRG-2152070. 
%The authors thank Editor Prof. John Stufken, the Associate Editor, and the two referees for their constructive comments, which significantly improved the paper.

\newpage
\appendix
\section*{Supplementary Materials}
\addcontentsline{toc}{section}{Supplementary Materials}

\section{Algorithm}
The outline of the proposed method in Section \ref{Chapter3-gmm-Section:method} of the main article is presented in the following Algorithm \ref{Chapter3-gmm-Algo:estimation}.

\begin{algorithm}[h]
\caption{Estimation of $\bbeta(s): s\in \sS$}
\label{Chapter3-gmm-Algo:estimation}
\begin{algorithmic}[1]
\Require $\{Y_{i}(s_{j}), X_{i}, s_{j}\}$, for $j = 1, \cdots, r; i = 1, \cdots, n$
\State Calculate optimal $h$: $\widehat{h}_{init} \gets \arg\min\limits_{h\in \sH} ({nr})^{-1}\sum\limits_{i=1}^{n}\sum\limits_{r=1}^{r}\left\{Y_{ij} - \bX_{i}^{\tp}\breve{\bbeta}^{-i}(s_{r}; h)\right\}^{2}$.
\State Calculate $\breve{\bgamma}(s; \widehat{h}_{init})$.
\State Compute 
$$\bg_{i}\{\bgamma(s)\} = r^{-1}\sum\limits_{j=1}^{r}K_{\widehat{h}_{init}}(s_{j}-s)\bQ_{ij}(s; \widehat{h}_{init})\{ Y_{ij} - \bW_{ij}(s; \widehat{h}_{init})^{\tp}\breve{\bgamma}(s; \widehat{h}_{init})\}.$$
\State Determine the instrument variables $\fM(\bX)$.
\State Compute eigen-components $\widehat{\lambda}_{k}, \widehat{\bphi}_{k}(s)$ and get the value of $\alpha$ using the condition \ref{Chapter3-gmm-Cond:eigen}. 
\State Calculate optimal $h$: $\widehat{h}_{opt} \gets \arg\min_{h\in \sH} ({nr})^{-1}\sum\limits_{i=1}^{n}\sum\limits_{r=1}^{r}\left\{Y_{ij} - \bX_{i}^{\tp}\widehat{\bbeta}^{-i}(s_{r}; h)\right\}^{2}$.
\State Calculate $\widehat{\bbeta}(s; h_{opt})$.
\end{algorithmic}
\end{algorithm}

\newpage
{
\section{Outline of the proposed method in a multi-dim spatial domain}
\label{sec:extension}
In the manuscript, we focus on a functional varying coefficient model on a univariate spatial domain $\sS\in [0,1]$. In this Section, we will extend the proposed method to a multi-dimensional spatial domain defined by $\sS = [0,1]^d = \left\lbrace \abs = (s_{1}, \cdots, s_{d})^{\tp} : s_{k} \in \left[0, 1\right] ~ \forall ~ k = 1, \cdots, d \right\rbrace $. Let $\{Y_{i}(\abs), \bX_{i}\}$ for $i =1, \cdots , n$ be independent copies of $\{Y(\abs), \bX\}$. For each curve, we observe $Y(\abs)$ on the discrete spatial grid $\{\abs_{1}, \cdots, \abs_{r}\}\in [0,1]^d$ on the functional domain $\sS$.
\par
To obtain an initial estimation for the coefficient function vector
$\bbeta(\abs)=(\beta_1(\abs),\cdots,\beta_p(\abs))^{\tp}$,
we approximate $\beta_{k}(\abs)$ at $\abs_0$ by the following the Taylor expansion of $\beta_{k}(\abs)$ at $\abs_0$, for any $\abs_{l}$ in a neighbourhood of $\abs$,
$$\beta_{k}(\abs_{l}) \approx \beta_{k}(\abs_0) + \sum_{a = 1}^{d}\frac{\partial \beta_{k}(\abs)}{\partial s_{a}}(s_{la} - s_{0a})\quad\mbox{for } k=1,\cdots,p,$$
where $\abs_{l}=(s_{l1},\cdots,s_{lp})^\tp$ and 
$\abs_{0}=(s_{01},\cdots,s_{0p})^\tp$.
Let $\bh_1=(h_{11},\cdots,h_{1d})^\tp$. In matrix notations, the above Taylor series expansion of the coefficient function becomes, 
\begin{equation}
    \bbeta(\abs) \approx \bA(\abs)\bz_{\bh_{1}}(\abs_{l} - \abs)   
\end{equation}
where 
\begin{align*}
\bz_{\bh_{1}}(\abs_{l} - \abs) &= \left\{ 1, (s_{l1} - s_{01})/{h_{11}}, \cdots, 
(s_{ld} - s_{0d})/{h_{1d}} \right\}^{\tp}\\
\bA(\abs_0) &= \left[\bbeta(\abs_0), h_{11}\dot{\bbeta}_{1}(\abs_0), \cdots, h_{1d}\dot{\bbeta}_{d}(\abs_0) \right].
\end{align*}
Here $\bA(\bss)$ is a $p\times (d+1)$ matrix and $\dot{\bbeta}_{a}=(\frac{\partial \beta_{1}(\abs)}{\partial s_{a}},\cdots, \frac{\partial \beta_{p}(\abs)}{\partial s_{a}})^\tp$ denotes the partial gradient of $\bbeta_{a}$ with respect to $s_{a}$. 
Define $\bW_{ij}(\abs_{0}) = [\bz_{h_1}(\abs_{j}-\abs_{0})\otimes \bX_{i}]$ and $\bgamma(\abs_{0}) = (\bbeta(\abs_{0})^{\tp}, h_1\dot{\bbeta}_1(s_{0})^{\tp},\cdots,
h_d\dot{\bbeta}_d(s_{0})^{\tp})^{\tp}$, both are vectors of length $p(d+1)$. 
The functional varying coefficient model can be written as
\begin{align*}
    \label{Chapter3-gmm-Eq:model-approx}
Y_{ij} &\approx \bW_{ij}(s_{0})^{\tp}\bgamma(s_{0}) + U_{ir},
    \numberthis
\end{align*}
such that $\abs_{j}$ are sufficiently close to $\abs_{0}$.

Define $K_{\bh_1}(\abs_{j}-\abs_{0})=\prod_{k=1}^d K_{h_{1k}}(s_{jk}-s_{0k})$. Based on the local linear smoother, we obtain the initial estimates 
$\Tilde{\bbeta}(\abs_{0})= [(1, \textbf{0}_d^\tp) \otimes \bI_{p}]\Tilde{\bgamma}(\abs_0)$ where $\textbf{0}_d$ is a column vector with all the $d$ element 0 and 
\begingroup
\begin{align*}
    \breve{\bgamma}(\abs_{0}) &= \left\{
        ({nr})^{-1}\sum_{i=1}^{n}\sum_{j=1}^{r}K_{\bh_1}(\abs_{j}-\abs_{0})
        \bW_{ij}(\abs_{0})\bW_{ij}(\abs_{0})^{\tp}
    \right\}^{-1}\\
    &\qquad \times \left\{
    ({nr})^{-1}\sum_{i=1}^{n}\sum_{j=1}^{r}K_{\bh_1}(\abs_{j}-\abs_{0})\bW_{ij}(\abs_{0})Y_{ij}
    \right\}.
    \numberthis
\end{align*}
\endgroup

Built on the discussion in Section \ref{sec:IV}, we can employ the same non-parametric regression framework as presented in Equation \eqref{Chapter3-gmm-Eq:Het} 
$$
\log R_i = \log \sigma^2(\bX_i) + \epsilon_i
$$
where $R_i= \int U^2_i(\abs)d\abs$. Similar to the univariate spatial domain, we can
replace $U_i(s)$ in $R_i$ by $\breve{U}_{i}(\abs) = Y_{i}(\abs) - \bX_{i}^{\tp}\breve{\bbeta}(\abs)$. Then, using a similar approach in Section \ref{sec:IV} of the main text, we can obtain an estimate of $\sigma^{2}(\bX_{i})$ as $\widehat{\sigma}^{2}(\bX_{i})$. 

Similar to main paper, assume that the instrument variable with dimension $q \geq p$ is
$$\bQ_{ij}(\abs_{0}) = (\fM(\bX_{i}), \fM(\bX_{i})(\abs_{j}-\abs_{0})/h)^{\tp},$$
where $\fM(\bX_{i}) = \left(\bX_{i}, \bX_{i}/\widehat{\sigma}^{2}(\bX_{i})\right)^{\tp}$.
Then we can define 
\begin{align}
    \bg_{i}\{ \bgamma(\abs_{0}) \} &= r^{-1}\sum_{j=1}^{r}K_{\bh}(\abs_{j}-\abs_{0})\bQ_{ij}(\abs_{0})\left\{
        Y_{ij} - \bW_{ij}(\abs_{0})^{\tp}\bgamma(\abs_{0})
    \right\}\nonumber\\
    &=r^{-1}\sum_{j=1}^{r}K_{\bh}(\abs_{j}-\abs_{0})\bz_{h}(\abs_{j}-\abs_{0})\otimes\bDelta_{ij}(\abs_{0})
    \label{ggamma1}
\end{align}
Note that the above $\bg_i\{\bgamma(\abs)\}=(g_{i1}\{\bgamma(\abs)\},\cdots, g_{i(2q)}\{\bgamma(\abs)\})^\tp$ is a $2q$-dim vector of mean zero functions.

The next step is to perform eigen-function decomposion of $\bC(\abs, \abs') = \E
\{\bg\{\bgamma(\abs)\}\bg\{\bgamma(\abs')\}^{\tp}\}$. 
Unlike the univariate spatial domain, for every $i\in\{1,\cdots,n\}$ and a set of chosen evenly distributed grid points $\abs_{1},\cdots,\abs_{M}\in [0,1]^d$,
the $M$-dim vector 
$g_{ij}\{\bgamma(\abs_1)\},\cdots, g_{ij}\{\bgamma(\abs_M)\}$ (for $ j=1,\cdots,2q$) cannot be ordered in a line naturally because $\abs_1,\cdots,\abs_M$
are in a $d$-dim spatial domain. So the method for univariate
spatial domain can not be applied directly.
To overcome the difficulty, we will firstly apply the 
stringing method proposed in \cite{Chenetal2011} to transform the $M$-dim vector 
$g_{ij}\{\bgamma(\abs_1)\},\cdots, g_{ij}\{\bgamma(\abs_M)\}$ on $d$-dim spatial domain to
functional data on 1-dim spatial domain so that it becomes a 
functional data $\{e^*_{ij}(\xi_1),\cdots,e^*_{ij}(\xi_{M^d})\}$ on a univariate domain in $\xi\in [0,1]$.
Then, we applied the fast algorithm in \cite{Xiaoetal20014} procedure to functional data $\be^*_{i}(\xi)=\{e^*_{i1}(\xi),\cdots,e^*_{i(2q)}(\xi)\}$ on a univariate spatial domain $\xi\in [0,1]$ to obtain the corresponding eigenvalues $\widehat{\lambda}_{k}$ and its eigenfunctions $\widehat{\bphi}_k(\xi)$. Lastly, one can convert them back to the original space to obtain eigenvalues $\widehat{\lambda}_{k}$ and eigenfunctions $\widehat{\bphi}_k(\abs)$.
\par
Finally, we will incorporate the covariance function into our estimation. For any positive $\alpha$ and given spatial location $\abs_0\in[0,1]^d$, an estimate of $\bgamma(\abs_{0})$ is given by minimizing the following objective function:
$$
\sJ\{\bgamma(\abs_{0})\} = \sum_{k=1}^{\infty}\frac{\widehat{\lambda}_{k}}{\widehat{\lambda}_{k}^{2} + \alpha}
        \left\{ 
             \overline{\bg}(\bgamma(\abs_{0}))^{\tp}\widehat{\bphi}_{k}(\abs_{0})
        \right\}^{2}.
$$
% More explicitly, $\sJ\{\bgamma(\abs_{0})\}=\sum_{k=1}^{\infty}\frac{\widehat{\lambda}_{k}}{\widehat{\lambda}_{k}^{2} + \alpha}
%         \left\{
%         \frac{1}{nr}\sum_{i=1}^{n}\sum_{j=1}^{r}
%         K_{\bh}(\abs_{j}-\abs_{0})\widehat{\bphi}_{k}(\abs_{0})^{\tp}\bQ_{ij}(\abs_{0})\left[
%             Y_{ij} - \bW_{ij}(\abs_{0})^{\tp}\bgamma(\abs_{0})
%         \right]
%         \right\}^{2}$.
By minimizing the above objective function, we obtain
\begingroup
\allowdisplaybreaks
\begin{align*}
        &\sum_{k=1}^{\infty}\frac{\widehat{\lambda}_{k}}{\widehat{\lambda}_{k}^{2} + \alpha}
        \left\{
            ({nr})^{-1}\sum_{i=1}^{n}\sum_{j=1}^{r}
            K_{\bh}(\abs_{j}-\abs_{0})\widehat{\bphi}_{k}(\abs_{0})^{\tp}\bQ_{ij}(\abs_{0})\bW_{ij}(\abs_{0})
        \right\}\\
        &\times
        \left\{
            ({nr})^{-1}\sum_{i=1}^{n}\sum_{j=1}^{r}
            K_{\bh}(\abs_{j}-\abs_{0})\widehat{\bphi}_{k}(\abs_{0})^{\tp}\bQ_{ij}(\abs_{0})
            \left[
                Y_{ij} - \bW_{ij}(\bss_{0})^{\tp}\bgamma(\bss_{0})
            \right]
        \right\}\\
        &:= \sum_{k=1}^{\infty}\frac{\widehat{\lambda}_{k}}{\widehat{\lambda}_{k}^{2} + \alpha}
        \sX_{k}(\abs_{0})\left\{
            \sY_{k}(\abs_{0}) - \sX_{k}(\abs_{0})^{\tp}\bgamma(\abs_{0})
        \right\},
    \numberthis
\end{align*}
\endgroup
where $\sX_{k}(\abs_{0}) = ({nr})^{-1}\sum\limits_{i=1}^{n}\sum\limits_{j=1}^{r}K_{h}(\abs_{j}-\abs_{0})\bW_{ij}(\abs_{0})\bQ_{ij}(\abs_{0})^{\tp}\widehat{\bphi}_{k}(\abs_{0})$ and $\sY_{k}(\abs_{0}) = ({nr})^{-1}\sum\limits_{i=1}^{n}\sum\limits_{j=1}^{r}K_{h}(\abs_{j}-\abs_{0})\widehat{\bphi}_{k}(\abs_{0})^{\tp}\bQ_{ij}(\abs_{0})Y_{ij}$. Therefore, the final estimate of the coefficient function is 
$\widehat{\bbeta}(\abs_{0}) = [(1,\textbf{0}_d^\tp)\otimes\bI_{p}]\widehat{\bgamma}(\bss_{0})$ 
where $\textbf{0}_d$ is a column vector with all the $d$ element 0 and 
\begin{equation*}
    \widehat{\bgamma}(\abs_{0}) = \left\{
    \sum_{k=1}^{\infty}\frac{\widehat{\lambda}_{k}}{\widehat{\lambda}_{k}^{2} + \alpha}
    \sX_{k}(\abs_{0})\sX_{k}(\abs_{0})^{\tp}
    \right\}^{-1}
    \left\{
    \sum_{k=1}^{\infty}\frac{\widehat{\lambda}_{k}}{\widehat{\lambda}_{k}^{2} + \alpha}
    \sX_{k}(\abs_{0})\sY_{k}(\abs_{0})
    \right\}.
\end{equation*}
}

{ 
\section{Some remarks on conditions}
\begin{remark}
Condition \ref{Chapter3-gmm-Cond:beta} is also common in functional data analysis literature \citep{wang2016functional}. This condition allows us to perform Taylor series expansion. 
Condition in \ref{Chapter3-gmm-Cond:donsker} avoids the smoothness condition of the sample path \citep{zhu2012multivariate, zhu2014spatially} which is commonly assumed in \citet{hall2006properties, zhang2007statistical, cardot2013confidence}.  
The smoothness of the coefficient functions may be checked
by comparing nonparametric functions estimated by the wavelet \citep{Amato2020,Antoniadis2007} approach and the proposed approach using the test procedure in \cite{Hardle1993}.
The wavelet method is popular for functions with a few discontinuities, sharp spikes and abrupt changes (\cite{Amato2020}). 
To check the smoothness of covariance functions, one could conduct a hypothesis test to compare two covariance matrices on observed grid points $s_1,\cdots, s_r$. Due to the large number of repeated measurements $r$, conventional sample covariance is not consistent and hence can not be applied for testing. Similar to \cite{CZZ2010}, the test can be constructed based on an estimator of a Frobenius norm between $\bSigma(s,s)$ and its smoothed version.
\end{remark}

\begin{remark}
To speed up the computation for the eigenfunction decomposition for multivariate covariance function with a large number of repeated measurements, conditions \ref{Chapter3-gmm-Cond:g} impose the continuity in the mean-zero function, which is equivalent to checking the mean square continuity of the process after lining up \citep{hadinejad2002karhunen}. Here, (a) shows the limits from right and remains always right; therefore, it involves only one process. A similar, but opposite, phenomenon occurs in (b). Moreover, if the vector process $\bg(s)$ is mean-square continuous then both approaches are equivalent, as a result, the covariance function is continuous after lining up the process.

This assumption facilitates the computation of the proposed method so that the standard packages in functional data analysis can be immediately applied. However, this assumption could be removed if we use fast algorithms for eigenfunction decomposition with a large number of repeated measurements (e.g. \citet{Xiaoetal20014, Zhong2023}).
\end{remark}
}

\section{Additional simulation results}
\subsection{Additional simulation results with single covariates}
Based on the simulation setup described in Section \ref{Chapter3-gmm-Section:simulation} of the main manuscript, we have presented in Table \ref{Chapter3-gmm-Table:univariate-1}. This includes the comparison of the proposed method with the local linear estimator and \cite{wei2017heteroskedasticity}'s approaches for signal-to-noise ratio as unity. Similar to Table \ref{Chapter3-gmm-Table:univariate-0.5}, we have seen that, the IMSE and IMAE are significantly reduced if we increase the sample size.

% \begin{landscape}
\begin{table}
\vspace{-1.5cm}
\tabcolsep 3pt
    \centering
    \caption{Comparison among the proposed LLGMM with the local linear estimator (LLE) and \citet{wei2017heteroskedasticity}'s approach (LLWS) for $\SNR_{\theta} = 1$.}
    \label{Chapter3-gmm-Table:univariate-1}
    \begin{tabular}{rrrrrrrrrrr}
    \toprule
    & \multicolumn{2}{c}{{n = 30}} & 
        \multicolumn{2}{c}{n = 50} &
        \multicolumn{2}{c}{n = 100} &
        \multicolumn{2}{c}{n = 200} &
        \multicolumn{2}{c}{n = 500}\\
    \midrule    
    Method & IMSE & IMAE &  
                    IMSE & IMAE & 
                    IMSE & IMAE &
                    IMSE & IMAE &
                    IMSE & IMAE \\
    \midrule    
    %homoskedastic         
    &\multicolumn{10}{c}{Case: S.0}\\       
    \texttt{LLE} & 0.0165 & 0.0962 & 0.0097 & 0.0729 & 0.0048 & 0.0515 & 0.0025 & 0.0372 & 0.0010 & 0.0237 \\ 
    & \scriptsize (0.0157) & \scriptsize (0.0472) & \scriptsize (0.0101) & \scriptsize (0.0380) & \scriptsize (0.0051) & \scriptsize (0.0273) & \scriptsize (0.0027) & \scriptsize (0.0198) & \scriptsize (0.0011) & \scriptsize (0.0129) \\ 
    
    {\texttt{LLWS}} & 0.0172 & 0.0980 & \blue 0.0099 & \blue 0.0738 & \blue 0.0051 & \blue 0.0526 & \blue 0.0025 & \blue 0.0373 & \blue 0.0010 & \blue 0.0238 \\ 
    & \scriptsize\blue (0.0128) & \scriptsize\blue (0.0094) & \scriptsize\blue (0.0104) & \scriptsize\blue (0.0738) & \scriptsize\blue (0.0053) & \scriptsize\blue (0.0526) & \scriptsize\blue (0.0028) & \scriptsize\blue (0.0373) & \scriptsize\blue (0.0011) & \scriptsize\blue (0.0238) \\ 
    
    \texttt{LLGMM} & 0.0172 & 0.0980 & 0.0101 & 0.0741 & 0.0052 & 0.0532 & 0.0025 & 0.0374 & 0.0010 & 0.0238 \\ 
    & \scriptsize (0.0164) & \scriptsize (0.0980) & \scriptsize (0.0109) & \scriptsize (0.0741) & \scriptsize (0.0054) & \scriptsize (0.0532) & \scriptsize (0.0028) & \scriptsize (0.0374) & \scriptsize (0.0011) & \scriptsize (0.0238) \\ 
    \midrule
    
    &\multicolumn{10}{c}{Case: S.1}\\
    \texttt{LLE} & 0.0394 & 0.1487 & 0.0248 & 0.1169 & 0.0135 & 0.0860 & 0.0068 & 0.0608 & 0.0027 & 0.0387 \\ 
    & \scriptsize (0.0370) & \scriptsize (0.0743) & \scriptsize (0.0253) & \scriptsize (0.0598) & \scriptsize (0.0138) & \scriptsize (0.0455) & \scriptsize (0.0073) & \scriptsize (0.0329) & \scriptsize (0.0027) & \scriptsize (0.0204) \\ 
    
    {\texttt{LLWS}} & \blue 0.0280 & \blue 0.1238 &\blue  0.0142 & \blue 0.0887 & \blue 0.0070 &\blue  0.0617 &\blue  0.0034 &\blue  0.0430 &\blue  0.0013 &\blue  0.0269 \\ 
    & \scriptsize\blue  (0.0832) & \scriptsize\blue  (0.1255) & \scriptsize\blue  (0.0142) & \scriptsize\blue  (0.0887) & \scriptsize\blue  (0.0074) & \scriptsize\blue  (0.0617) & \scriptsize\blue  (0.0037) & \scriptsize\blue  (0.0430) & \scriptsize\blue  (0.0014) & \scriptsize\blue  (0.0269)\\
    
    \texttt{LLGMM} & 0.0270 & 0.1211 & 0.0126 & 0.0836 & 0.0062 & 0.0573 & 0.0029 & 0.0403 & 0.0012 & 0.0253 \\ 
    & \scriptsize (0.0265) & \scriptsize (0.1211) & \scriptsize (0.0126) & \scriptsize (0.0836) & \scriptsize (0.0066) & \scriptsize (0.0573) & \scriptsize (0.0031) & \scriptsize (0.0403) & \scriptsize (0.0013) & \scriptsize (0.0253) \\ 
    \midrule

    &\multicolumn{10}{c}{Case: S.2}\\
    \texttt{LLE} & 0.0518 & 0.1726 & 0.0363 & 0.1440 & 0.0215 & 0.1117 & 0.0124 & 0.0842 & 0.0055 & 0.0560 \\ 
    & \scriptsize (0.0440) & \scriptsize (0.0820) & \scriptsize (0.0321) & \scriptsize (0.0687) & \scriptsize (0.0184) & \scriptsize (0.0522) & \scriptsize (0.0107) & \scriptsize (0.0399) & \scriptsize (0.0045) & \scriptsize (0.0264) \\ 
    
    {\texttt{LLWS}} & 0.0215 & 0.1069 & \blue 0.0103 & \blue 0.0734 &\blue 0.0046 & \blue 0.0480 & \blue 0.0019 & \blue 0.0304 & \blue 0.0006 & \blue 0.0172 \\ 
    & \scriptsize (0.0226) & \scriptsize (0.1590) & \scriptsize\blue (0.0108) & \scriptsize\blue (0.0734) & \scriptsize\blue (0.0056) & \scriptsize\blue (0.0480) & \scriptsize\blue (0.0025) & \scriptsize\blue (0.0304) & \scriptsize\blue (0.0007) & \scriptsize\blue (0.0172) \\
    
    \texttt{LLGMM} & 0.0165 & 0.0918 & 0.0069 & 0.0589 & 0.0029 & 0.0376 & 0.0012 & 0.0239 & 0.0004 & 0.0142 \\ 
    & \scriptsize (0.0186) & \scriptsize (0.0918) & \scriptsize (0.0083) & \scriptsize (0.0589) & \scriptsize (0.0038) & \scriptsize (0.0376) & \scriptsize (0.0020) & \scriptsize (0.0239) & \scriptsize (0.0005) & \scriptsize (0.0142) \\ 
    \midrule
    
    &\multicolumn{10}{c}{Case: S.3}\\
    \texttt{LLE} & 0.0658 & 0.2020 & 0.0466 & 0.1705 & 0.0274 & 0.1314 & 0.0157 & 0.0994 & 0.0071 & 0.0669 \\ 
    & \scriptsize (0.0442) & \scriptsize (0.0771) & \scriptsize (0.0297) & \scriptsize (0.0616) & \scriptsize (0.0169) & \scriptsize (0.0461) & \scriptsize (0.0091) & \scriptsize (0.0319) & \scriptsize (0.0041) & \scriptsize (0.0232) \\ 
    
    {\texttt{LLWS}}  & 0.0086 & 0.0547 & \blue 0.0050 & \blue 0.0398 & \blue 0.0025 & \blue 0.0273 & \blue 0.0011 & \blue 0.0168 & \blue 0.0004 & \blue 0.0101 \\ 
    & \scriptsize (0.0461) & \scriptsize (0.0601) & \scriptsize\blue (0.0079) & \scriptsize\blue (0.0398) & \scriptsize\blue (0.0044) & \scriptsize\blue (0.0273) & \scriptsize\blue (0.0021) & \scriptsize\blue (0.0168) & \scriptsize\blue (0.0007) & \scriptsize\blue (0.0101) \\
    
    \texttt{LLGMM} & 0.0021 & 0.0261 & 0.0006 & 0.0133 & 0.0003 & 0.0078 & 0.0002 & 0.0069 & 0.0001 & 0.0052 \\
    & \scriptsize (0.0055) & \scriptsize (0.0261) & \scriptsize (0.0014) & \scriptsize (0.0133) & \scriptsize (0.0013) & \scriptsize (0.0078) & \scriptsize (0.0006) & \scriptsize (0.0069) & \scriptsize (0.0003) & \scriptsize (0.0052) \\ 
    \midrule
    
    &\multicolumn{10}{c}{Case: S.4}\\
    \texttt{LLE} & 0.0257 & 0.1200 & 0.0155 & 0.0924 & 0.0080 & 0.0662 & 0.0041 & 0.0472 & 0.0016 & 0.0300 \\ 
    & \scriptsize (0.0243) & \scriptsize (0.0593) & \scriptsize (0.0161) & \scriptsize (0.0475) & \scriptsize (0.0084) & \scriptsize (0.0349) & \scriptsize (0.0044) & \scriptsize (0.0253) & \scriptsize (0.0017) & \scriptsize (0.0160) \\ 

    {\texttt{LLWS}} & 0.0268 & 0.1213 & \blue 0.0141 &\blue 0.0881 & \blue 0.0073 & \blue 0.0631 & \blue 0.0036 & \blue 0.0447 & \blue 0.0015 & \blue 0.0285 \\ 
    & \scriptsize (0.0224) & \scriptsize (0.1231) & \scriptsize\blue (0.0143) & \scriptsize\blue (0.0881) & \scriptsize\blue (0.0078) & \scriptsize\blue (0.0631) & \scriptsize\blue (0.0040) & \scriptsize\blue (0.0447) & \scriptsize\blue (0.0016) & \scriptsize\blue (0.0285)\\
    
    \texttt{LLGMM} & 0.0254 & 0.1185 & 0.0142 & 0.0883 & 0.0074 & 0.0634 & 0.0037 & 0.0449 & 0.0015 & 0.0285 \\ 
    & \scriptsize (0.0245) & \scriptsize (0.1185) & \scriptsize (0.0142) &  \scriptsize (0.0883) & \scriptsize (0.0078) & \scriptsize (0.0634) & \scriptsize (0.0039) & \scriptsize (0.0449) & \scriptsize (0.0016) & \scriptsize (0.0285) \\ 
    \bottomrule
    \end{tabular}
\end{table}
% \end{landscape}

{
\subsection{Additional simulation results with multiple covariates}
\label{sec:trisim}
In this simulation, the data are generated from the model 
\begin{equation}
    Y_{i}(s) = X_{i1}\bbeta_{1}(s)+ X_{i2}\bbeta_{2}(s) + X_{i3}\bbeta_{3}(s) + U_{i}(s)
\end{equation}
where we generate trajectories that are observed at $r$ spatial locations for $i$-th curve, $i = 1, \cdots, n$.  Assume the functional fixed effect $\beta_{1}(s) = 1 + \cos(2\pi s)$, $\beta_{2}(s) = 2 + \sin(2\pi s)$ and $\beta_{3}(s) = 3 + s^{4}$. The corresponding fixed effect covariates $X_{1}$ and $(X_{2}, X_{3})^{\tp}$ generated from uniform distribution on $[1, 2]$ and bivariate normal distribution with mean $(1, 2)^{\tp}$ and variance $\begin{pmatrix}
    1 & \rho\\
    \rho & 1
\end{pmatrix}$, respectively. To check the impact of dependence on the proposed method, we choose $\rho \in \{0, 0.5\}$. The different choices of true conditional variance functions are described as follows.
\begin{enumerate}[label={T.\arabic*}]\addtocounter{enumi}{-1}
    \item $\sigma^{2}(x_{1}, x_{2}, x_{3}) = 1$ (homoskedastic)
    \item $\sigma^{2}(x_{1}, x_{2}, x_{3}) = x_{2}^{2} + x_{3}^{2}$
    \item $\sigma^{2}(x_{1}, x_{2}, x_{3}) = x_{2}^{2}\times x_{3}^{2}$
    \item $\sigma^{2}(x_{1}, x_{2}, x_{3}) = x_{1}^2 + x_{2}^2 + x_{3}^2$
    \item $\sigma^{2}(x_{1}, x_{2}, x_{3}) = x_{1}^2\times x_{2}^2 \times x_{3}^2$
    \item $\sigma^{2}(x_{1}, x_{2}, x_{3}) = 1 +  |x_{2}| + x_{3}^{2}/2$
\end{enumerate}
The definition of $U_{i}$s remains the same as in Section \ref{Chapter3-gmm-Section:simulation}. We choose the signal-to-noise ratio of 0.5. Based on 500 replications, we obtain IMSE, IMAE, and their standard deviation in Tables \ref{tab:3beta1-0}, \ref{tab:3beta2-0}, \ref{tab:3beta3-0} for $\rho = 0$ and \ref{tab:3beta1-0.5}, \ref{tab:3beta2-0.5}, \ref{tab:3beta3-0.5} for $\rho = 0.5$. We observe that, for all simulation situations, IMSE and IMAE of the proposed method are smaller than that the local linear estimator and the method discussed in \citet{wei2017heteroskedasticity}. 
}

{\footnotesize
\begin{table}
\vspace{-2.8cm}
\tabcolsep 2pt
%\rowcolors{1}{lightgray}{lightgray}
\centering
\caption{{ Comparison among the proposed LLGMM with the local linear (LLE) and \citet{wei2017heteroskedasticity}'s estimators (LLWS) for $\beta_{1}(\cdot)$ with $\rho = 0$.}}
\label{tab:3beta1-0}
{\begin{tabular}{rrrrrrrrrrrr}
 \toprule
    & & \multicolumn{2}{c}{n = 30} &
        \multicolumn{2}{c}{n = 50} &
        \multicolumn{2}{c}{n = 100} &
        \multicolumn{2}{c}{n = 200} &
        \multicolumn{2}{c}{n = 500}\\
    \midrule    
    & Method & IMSE & IMAE &  
                    IMSE & IMAE & 
                    IMSE & IMAE &
                    IMSE & IMAE &
                    IMSE & IMAE \\
    \midrule            
  &\multicolumn{10}{c}{Case: T.0}\\
  &\texttt{LLE} & 3.2568 & 1.3777 & 2.1728 & 1.1536 & 1.1894 & 0.8615 & 0.7139 & 0.6803 & 0.3897 & 0.5033 \\ 
  && \scriptsize{(3.8318)} & \scriptsize{(0.7851)} & \scriptsize{(2.3065)} & \scriptsize{(0.5904)} & \scriptsize{(1.0737)} & \scriptsize{(0.3757)} & \scriptsize{(0.5511)} & \scriptsize{(0.2417)} & \scriptsize{(0.2525)} & \scriptsize{(0.1533)} \\ 
  &\texttt{LLWS} & 3.2272 & 1.3731 & 2.1738 & 1.1523 & 1.1804 & 0.8577 & 0.7093 & 0.6782 & 0.3878 & 0.5007 \\ 
  && \scriptsize{(3.7593)} & \scriptsize{(0.7781)} & \scriptsize{(2.3238)} & \scriptsize{(0.5955)} & \scriptsize{(1.0617)} & \scriptsize{(0.3734)} & \scriptsize{(0.5494)} & \scriptsize{(0.2415)} & \scriptsize{(0.2544)} & \scriptsize{(0.1533)} \\ 
  &\texttt{LLGMM} & 2.9915 & 1.3574 & 2.0021 & 1.1306 & 1.1635 & 0.8682 & 0.7095 & 0.6762 & 0.3894 & 0.5017 \\ 
  && \scriptsize{(3.6036)} & \scriptsize{(0.8246)} & \scriptsize{(2.1672)} & \scriptsize{(0.6111)} & \scriptsize{(1.0652)} & \scriptsize{(0.3843)} & \scriptsize{(0.5202)} & \scriptsize{(0.2257)} & \scriptsize{(0.2411)} & \scriptsize{(0.1434)} \\   
   
   \midrule
   &\multicolumn{10}{c}{Case: T.1}\\
    &\texttt{LLE} & 2.6659 & 1.2523 & 1.7693 & 1.0422 & 1.0754 & 0.8235 & 0.6567 & 0.6568 & 0.3930 & 0.5100 \\
    && \scriptsize{(2.9965)} & \scriptsize{(0.6912)} & \scriptsize{(1.8162)} & \scriptsize{(0.5034)} & \scriptsize{(0.9595)} & \scriptsize{(0.3356)} & \scriptsize{(0.4589)} & \scriptsize{(0.2059)} & \scriptsize{(0.2118)} & \scriptsize{(0.1323)} \\ 
    &\texttt{LLWS} & 2.6047 & 1.2398 & 1.7710 & 1.0425 & 1.0686 & 0.8204 & 0.6560 & 0.6544 & 0.3973 & 0.5138 \\
    && \scriptsize{(2.9044)} & \scriptsize{(0.6797)} & \scriptsize{(1.8124)} & \scriptsize{(0.5057)} & \scriptsize{(0.9652)} & \scriptsize{(0.3368)} & \scriptsize{(0.4692)} & \scriptsize{(0.2090)} & \scriptsize{(0.2108)} & \scriptsize{(0.1304)} \\ 
    &\texttt{LLGMM} & 2.3945 & 1.2079 & 1.6664 & 1.0235 & 1.0646 & 0.8203 & 0.6511 & 0.6488 & 0.3215 & 0.5079 \\
    && \scriptsize{(2.8063)} & \scriptsize{(0.6940)} & \scriptsize{(1.7940)} & \scriptsize{(0.5143)} & \scriptsize{(0.9408)} & \scriptsize{(0.3407)} & \scriptsize{(0.4271)} & \scriptsize{(0.1872)} & \scriptsize{(0.2020)} & \scriptsize{(0.1289)} \\ 

    \midrule
    &\multicolumn{10}{c}{Case: T.2}\\
    &\texttt{LLE} & 3.2029 & 1.3848 & 2.0038 & 1.0999 & 1.2863 & 0.8933 & 0.7828 & 0.7084 & 0.4607 & 0.5539 \\ 
    && \scriptsize{(3.5392)} & \scriptsize{(0.7711)} & \scriptsize{(2.1383)} & \scriptsize{(0.5643)} & \scriptsize{(1.3877)} & \scriptsize{(0.4003)} & \scriptsize{(0.6149)} & \scriptsize{(0.2484)} & \scriptsize{(0.2607)} & \scriptsize{(0.1487)} \\ 
    &\texttt{LLWS} & 3.1422 & 1.3703 & 2.0269 & 1.1057 & 1.2776 & 0.8916 & 0.7822 & 0.7067 & 0.4630 & 0.5553 \\ 
    && \scriptsize{(3.4934)} & \scriptsize{(0.7670)} & \scriptsize{(2.1689)} & \scriptsize{(0.5675)} & \scriptsize{(1.3508)} & \scriptsize{(0.3972)} & \scriptsize{(0.6288)} & \scriptsize{(0.2523)} & \scriptsize{(0.2553)} & \scriptsize{(0.1486)} \\ 
    &\texttt{LLGMM} & 2.0304 & 1.3647 & 1.7920 & 1.0462 & 1.2212 & 0.8442 & 0.7740 & 0.5648 & 0.1772 & 0.2466 \\
    && \scriptsize{(2.0394)} & \scriptsize{(1.1751)} & \scriptsize{(2.0206)} & \scriptsize{(0.5583)} & \scriptsize{(1.7906)} & \scriptsize{(0.4287)} & \scriptsize{(4.2720)} & \scriptsize{(0.3577)} & \scriptsize{(0.6441)} & \scriptsize{(0.2205)} \\

    \midrule
    &\multicolumn{10}{c}{Case: T.3}\\
    &\texttt{LLE} & 2.8732 & 1.2986 & 1.9363 & 1.0928 & 1.1326 & 0.8439 & 0.6834 & 0.6685 & 0.3980 & 0.5119 \\
    && \scriptsize{(3.2852)} & \scriptsize{(0.7229)} & \scriptsize{(1.9551)} & \scriptsize{(0.5328)} & \scriptsize{(1.0146)} & \scriptsize{(0.3569)} & \scriptsize{(0.4968)} & \scriptsize{(0.2207)} & \scriptsize{(0.2278)} & \scriptsize{(0.1396)} \\ 
    &\texttt{LLWS} & 2.8318 & 1.2914 & 1.9370 & 1.0917 & 1.1278 & 0.8420 & 0.6825 & 0.6665 & 0.3984 & 0.5121 \\
    && \scriptsize{(3.2071)} & \scriptsize{(0.7140)} & \scriptsize{(1.9604)} & \scriptsize{(0.5363)} & \scriptsize{(1.0216)} & \scriptsize{(0.3569)} & \scriptsize{(0.5059)} & \scriptsize{(0.2232)} & \scriptsize{(0.2302)} & \scriptsize{(0.1396)} \\ 
    &\texttt{LLGMM} & 2.5389 & 1.2471 & 1.7758 & 1.0585 & 1.1097 & 0.8452 & 0.6544 & 0.6910 & 0.3691 & 0.5008 \\
    && \scriptsize{(3.0573)} & \scriptsize{(0.7282)} & \scriptsize{(1.9194)} & \scriptsize{(0.5519)} & \scriptsize{(1.0182)} & \scriptsize{(0.3669)} & \scriptsize{(0.4746)} & \scriptsize{(0.2052)} & \scriptsize{(0.2282)} & \scriptsize{(0.1444)} \\ 

    \midrule
    &\multicolumn{10}{c}{Case: T.4}\\
    & \texttt{LLE} & 2.8747 & 1.3048 & 1.7945 & 1.0425 & 1.1675 & 0.8531 & 0.7128 & 0.6792 & 0.4244 & 0.5323 \\
    && \scriptsize{(3.3002)} & \scriptsize{(0.7290)} & \scriptsize{(1.8842)} & \scriptsize{(0.5206)} & \scriptsize{(1.2541)} & \scriptsize{(0.3689)} & \scriptsize{(0.5366)} & \scriptsize{(0.2251)} & \scriptsize{(0.2256)} & \scriptsize{(0.1361)} \\ 
    &\texttt{LLWS} & 2.8387 & 1.2946 & 1.8209 & 1.0509 & 1.1658 & 0.8529 & 0.7099 & 0.6759 & 0.4257 & 0.5344 \\
    && \scriptsize{(3.3117)} & \scriptsize{(0.7298)} & \scriptsize{(1.9036)} & \scriptsize{(0.5229)} & \scriptsize{(1.2509)} & \scriptsize{(0.3688)} & \scriptsize{(0.5450)} & \scriptsize{(0.2291)} & \scriptsize{(0.2211)} & \scriptsize{(0.1343)} \\ 
    &\texttt{LLGMM} & 2.6306 & 1.2161 & 1.6996 & 1.0124 & 1.0941 & 0.8075 & 0.5686 & 0.5456 & 0.1445 & 0.2421 \\
    && \scriptsize{(4.3648)} & \scriptsize{(0.7294)} & \scriptsize{(1.9075)} & \scriptsize{(0.5220)} & \scriptsize{(1.3083)} & \scriptsize{(0.3978)} & \scriptsize{(0.7412)} & \scriptsize{(0.3043)} & \scriptsize{(0.2579)} & \scriptsize{(0.2091)} \\ 

     \midrule
    &\multicolumn{10}{c}{Case: T.5}\\
    & \texttt{LLE} & 2.8335 & 1.2865 & 1.8836 & 1.0774 & 1.1054 & 0.8340 & 0.6714 & 0.6632 & 0.3898 & 0.5063 \\
    && \scriptsize{(3.2835)} & \scriptsize{(0.7223)} & \scriptsize{(1.8885)} & \scriptsize{(0.5218)} & \scriptsize{(0.9825)} & \scriptsize{(0.3485)} & \scriptsize{(0.4825)} & \scriptsize{(0.2164)} & \scriptsize{(0.2228)} & \scriptsize{(0.1386)} \\ 
    &\texttt{LLGMM} & 2.7871 & 1.2783 & 1.8836 & 1.0763 & 1.1008 & 0.8317 & 0.6707 & 0.6615 & 0.3912 & 0.5072 \\
    && \scriptsize{(3.2161)} & \scriptsize{(0.7137)} & \scriptsize{(1.8877)} & \scriptsize{(0.5250)} & \scriptsize{(0.9924)} & \scriptsize{(0.3490)} & \scriptsize{(0.4904)} & \scriptsize{(0.2184)} & \scriptsize{(0.2249)} & \scriptsize{(0.1381)} \\ 
    &\texttt{LLGMM} & 2.5307 & 1.2377 & 1.7261 & 1.0476 & 1.0885 & 0.8319 & 0.6619 & 0.6542 & 0.3291 & 0.5042 \\
    && \scriptsize{(3.1371)} & \scriptsize{(0.7243)} & \scriptsize{(1.8136)} & \scriptsize{(0.5321)} & \scriptsize{(0.9762)} & \scriptsize{(0.3580)} & \scriptsize{(0.4642)} & \scriptsize{(0.2059)} & \scriptsize{(0.2141)} & \scriptsize{(0.1329)} \\ 
   \bottomrule
\end{tabular}}
\end{table}

\begin{table}
\vspace{-2.8cm}
\tabcolsep 2pt
\centering
\caption{{
Comparison among the proposed LLGMM with the local linear (LLE) and \citet{wei2017heteroskedasticity}'s estimators (LLWS) for $\beta_{2}(\cdot)$ with $\rho = 0$.}}
\label{tab:3beta2-0}
{\begin{tabular}{rrrrrrrrrrrr}
  \toprule
    & & \multicolumn{2}{c}{n = 30} &
        \multicolumn{2}{c}{n = 50} &
        \multicolumn{2}{c}{n = 100} &
        \multicolumn{2}{c}{n = 200} &
        \multicolumn{2}{c}{n = 500}\\
    \midrule    
    & Method & IMSE & IMAE &  
                    IMSE & IMAE & 
                    IMSE & IMAE &
                    IMSE & IMAE &
                    IMSE & IMAE \\
    \midrule            
  &\multicolumn{10}{c}{Case: T.0}\\
  &\texttt{LLE}  & 1.5635 & 0.9661 & 0.9508 & 0.7626 & 0.5528 & 0.5889 & 0.3634 & 0.4887 & 0.2420 & 0.4098 \\ 
  && \scriptsize{(1.7718)} & \scriptsize{(0.5549)} & \scriptsize{(0.9587)} & \scriptsize{(0.3780)} & \scriptsize{(0.4757)} & \scriptsize{(0.2383)} & \scriptsize{(0.2113)} & \scriptsize{(0.1304)} & \scriptsize{(0.0903)} & \scriptsize{(0.0649)} \\ 
  &\texttt{LLWS} & 1.5540 & 0.9640 & 0.9511 & 0.7630 & 0.5532 & 0.5880 & 0.3619 & 0.4880 & 0.2422 & 0.4102 \\ 
  && \scriptsize{(1.7481)} & \scriptsize{(0.5530)} & \scriptsize{(0.9529)} & \scriptsize{(0.3775)} & \scriptsize{(0.4764)} & \scriptsize{(0.2389)} & \scriptsize{(0.2114)} & \scriptsize{(0.1297)} & \scriptsize{(0.0886)} & \scriptsize{(0.0633)} \\ 
  &\texttt{LLGMM} & 2.2415 & 1.0462 & 1.0377 & 0.8200 & 0.5784 & 0.6090 & 0.3349 & 0.4676 & 0.2144 & 0.3807 \\ 
  && \scriptsize{(1.7499)} & \scriptsize{(0.7163)} & \scriptsize{(0.9056)} & \scriptsize{(0.3582)} & \scriptsize{(0.4715)} & \scriptsize{(0.2469)} & \scriptsize{(0.2236)} & \scriptsize{(0.1484)} & \scriptsize{(0.1006)} & \scriptsize{(0.0853)} \\  

  \midrule
  &\multicolumn{10}{c}{Case: T.1}\\
  &\texttt{LLE} & 1.7851 & 1.0324 & 1.1363 & 0.8286 & 0.6596 & 0.6402 & 0.4156 & 0.5200 & 0.2695 & 0.4289 \\
  && \scriptsize{(2.0175)} & \scriptsize{(0.5972)} & \scriptsize{(1.1973)} & \scriptsize{(0.4346)} & \scriptsize{(0.5973)} & \scriptsize{(0.2816)} & \scriptsize{(0.2688)} & \scriptsize{(0.1605)} & \scriptsize{(0.1207)} & \scriptsize{(0.0834)} \\ 
  &\texttt{LLWS} & 1.7721 & 1.0285 & 1.1367 & 0.8285 & 0.6590 & 0.6395 & 0.4138 & 0.5190 & 0.2707 & 0.4303 \\
  && \scriptsize{(2.0068)} & \scriptsize{(0.5942)} & \scriptsize{(1.2065)} & \scriptsize{(0.4361)} & \scriptsize{(0.5963)} & \scriptsize{(0.2815)} & \scriptsize{(0.2668)} & \scriptsize{(0.1592)} & \scriptsize{(0.1191)} & \scriptsize{(0.0831)} \\ 
  &\texttt{LLGMM} & 1.6199 & 1.0147 & 1.1238 & 0.8029 & 0.6201 & 0.6304 & 0.4014 & 0.5178 & 0.2388 & 0.3961 \\
  && \scriptsize{(1.9936)} & \scriptsize{(0.5702)} & \scriptsize{(1.1356)} & \scriptsize{(0.4096)} & \scriptsize{(0.6133)} & \scriptsize{(0.2755)} & \scriptsize{(0.2816)} & \scriptsize{(0.1745)} & \scriptsize{(0.1369)} & \scriptsize{(0.1096)} \\ 

  \midrule
  &\multicolumn{10}{c}{Case: T.2}\\
  &\texttt{LLE} & 2.2957 & 1.1815 & 1.5187 & 0.9548 & 0.8927 & 0.7432 & 0.5187 & 0.5766 & 0.3155 & 0.4600 \\ 
  && \scriptsize{(2.5891)} & \scriptsize{(0.6742)} & \scriptsize{(1.7133)} & \scriptsize{(0.5301)} & \scriptsize{(0.8275)} & \scriptsize{(0.3519)} & \scriptsize{(0.3873)} & \scriptsize{(0.2140)} & \scriptsize{(0.1750)} & \scriptsize{(0.1143)} \\ 
  &\texttt{LLWS} & 2.2731 & 1.1737 & 1.5223 & 0.9543 & 0.8945 & 0.7440 & 0.5157 & 0.5748 & 0.3153 & 0.4604 \\ 
  && \scriptsize{(2.5888)} & \scriptsize{(0.6738)} & \scriptsize{(1.7277)} & \scriptsize{(0.5298)} & \scriptsize{(0.8212)} & \scriptsize{(0.3516)} & \scriptsize{(0.3859)} & \scriptsize{(0.2141)} & \scriptsize{(0.1755)} & \scriptsize{(0.1148)} \\ 
  &\texttt{LLGMM} & 2.7597 & 1.2749 & 1.6352 & 1.0209 & 1.3148 & 0.8134 & 0.5747 & 0.5636 & 0.1945 & 0.3224 \\ 
  && \scriptsize{(3.6677)} & \scriptsize{(0.7070)} & \scriptsize{(1.6155)} & \scriptsize{(0.4973)} & \scriptsize{(6.3750)} & \scriptsize{(0.4270)} & \scriptsize{(1.1856)} & \scriptsize{(0.2576)} & \scriptsize{(0.2241)} & \scriptsize{(0.1846)} \\

  \midrule
  &\multicolumn{10}{c}{Case: T.3}\\
  &\texttt{LLE} & 1.7381 & 1.0186 & 1.0925 & 0.8137 & 0.6343 & 0.6281 & 0.4019 & 0.5121 & 0.2641 & 0.4253 \\
  && \scriptsize{(1.9767)} & \scriptsize{(0.5890)} & \scriptsize{(1.1372)} & \scriptsize{(0.4213)} & \scriptsize{(0.5718)} & \scriptsize{(0.2729)} & \scriptsize{(0.2535)} & \scriptsize{(0.1531)} & \scriptsize{(0.1128)} & \scriptsize{(0.0785)} \\ 
  &\texttt{LLWS} & 1.7257 & 1.0149 & 1.0912 & 0.8134 & 0.6349 & 0.6278 & 0.4004 & 0.5110 & 0.2641 & 0.4256 \\
  && \scriptsize{(1.9553)} & \scriptsize{(0.5847)} & \scriptsize{(1.1357)} & \scriptsize{(0.4212)} & \scriptsize{(0.5709)} & \scriptsize{(0.2728)} & \scriptsize{(0.2526)} & \scriptsize{(0.1522)} & \scriptsize{(0.1107)} & \scriptsize{(0.0779)} \\ 
  &\texttt{LLGMM} & 1.8121 & 1.0767 & 1.1705 & 0.8681 & 0.6859 & 0.6651 & 0.3872 & 0.5020 & 0.2303 & 0.3884 \\
  && \scriptsize{(1.8376)} & \scriptsize{(0.5605)} & \scriptsize{(1.0582)} & \scriptsize{(0.3965)} & \scriptsize{(0.5585)} & \scriptsize{(0.2717)} & \scriptsize{(0.2689)} & \scriptsize{(0.1723)} & \scriptsize{(0.1281)} & \scriptsize{(0.1091)} \\ 

  \midrule
  &\multicolumn{10}{c}{Case: T.4}\\
  &\texttt{LLE} & 2.2278 & 1.1600 & 1.4627 & 0.9377 & 0.8537 & 0.7259 & 0.4973 & 0.5655 & 0.3058 & 0.4529 \\
  && \scriptsize{(2.5512)} & \scriptsize{(0.6650)} & \scriptsize{(1.6777)} & \scriptsize{(0.5198)} & \scriptsize{(0.8077)} & \scriptsize{(0.3449)} & \scriptsize{(0.3576)} & \scriptsize{(0.2008)} & \scriptsize{(0.1649)} & \scriptsize{(0.1084)} \\ 
  &\texttt{LLWS} & 2.2169 & 1.1566 & 1.4706 & 0.9389 & 0.8546 & 0.7266 & 0.4944 & 0.5642 & 0.3051 & 0.4539 \\
  && \scriptsize{(2.5526)} & \scriptsize{(0.6649)} & \scriptsize{(1.6887)} & \scriptsize{(0.5194)} & \scriptsize{(0.7984)} & \scriptsize{(0.3427)} & \scriptsize{(0.3564)} & \scriptsize{(0.2009)} & \scriptsize{(0.1619)} & \scriptsize{(0.1063)} \\ 
  &\texttt{LLGMM} & 2.2150 & 1.1312 & 0.8492 & 0.0109 & 0.7889 & 0.7242 & 0.4557 & 0.5557 & 0.1915 & 0.3234 \\
  && \scriptsize{(2.1676)} & \scriptsize{(0.6740)} & \scriptsize{(1.7830)} & \scriptsize{(0.4886)} & \scriptsize{(1.3157)} & \scriptsize{(0.3630)} & \scriptsize{(0.7496)} & \scriptsize{(0.2499)} & \scriptsize{(0.2053)} & \scriptsize{(0.1788)} \\

  \midrule
  &\multicolumn{10}{c}{Case: T.5}\\
  &\texttt{LLE} & 1.6286 & 0.9876 & 1.0141 & 0.7833 & 0.5971 & 0.6111 & 0.3820 & 0.5003 & 0.2543 & 0.4185 \\
  && \scriptsize{(1.8335)} & \scriptsize{(0.5654)} & \scriptsize{(1.0469)} & \scriptsize{(0.4011)} & \scriptsize{(0.5250)} & \scriptsize{(0.2561)} & \scriptsize{(0.2303)} & \scriptsize{(0.1414)} & \scriptsize{(0.1014)} & \scriptsize{(0.0718)} \\ 
  &\texttt{LLWS} & 1.6170 & 0.9843 & 1.0130 & 0.7832 & 0.5967 & 0.6101 & 0.3805 & 0.4994 & 0.2548 & 0.4194 \\
  && \scriptsize{(1.8176)} & \scriptsize{(0.5622)} & \scriptsize{(1.0455)} & \scriptsize{(0.4010)} & \scriptsize{(0.5245)} & \scriptsize{(0.2559)} & \scriptsize{(0.2291)} & \scriptsize{(0.1403)} & \scriptsize{(0.0995)} & \scriptsize{(0.0713)} \\ 
  &\texttt{LLGMM} & 1.5404 & 0.9530 & 1.0002 & 0.7422 & 0.5874 & 0.6165 & 0.3689 & 0.4913 & 0.2210 & 0.3833 \\
  && \scriptsize{(1.7433)} & \scriptsize{(0.5378)} & \scriptsize{(0.9761)} & \scriptsize{(0.3753)} & \scriptsize{(0.5155)} & \scriptsize{(0.2538)} & \scriptsize{(0.2442)} & \scriptsize{(0.1597)} & \scriptsize{(0.1133)} & \scriptsize{(0.0979)} \\ 
    \bottomrule
\end{tabular}}
\end{table}

\begin{table}
\vspace{-2.8cm}
\tabcolsep 2pt
\centering
\caption{{
Comparison among the proposed LLGMM with the local linear (LLE) and \citet{wei2017heteroskedasticity}'s estimators (LLWS) for $\beta_{3}(\cdot)$ with $\rho = 0$.}}
\label{tab:3beta3-0}
{\begin{tabular}{rrrrrrrrrrrr}
  \toprule
    & & \multicolumn{2}{c}{n = 30} &
        \multicolumn{2}{c}{n = 50} &
        \multicolumn{2}{c}{n = 100} &
        \multicolumn{2}{c}{n = 200} &
        \multicolumn{2}{c}{n = 500}\\
    \midrule    
    & Method & IMSE & IMAE &  
                    IMSE & IMAE & 
                    IMSE & IMAE &
                    IMSE & IMAE &
                    IMSE & IMAE \\
    \midrule            
  &\multicolumn{10}{c}{Case: T.0}\\
  &\texttt{LLE}  & 1.2156 & 0.8453 & 0.7101 & 0.6318 & 0.3443 & 0.4523 & 0.1829 & 0.3219 & 0.0665 & 0.1954 \\ 
  && \scriptsize{(1.6491)} & \scriptsize{(0.5471)} & \scriptsize{(0.9631)} & \scriptsize{(0.4298)} & \scriptsize{(0.4159)} & \scriptsize{(0.2796)} & \scriptsize{(0.2299)} & \scriptsize{(0.2096)} & \scriptsize{(0.0773)} & \scriptsize{(0.1154)} \\ 
  &\texttt{LLWS} & 1.2100 & 0.8442 & 0.7132 & 0.6326 & 0.3467 & 0.4534 & 0.1812 & 0.3202 & 0.0668 & 0.1959 \\ 
  && \scriptsize{(1.6346)} & \scriptsize{(0.5435)} & \scriptsize{(0.9781)} & \scriptsize{(0.4327)} & \scriptsize{(0.4180)} & \scriptsize{(0.2795)} & \scriptsize{(0.2258)} & \scriptsize{(0.2066)} & \scriptsize{(0.0779)} & \scriptsize{(0.1159)} \\ 
  &\texttt{LLGMM} & 1.2128 & 0.8505 & 0.6865 & 0.6443 & 0.3608 & 0.4777 & 0.2078 & 0.3522 & 0.0885 & 0.2302 \\ 
  && \scriptsize{(1.7499)} & \scriptsize{(0.7163)} & \scriptsize{(0.9056)} & \scriptsize{(0.3582)} & \scriptsize{(0.4715)} & \scriptsize{(0.2469)} & \scriptsize{(0.2236)} & \scriptsize{(0.1484)} & \scriptsize{(0.1006)} & \scriptsize{(0.0853)} \\ 

  \midrule
  &\multicolumn{10}{c}{Case: T.1}\\
  &\texttt{LLE} & 1.5328 & 0.9466 & 0.8888 & 0.7080 & 0.4609 & 0.5207 & 0.2327 & 0.3639 & 0.0906 & 0.2284 \\
  && \scriptsize{(2.0942)} & \scriptsize{(0.6205)} & \scriptsize{(1.2402)} & \scriptsize{(0.4815)} & \scriptsize{(0.5625)} & \scriptsize{(0.3289)} & \scriptsize{(0.3106)} & \scriptsize{(0.2364)} & \scriptsize{(0.1049)} & \scriptsize{(0.1365)} \\ 
  &\texttt{LLWS} & 1.5122 & 0.9410 & 0.8860 & 0.7069 & 0.4628 & 0.5216 & 0.2298 & 0.3616 & 0.0911 & 0.2295 \\
  && \scriptsize{(2.0843)} & \scriptsize{(0.6183)} & \scriptsize{(1.2433)} & \scriptsize{(0.4822)} & \scriptsize{(0.5633)} & \scriptsize{(0.3293)} & \scriptsize{(0.3084)} & \scriptsize{(0.2336)} & \scriptsize{(0.1059)} & \scriptsize{(0.1369)} \\ 
  &\texttt{LLGMM} & 1.4442 & 0.9293 & 0.8817 & 0.7256 & 0.4611 & 0.5237 & 0.2237 & 0.3563 & 0.0821 & 0.2192 \\
  && \scriptsize{(1.9936)} & \scriptsize{(0.5702)} & \scriptsize{(1.1356)} & \scriptsize{(0.4096)} & \scriptsize{(0.6133)} & \scriptsize{(0.2755)} & \scriptsize{(0.2816)} & \scriptsize{(0.1745)} & \scriptsize{(0.1369)} & \scriptsize{(0.1096)} \\   

  \midrule
  &\multicolumn{10}{c}{Case: T.2}\\
  &\texttt{LLE} & 1.4960 & 0.9446 & 0.8248 & 0.6864 & 0.4740 & 0.5181 & 0.2352 & 0.3605 & 0.0923 & 0.2292 \\ 
  && \scriptsize{(1.7908)} & \scriptsize{(0.6054)} & \scriptsize{(1.1366)} & \scriptsize{(0.4589)} & \scriptsize{(0.6504)} & \scriptsize{(0.3507)} & \scriptsize{(0.3445)} & \scriptsize{(0.2452)} & \scriptsize{(0.1104)} & \scriptsize{(0.1371)} \\ 
  &\texttt{LLWS} & 1.4706 & 0.9351 & 0.8295 & 0.6884 & 0.4743 & 0.5189 & 0.2303 & 0.3565 & 0.0927 & 0.2301 \\ 
  && \scriptsize{(1.7664)} & \scriptsize{(0.5993)} & \scriptsize{(1.1397)} & \scriptsize{(0.4600)} & \scriptsize{(0.6314)} & \scriptsize{(0.3483)} & \scriptsize{(0.3390)} & \scriptsize{(0.2415)} & \scriptsize{(0.1080)} & \scriptsize{(0.1364)} \\ 
  &\texttt{LLGMM} & 1.4194 & 0.9387 & 0.8176 & 0.6898 & 0.4477 & 0.5100 & 0.2265 & 0.3576 & 0.0939 & 0.2240 \\
  && \scriptsize{(3.6677)} & \scriptsize{(0.7070)} & \scriptsize{(1.6155)} & \scriptsize{(0.4973)} & \scriptsize{(6.3750)} & \scriptsize{(0.4270)} & \scriptsize{(1.1856)} & \scriptsize{(0.2576)} & \scriptsize{(0.2241)} & \scriptsize{(0.1846)} \\ 

  \midrule
  &\multicolumn{10}{c}{Case: T.3}\\
  &\texttt{LLE} & 1.4474 & 0.9208 & 0.8396 & 0.6883 & 0.4305 & 0.5031 & 0.2210 & 0.3550 & 0.0845 & 0.2206 \\
  && \scriptsize{(1.9870)} & \scriptsize{(0.6003)} & \scriptsize{(1.1473)} & \scriptsize{(0.4683)} & \scriptsize{(0.5282)} & \scriptsize{(0.3183)} & \scriptsize{(0.2930)} & \scriptsize{(0.2296)} & \scriptsize{(0.0977)} & \scriptsize{(0.1316)} \\ 
  &\texttt{LLWS} & 1.4327 & 0.9172 & 0.8364 & 0.6868 & 0.4336 & 0.5044 & 0.2187 & 0.3531 & 0.0851 & 0.2215 \\
  && \scriptsize{(1.9797)} & \scriptsize{(0.5991)} & \scriptsize{(1.1544)} & \scriptsize{(0.4697)} & \scriptsize{(0.5315)} & \scriptsize{(0.3192)} & \scriptsize{(0.2904)} & \scriptsize{(0.2268)} & \scriptsize{(0.0992)} & \scriptsize{(0.1322)} \\ 
  &\texttt{LLGMM} & 1.3549 & 0.9076 & 0.8076 & 0.6837 & 0.4337 & 0.5017 & 0.2103 & 0.3524 & 0.1075 & 0.2144 \\
  && \scriptsize{(1.8376)} & \scriptsize{(0.5605)} & \scriptsize{(1.0582)} & \scriptsize{(0.3965)} & \scriptsize{(0.5585)} & \scriptsize{(0.2717)} & \scriptsize{(0.2689)} & \scriptsize{(0.1723)} & \scriptsize{(0.1281)} & \scriptsize{(0.1091)} \\ 
  
  \midrule
  &\multicolumn{10}{c}{Case: T.4}\\
  &\texttt{LLE} & 1.3624 & 0.8970 & 0.7165 & 0.6409 & 0.4215 & 0.4887 & 0.2065 & 0.3364 & 0.0802 & 0.2126 \\
  && \scriptsize{(1.6941)} & \scriptsize{(0.5814)} & \scriptsize{(0.9798)} & \scriptsize{(0.4261)} & \scriptsize{(0.5797)} & \scriptsize{(0.3288)} & \scriptsize{(0.3120)} & \scriptsize{(0.2306)} & \scriptsize{(0.0959)} & \scriptsize{(0.1274)} \\ 
  &\texttt{LLWS} & 1.3458 & 0.8887 & 0.7241 & 0.6441 & 0.4251 & 0.4903 & 0.2029 & 0.3330 & 0.0801 & 0.2128 \\
  && \scriptsize{(1.6794)} & \scriptsize{(0.5794)} & \scriptsize{(0.9879)} & \scriptsize{(0.4280)} & \scriptsize{(0.5770)} & \scriptsize{(0.3287)} & \scriptsize{(0.3080)} & \scriptsize{(0.2284)} & \scriptsize{(0.0936)} & \scriptsize{(0.1264)} \\ 
  &\texttt{LLGMM} & 1.3221 & 0.8893 & 0.7254 & 0.6640 & 0.4199 & 0.4383 & 0.2080 & 0.3276 & 0.0805 & 0.2100 \\
  && \scriptsize{(1.1676)} & \scriptsize{(0.6740)} & \scriptsize{(1.7830)} & \scriptsize{(0.4886)} & \scriptsize{(1.3157)} & \scriptsize{(0.3630)} & \scriptsize{(0.7496)} & \scriptsize{(0.2499)} & \scriptsize{(0.2053)} & \scriptsize{(0.1788)} \\ 

  \midrule
  &\multicolumn{10}{c}{Case: T.5}\\
  &\texttt{LLE} & 1.4437 & 0.9181 & 0.8421 & 0.6886 & 0.4304 & 0.5041 & 0.2213 & 0.3559 & 0.0840 & 0.2203 \\
  && \scriptsize{(1.9620)} & \scriptsize{(0.6026)} & \scriptsize{(1.1569)} & \scriptsize{(0.4696)} & \scriptsize{(0.5233)} & \scriptsize{(0.3160)} & \scriptsize{(0.2908)} & \scriptsize{(0.2292)} & \scriptsize{(0.0971)} & \scriptsize{(0.1308)} \\ 
  &\texttt{LLWS} & 1.4250 & 0.9135 & 0.8414 & 0.6880 & 0.4323 & 0.5048 & 0.2190 & 0.3539 & 0.0845 & 0.2210 \\
  && \scriptsize{(1.9473)} & \scriptsize{(0.6006)} & \scriptsize{(1.1661)} & \scriptsize{(0.4715)} & \scriptsize{(0.5262)} & \scriptsize{(0.3169)} & \scriptsize{(0.2874)} & \scriptsize{(0.2260)} & \scriptsize{(0.0980)} & \scriptsize{(0.1313)} \\ 
  &\texttt{LLGMM} & 1.3465 & 0.9059 & 0.8026 & 0.6921 & 0.4370 & 0.5044 & 0.2197 & 0.3534 & 0.0860 & 0.2234 \\
  && \scriptsize{(1.7433)} & \scriptsize{(0.5378)} & \scriptsize{(0.9761)} & \scriptsize{(0.3753)} & \scriptsize{(0.5155)} & \scriptsize{(0.2538)} & \scriptsize{(0.2442)} & \scriptsize{(0.1597)} & \scriptsize{(0.1133)} & \scriptsize{(0.0979)} \\ 
    \bottomrule
\end{tabular}}
\end{table}

\begin{table}
\vspace{-2.8cm}
\tabcolsep 2pt
\centering
\caption{{Comparison among the proposed LLGMM with the local linear (LLE) and \citet{wei2017heteroskedasticity}'s estimators (LLWS) for $\beta_{1}(\cdot)$ with $\rho = 0.5$}}
\label{tab:3beta1-0.5}
{\begin{tabular}{rrrrrrrrrrr}
\toprule
& \multicolumn{2}{c}{n = 30} &
  \multicolumn{2}{c}{n = 50} &
  \multicolumn{2}{c}{n = 100} &
  \multicolumn{2}{c}{n = 200} &
  \multicolumn{2}{c}{n = 500}\\
\midrule    
Method & IMSE & IMAE &  
  IMSE & IMAE & 
  IMSE & IMAE &
  IMSE & IMAE &
  IMSE & IMAE \\
\midrule 
\multicolumn{10}{c}{Case: T.0}\\
\texttt{LLE} & 4.0246 & 1.5247 & 2.6216 & 1.2612 & 1.4204 & 0.9363 & 0.8433 & 0.7371 & 0.4629 & 0.5528 \\ 
& \scriptsize{(4.9122)} & \scriptsize{(0.9052)} & \scriptsize{(2.9427)} & \scriptsize{(0.6768)} & \scriptsize{(1.3627)} & \scriptsize{(0.4336)} & \scriptsize{(0.6828)} & \scriptsize{(0.2761)} & \scriptsize{(0.2973)} & \scriptsize{(0.1649)} \\ 
\texttt{LLWS} & 3.9945 & 1.5218 & 2.6262 & 1.2607 & 1.4116 & 0.9333 & 0.8394 & 0.7359 & 0.4608 & 0.5509 \\ 
& \scriptsize{(4.8446)} & \scriptsize{(0.8989)} & \scriptsize{(2.9628)} & \scriptsize{(0.6818)} & \scriptsize{(1.3366)} & \scriptsize{(0.4287)} & \scriptsize{(0.6809)} & \scriptsize{(0.2762)} & \scriptsize{(0.2967)} & \scriptsize{(0.1636)} \\ 
\texttt{LLGMM} & 3.5110 & 1.4698 & 2.3892 & 1.2275 & 1.3597 & 0.9358 & 0.8577 & 0.7527 & 0.4726 & 0.5601 \\ 
& \scriptsize{(4.5283)} & \scriptsize{(0.9100)} & \scriptsize{(2.7873)} & \scriptsize{(0.6987)} & \scriptsize{(1.3248)} & \scriptsize{(0.4440)} & \scriptsize{(0.6519)} & \scriptsize{(0.2681)} & \scriptsize{(0.2974)} & \scriptsize{(0.1753)} \\ 
\midrule

\multicolumn{10}{c}{Case: T.1}\\
\texttt{LLE}  & 3.1961 & 1.3651 & 2.0250 & 1.1116 & 1.2346 & 0.8790 & 0.7430 & 0.6958 & 0.4547 & 0.5528 \\ 
& \scriptsize{(3.7864)} & \scriptsize{(0.7877)} & \scriptsize{(2.1379)} & \scriptsize{(0.5571)} & \scriptsize{(1.1571)} & \scriptsize{(0.3746)} & \scriptsize{(0.5339)} & \scriptsize{(0.2237)} & \scriptsize{(0.2337)} & \scriptsize{(0.1371)} \\ 
\texttt{LLWS}& 3.1227 & 1.3519 & 2.0374 & 1.1141 & 1.2178 & 0.8743 & 0.7443 & 0.6944 & 0.4589 & 0.5556 \\ 
& \scriptsize{(3.6681)} & \scriptsize{(0.7729)} & \scriptsize{(2.1405)} & \scriptsize{(0.5591)} & \scriptsize{(1.1468)} & \scriptsize{(0.3711)} & \scriptsize{(0.5508)} & \scriptsize{(0.2286)} & \scriptsize{(0.2351)} & \scriptsize{(0.1369)} \\ 
\texttt{LLGMM} & 2.6968 & 1.2834 & 1.3503 & 1.0721 & 1.1188 & 0.8659 & 0.7406 & 0.6908 & 0.4524 & 0.5457 \\ 
& \scriptsize{(3.3766)} & \scriptsize{(0.7542)} & \scriptsize{(2.1055)} & \scriptsize{(0.5701)} & \scriptsize{(1.1064)} & \scriptsize{(0.3743)} & \scriptsize{(0.5045)} & \scriptsize{(0.2103)} & \scriptsize{(0.2378)} & \scriptsize{(0.1451)} \\ 
\midrule

\multicolumn{10}{c}{Case: T.2}\\
\texttt{LLE} & 3.6110 & 1.4553 & 2.2306 & 1.1547 & 1.4132 & 0.9351 & 0.8637 & 0.7427 & 0.5159 & 0.5895 \\ 
& \scriptsize{(4.2514)} & \scriptsize{(0.8505)} & \scriptsize{(2.4709)} & \scriptsize{(0.6127)} & \scriptsize{(1.5061)} & \scriptsize{(0.4266)} & \scriptsize{(0.6772)} & \scriptsize{(0.2632)} & \scriptsize{(0.2649)} & \scriptsize{(0.1443)} \\ 
\texttt{LLWS}& 3.5239 & 1.4347 & 2.2264 & 1.1516 & 1.4034 & 0.9323 & 0.8631 & 0.7406 & 0.5259 & 0.5948 \\ 
& \scriptsize{(4.2236)} & \scriptsize{(0.8441)} & \scriptsize{(2.4801)} & \scriptsize{(0.6157)} & \scriptsize{(1.4922)} & \scriptsize{(0.4262)} & \scriptsize{(0.6898)} & \scriptsize{(0.2657)} & \scriptsize{(0.2857)} & \scriptsize{(0.1498)} \\ 
\texttt{LLGMM} & 2.9741 & 1.3315 & 2.2427 & 1.1195 & 1.3349 & 0.9111 & 0.8003 & 0.7105 & 0.4527 & 0.5103 \\ 
& \scriptsize{(3.8570)} & \scriptsize{(0.7977)} & \scriptsize{(4.9414)} & \scriptsize{(0.6636)} & \scriptsize{(1.3713)} & \scriptsize{(0.4105)} & \scriptsize{(0.6655)} & \scriptsize{(0.2702)} & \scriptsize{(0.3940)} & \scriptsize{(0.2311)} \\ 
\midrule

\multicolumn{10}{c}{Case: T.3}\\
\texttt{LLE} & 3.4697 & 1.4212 & 2.2654 & 1.1808 & 1.3171 & 0.9072 & 0.7842 & 0.7134 & 0.4631 & 0.5559 \\ 
& \scriptsize{(4.1178)} & \scriptsize{(0.8256)} & \scriptsize{(2.3604)} & \scriptsize{(0.5930)} & \scriptsize{(1.2270)} & \scriptsize{(0.3999)} & \scriptsize{(0.5918)} & \scriptsize{(0.2435)} & \scriptsize{(0.2574)} & \scriptsize{(0.1470)} \\ 
\texttt{LLWS}& 3.4199 & 1.4134 & 2.2759 & 1.1826 & 1.3060 & 0.9036 & 0.7843 & 0.7118 & 0.4654 & 0.5576 \\ 
& \scriptsize{(4.0162)} & \scriptsize{(0.8133)} & \scriptsize{(2.3668)} & \scriptsize{(0.5964)} & \scriptsize{(1.2214)} & \scriptsize{(0.3971)} & \scriptsize{(0.6033)} & \scriptsize{(0.2481)} & \scriptsize{(0.2604)} & \scriptsize{(0.1462)} \\ 
\texttt{LLGMM} & 3.0901 & 1.3661 & 2.0644 & 1.1429 & 1.2589 & 0.9031 & 0.7828 & 0.7126 & 0.4615 & 0.5515 \\ 
& \scriptsize{(3.9654)} & \scriptsize{(0.8356)} & \scriptsize{(2.2492)} & \scriptsize{(0.6111)} & \scriptsize{(1.1749)} & \scriptsize{(0.4036)} & \scriptsize{(0.5555)} & \scriptsize{(0.2328)} & \scriptsize{(0.2628)} & \scriptsize{(0.1605)} \\ 
\midrule

\multicolumn{10}{c}{Case: T.4}\\
\texttt{LLE} & 3.2962 & 1.3851 & 2.0131 & 1.0992 & 1.2988 & 0.8984 & 0.7966 & 0.7155 & 0.4796 & 0.5698 \\ 
& \scriptsize{(3.9847)} & \scriptsize{(0.8087)} & \scriptsize{(2.1997)} & \scriptsize{(0.5720)} & \scriptsize{(1.3619)} & \scriptsize{(0.3994)} & \scriptsize{(0.5973)} & \scriptsize{(0.2399)} & \scriptsize{(0.2249)} & \scriptsize{(0.1307)} \\ 
\texttt{LLWS}& 3.2194 & 1.3693 & 2.0182 & 1.0988 & 1.2880 & 0.8951 & 0.7870 & 0.7103 & 0.5172 & 0.5822 \\ 
& \scriptsize{(3.9197)} & \scriptsize{(0.8011)} & \scriptsize{(2.2185)} & \scriptsize{(0.5759)} & \scriptsize{(1.3443)} & \scriptsize{(0.3970)} & \scriptsize{(0.6039)} & \scriptsize{(0.2416)} & \scriptsize{(0.5771)} & \scriptsize{(0.1889)} \\ 
\texttt{LLGMM} & 2.7381 & 1.2654 & 1.9922 & 1.0898 & 1.2332 & 0.8774 & 0.7609 & 0.6962 & 0.4069 & 0.4914 \\ 
& \scriptsize{(3.5198)} & \scriptsize{(0.7577)} & \scriptsize{(2.3459)} & \scriptsize{(0.5875)} & \scriptsize{(1.2710)} & \scriptsize{(0.3877)} & \scriptsize{(0.5922)} & \scriptsize{(0.2339)} & \scriptsize{(0.3120)} & \scriptsize{(0.2181)} \\ 
\midrule

\multicolumn{10}{c}{Case: T.5}\\
\texttt{LLE} & 3.4186 & 1.4066 & 2.2018 & 1.1641 & 1.2788 & 0.8948 & 0.7688 & 0.7076 & 0.4536 & 0.5501 \\ 
& \scriptsize{(4.1108)} & \scriptsize{(0.8256)} & \scriptsize{(2.2746)} & \scriptsize{(0.5801)} & \scriptsize{(1.1756)} & \scriptsize{(0.3887)} & \scriptsize{(0.5693)} & \scriptsize{(0.2374)} & \scriptsize{(0.2515)} & \scriptsize{(0.1453)} \\ 
\texttt{LLWS}& 3.3813 & 1.4002 & 2.2097 & 1.1641 & 1.2703 & 0.8916 & 0.7688 & 0.7055 & 0.4567 & 0.5524 \\ 
& \scriptsize{(4.0530)} & \scriptsize{(0.8188)} & \scriptsize{(2.2873)} & \scriptsize{(0.5856)} & \scriptsize{(1.1762)} & \scriptsize{(0.3873)} & \scriptsize{(0.5828)} & \scriptsize{(0.2422)} & \scriptsize{(0.2536)} & \scriptsize{(0.1446)} \\ 
\texttt{LLGMM} & 2.9336 & 1.3323 & 2.0072 & 1.1272 & 1.2227 & 0.8891 & 0.7678 & 0.7047 & 0.4528 & 0.5504 \\ 
& \scriptsize{(3.7628)} & \scriptsize{(0.8167)} & \scriptsize{(2.1937)} & \scriptsize{(0.5943)} & \scriptsize{(1.1367)} & \scriptsize{(0.3929)} & \scriptsize{(0.5466)} & \scriptsize{(0.2296)} & \scriptsize{(0.2522)} & \scriptsize{(0.1521)} \\ 
\bottomrule
\end{tabular}}
\end{table}

\begin{table}
\vspace{-2.8cm}
\tabcolsep 2pt
\centering
\caption{{Comparison among the proposed LLGMM with the local linear (LLE) and \citet{wei2017heteroskedasticity}'s estimators (LLWS) for $\beta_{2}(\cdot)$ with $\rho = 0.5$}}
\label{tab:3beta2-0.5}
{\begin{tabular}{rrrrrrrrrrr}
\toprule
 & \multicolumn{2}{c}{n = 30} &
  \multicolumn{2}{c}{n = 50} &
  \multicolumn{2}{c}{n = 100} &
  \multicolumn{2}{c}{n = 200} &
  \multicolumn{2}{c}{n = 500}\\
\midrule    
 Method & IMSE & IMAE &  
  IMSE & IMAE & 
  IMSE & IMAE &
  IMSE & IMAE &
  IMSE & IMAE \\
\midrule 

\multicolumn{10}{c}{Case: T.0}\\
\texttt{LLE}& 3.5540 & 1.4543 & 2.0623 & 1.1037 & 1.0305 & 0.7908 & 0.6318 & 0.6268 & 0.3416 & 0.4745 \\ 
& \scriptsize{(4.5008)} & \scriptsize{(0.9030)} & \scriptsize{(2.4326)} & \scriptsize{(0.6529)} & \scriptsize{(1.0705)} & \scriptsize{(0.4072)} & \scriptsize{(0.5851)} & \scriptsize{(0.2762)} & \scriptsize{(0.2075)} & \scriptsize{(0.1268)} \\ 
\texttt{LLWS} & 3.5376 & 1.4503 & 2.0661 & 1.1055 & 1.0308 & 0.7901 & 0.6310 & 0.6268 & 0.3424 & 0.4752 \\ 
& \scriptsize{(4.5203)} & \scriptsize{(0.8983)} & \scriptsize{(2.4252)} & \scriptsize{(0.6513)} & \scriptsize{(1.0676)} & \scriptsize{(0.4057)} & \scriptsize{(0.5772)} & \scriptsize{(0.2730)} & \scriptsize{(0.2072)} & \scriptsize{(0.1265)} \\ 
\texttt{LLGMM} & 3.4935 & 1.4702 & 2.0134 & 1.1347 & 1.0884 & 0.8348 & 0.6630 & 0.6577 & 0.3270 & 0.4626 \\ 
& \scriptsize{(4.6278)} & \scriptsize{(0.9365)} & \scriptsize{(2.1829)} & \scriptsize{(0.6307)} & \scriptsize{(1.0182)} & \scriptsize{(0.3937)} & \scriptsize{(0.5646)} & \scriptsize{(0.2735)} & \scriptsize{(0.2264)} & \scriptsize{(0.1460)} \\ 
\midrule

\multicolumn{10}{c}{Case: T.1}\\
\texttt{LLE}& 3.8754 & 1.5072 & 2.3521 & 1.1766 & 1.1843 & 0.8451 & 0.7093 & 0.6634 & 0.3776 & 0.4953 \\ 
& \scriptsize{(4.9489)} & \scriptsize{(0.9466)} & \scriptsize{(2.7671)} & \scriptsize{(0.7136)} & \scriptsize{(1.2803)} & \scriptsize{(0.4565)} & \scriptsize{(0.6832)} & \scriptsize{(0.3037)} & \scriptsize{(0.2537)} & \scriptsize{(0.1496)} \\ 
\texttt{LLWS} & 3.8569 & 1.5019 & 2.3575 & 1.1790 & 1.1853 & 0.8455 & 0.7067 & 0.6622 & 0.3775 & 0.4953 \\ 
& \scriptsize{(4.9550)} & \scriptsize{(0.9431)} & \scriptsize{(2.7801)} & \scriptsize{(0.7137)} & \scriptsize{(1.2717)} & \scriptsize{(0.4564)} & \scriptsize{(0.6788)} & \scriptsize{(0.3017)} & \scriptsize{(0.2537)} & \scriptsize{(0.1499)} \\ 
\texttt{LLGMM} & 3.5657 & 1.4722 & 2.2346 & 1.1643 & 1.1852 & 0.8355 & 0.6910 & 0.6524 & 0.3752 & 0.4944 \\ 
& \scriptsize{(4.8701)} & \scriptsize{(0.9426)} & \scriptsize{(2.7355)} & \scriptsize{(0.6987)} & \scriptsize{(1.2366)} & \scriptsize{(0.4327)} & \scriptsize{(0.6451)} & \scriptsize{(0.2765)} & \scriptsize{(0.2822)} & \scriptsize{(0.1677)} \\ 
\midrule 

\multicolumn{10}{c}{Case: T.2}\\
\texttt{LLE}& 3.9219 & 1.4984 & 2.6610 & 1.2653 & 1.3586 & 0.9023 & 0.8066 & 0.7048 & 0.4164 & 0.5174 \\ 
& \scriptsize{(5.2307)} & \scriptsize{(0.9739)} & \scriptsize{(2.9867)} & \scriptsize{(0.7500)} & \scriptsize{(1.5197)} & \scriptsize{(0.5028)} & \scriptsize{(0.8089)} & \scriptsize{(0.3421)} & \scriptsize{(0.3123)} & \scriptsize{(0.1760)} \\ 
\texttt{LLWS} & 3.8904 & 1.4846 & 2.6496 & 1.2623 & 1.3594 & 0.9028 & 0.8053 & 0.7039 & 0.4154 & 0.5167 \\ 
& \scriptsize{(5.3176)} & \scriptsize{(0.9768)} & \scriptsize{(2.9675)} & \scriptsize{(0.7460)} & \scriptsize{(1.5175)} & \scriptsize{(0.5040)} & \scriptsize{(0.8099)} & \scriptsize{(0.3433)} & \scriptsize{(0.3122)} & \scriptsize{(0.1756)} \\ 
\texttt{LLGMM} & 3.8468 & 1.4146 & 2.5303 & 1.2154 & 1.3000 & 0.8976 & 0.7953 & 0.7028 & 0.4144 & 0.5154 \\ 
& \scriptsize{(5.6730)} & \scriptsize{(0.9750)} & \scriptsize{(3.3317)} & \scriptsize{(0.7419)} & \scriptsize{(1.5252)} & \scriptsize{(0.4698)} & \scriptsize{(0.7880)} & \scriptsize{(0.3133)} & \scriptsize{(0.3823)} & \scriptsize{(0.2244)} \\ 
\midrule 

\multicolumn{10}{c}{Case: T.3}\\
\texttt{LLE}& 3.8508 & 1.5034 & 2.2722 & 1.1590 & 1.1428 & 0.8297 & 0.6894 & 0.6541 & 0.3706 & 0.4916 \\ 
& \scriptsize{(4.9925)} & \scriptsize{(0.9442)} & \scriptsize{(2.6654)} & \scriptsize{(0.6950)} & \scriptsize{(1.2399)} & \scriptsize{(0.4462)} & \scriptsize{(0.6550)} & \scriptsize{(0.2971)} & \scriptsize{(0.2425)} & \scriptsize{(0.1445)} \\ 
\texttt{LLWS} & 3.8340 & 1.4997 & 2.2762 & 1.1609 & 1.1434 & 0.8300 & 0.6877 & 0.6534 & 0.3705 & 0.4916 \\ 
& \scriptsize{(4.9855)} & \scriptsize{(0.9402)} & \scriptsize{(2.6605)} & \scriptsize{(0.6935)} & \scriptsize{(1.2262)} & \scriptsize{(0.4453)} & \scriptsize{(0.6506)} & \scriptsize{(0.2950)} & \scriptsize{(0.2423)} & \scriptsize{(0.1445)} \\ 
\texttt{LLGMM} & 3.7622 & 1.4137 & 2.2216 & 1.1576 & 1.2392 & 0.8290 & 0.6704 & 0.6513 & 0.3701 & 0.4912 \\ 
& \scriptsize{(5.1293)} & \scriptsize{(0.9651)} & \scriptsize{(2.4444)} & \scriptsize{(0.6786)} & \scriptsize{(1.1947)} & \scriptsize{(0.4277)} & \scriptsize{(0.6291)} & \scriptsize{(0.2795)} & \scriptsize{(0.2670)} & \scriptsize{(0.1639)} \\ 
\midrule 

\multicolumn{10}{c}{Case: T.4}\\
\texttt{LLE}& 4.0527 & 1.5166 & 2.6907 & 1.2700 & 1.3481 & 0.8962 & 0.8114 & 0.7087 & 0.4206 & 0.5197 \\ 
& \scriptsize{(5.5110)} & \scriptsize{(1.0066)} & \scriptsize{(3.0123)} & \scriptsize{(0.7582)} & \scriptsize{(1.5300)} & \scriptsize{(0.5026)} & \scriptsize{(0.7705)} & \scriptsize{(0.3398)} & \scriptsize{(0.3113)} & \scriptsize{(0.1762)} \\ 
\texttt{LLWS} & 4.0469 & 1.5112 & 2.6848 & 1.2677 & 1.3505 & 0.8982 & 0.8081 & 0.7069 & 0.4182 & 0.5187 \\ 
& \scriptsize{(5.5581)} & \scriptsize{(1.0080)} & \scriptsize{(3.0041)} & \scriptsize{(0.7545)} & \scriptsize{(1.5160)} & \scriptsize{(0.5023)} & \scriptsize{(0.7699)} & \scriptsize{(0.3393)} & \scriptsize{(0.3099)} & \scriptsize{(0.1751)} \\ 
\texttt{LLGMM} & 3.9732 & 1.5366 & 2.7430 & 1.3131 & 1.5087 & 0.9630 & 0.9692 & 0.7872 & 0.4665 & 0.5290 \\ 
& \scriptsize{(3.6553)} & \scriptsize{(1.0011)} & \scriptsize{(2.9492)} & \scriptsize{(0.7331)} & \scriptsize{(1.5978)} & \scriptsize{(0.4878)} & \scriptsize{(0.7624)} & \scriptsize{(0.3092)} & \scriptsize{(0.3807)} & \scriptsize{(0.2279)} \\ 
\midrule 

\multicolumn{10}{c}{Case: T.5}\\
\texttt{LLE}& 3.6722 & 1.4702 & 2.1675 & 1.1302 & 1.0886 & 0.8118 & 0.6598 & 0.6403 & 0.3562 & 0.4830 \\ 
& \scriptsize{(4.7035)} & \scriptsize{(0.9218)} & \scriptsize{(2.5478)} & \scriptsize{(0.6788)} & \scriptsize{(1.1530)} & \scriptsize{(0.4285)} & \scriptsize{(0.6201)} & \scriptsize{(0.2855)} & \scriptsize{(0.2249)} & \scriptsize{(0.1358)} \\ 
\texttt{LLWS} & 3.6558 & 1.4658 & 2.1686 & 1.1320 & 1.0893 & 0.8114 & 0.6580 & 0.6398 & 0.3566 & 0.4834 \\ 
& \scriptsize{(4.7013)} & \scriptsize{(0.9181)} & \scriptsize{(2.5422)} & \scriptsize{(0.6778)} & \scriptsize{(1.1447)} & \scriptsize{(0.4279)} & \scriptsize{(0.6144)} & \scriptsize{(0.2830)} & \scriptsize{(0.2248)} & \scriptsize{(0.1361)} \\ 
\texttt{LLGMM} & 3.5403 & 1.4808 & 2.1296 & 1.1557 & 1.1967 & 0.8707 & 0.7413 & 0.6980 & 0.3502 & 0.4790 \\ 
& \scriptsize{(4.5039)} & \scriptsize{(0.9236)} & \scriptsize{(2.4036)} & \scriptsize{(0.6701)} & \scriptsize{(1.1644)} & \scriptsize{(0.4136)} & \scriptsize{(0.5899)} & \scriptsize{(0.2691)} & \scriptsize{(0.2418)} & \scriptsize{(0.1520)} \\ 
\bottomrule
\end{tabular}}
\end{table}

\begin{table}
\vspace{-2.8cm}
\tabcolsep 2pt
\centering
\caption{{Comparison among the proposed LLGMM with the local linear (LLE) and \citet{wei2017heteroskedasticity}'s estimators (LLWS) for $\beta_{3}(\cdot)$ with $\rho = 0.5$}}
\label{tab:3beta3-0.5}
{\begin{tabular}{rrrrrrrrrrrr}
\toprule
& & \multicolumn{2}{c}{n = 30} &
  \multicolumn{2}{c}{n = 50} &
  \multicolumn{2}{c}{n = 100} &
  \multicolumn{2}{c}{n = 200} &
  \multicolumn{2}{c}{n = 500}\\
\midrule    
& Method & IMSE & IMAE &  
  IMSE & IMAE & 
  IMSE & IMAE &
  IMSE & IMAE &
  IMSE & IMAE \\
\midrule 
&\multicolumn{10}{c}{Case: T.0}\\
&\texttt{LLE} & 2.9058 & 1.2975 & 1.6299 & 0.9565 & 0.7949 & 0.6843 & 0.4162 & 0.4843 & 0.1481 & 0.2909 \\ 
&& \scriptsize{(4.1942)} & \scriptsize{(0.8670)} & \scriptsize{(2.2539)} & \scriptsize{(0.6614)} & \scriptsize{(1.0063)} & \scriptsize{(0.4405)} & \scriptsize{(0.5441)} & \scriptsize{(0.3296)} & \scriptsize{(0.1835)} & \scriptsize{(0.1829)} \\ 
&\texttt{LLWS}& 2.8981 & 1.2965 & 1.6386 & 0.9585 & 0.8005 & 0.6855 & 0.4134 & 0.4823 & 0.1489 & 0.2914 \\ 
&& \scriptsize{(4.2126)} & \scriptsize{(0.8643)} & \scriptsize{(2.2978)} & \scriptsize{(0.6654)} & \scriptsize{(1.0124)} & \scriptsize{(0.4399)} & \scriptsize{(0.5371)} & \scriptsize{(0.3257)} & \scriptsize{(0.1855)} & \scriptsize{(0.1842)} \\ 
&\texttt{LLGMM}& 2.6330 & 1.2561 & 1.5015 & 0.9467 & 0.7889 & 0.7044 & 0.4367 & 0.5220 & 0.1934 & 0.3460 \\ 
&& \scriptsize{(4.6278)} & \scriptsize{(0.9365)} & \scriptsize{(2.1829)} & \scriptsize{(0.6307)} & \scriptsize{(1.0182)} & \scriptsize{(0.3937)} & \scriptsize{(0.5646)} & \scriptsize{(0.2735)} & \scriptsize{(0.2264)} & \scriptsize{(0.1460)} \\ 
\midrule

&\multicolumn{10}{c}{Case: T.1}\\
&\texttt{LLE} & 3.4259 & 1.3982 & 1.8563 & 1.0235 & 0.9818 & 0.7578 & 0.4790 & 0.5206 & 0.1879 & 0.3277 \\ 
&& \scriptsize{(5.1517)} & \scriptsize{(0.9554)} & \scriptsize{(2.6287)} & \scriptsize{(0.7038)} & \scriptsize{(1.2684)} & \scriptsize{(0.4953)} & \scriptsize{(0.6689)} & \scriptsize{(0.3498)} & \scriptsize{(0.2318)} & \scriptsize{(0.2071)} \\ 
&\texttt{LLWS}& 3.3861 & 1.3901 & 1.8511 & 1.0212 & 0.9856 & 0.7586 & 0.4735 & 0.5166 & 0.1903 & 0.3300 \\ 
&& \scriptsize{(5.1381)} & \scriptsize{(0.9525)} & \scriptsize{(2.6382)} & \scriptsize{(0.7059)} & \scriptsize{(1.2680)} & \scriptsize{(0.4941)} & \scriptsize{(0.6656)} & \scriptsize{(0.3467)} & \scriptsize{(0.2361)} & \scriptsize{(0.2086)} \\ 
&\texttt{LLGMM}& 2.8195 & 1.2590 & 1.6503 & 0.9728 & 0.9681 & 0.7458 & 0.4685 & 0.5111 & 0.1873 & 0.3248 \\ 
&& \scriptsize{(4.8701)} & \scriptsize{(0.9426)} & \scriptsize{(2.7355)} & \scriptsize{(0.6987)} & \scriptsize{(1.2366)} & \scriptsize{(0.4327)} & \scriptsize{(0.6451)} & \scriptsize{(0.2765)} & \scriptsize{(0.2822)} & \scriptsize{(0.1677)} \\ 
\midrule

&\multicolumn{10}{c}{Case: T.2}\\
&\texttt{LLE} & 3.1823 & 1.3546 & 1.6713 & 0.9685 & 0.9461 & 0.7306 & 0.4524 & 0.5015 & 0.1776 & 0.3176 \\ 
&& \scriptsize{(4.5371)} & \scriptsize{(0.9171)} & \scriptsize{(2.3302)} & \scriptsize{(0.6752)} & \scriptsize{(1.2673)} & \scriptsize{(0.5037)} & \scriptsize{(0.6760)} & \scriptsize{(0.3464)} & \scriptsize{(0.2194)} & \scriptsize{(0.1989)} \\ 
&\texttt{LLWS}& 3.1297 & 1.3359 & 1.6620 & 0.9624 & 0.9451 & 0.7302 & 0.4483 & 0.4987 & 0.1805 & 0.3206 \\ 
&& \scriptsize{(4.5271)} & \scriptsize{(0.9165)} & \scriptsize{(2.3358)} & \scriptsize{(0.6779)} & \scriptsize{(1.2495)} & \scriptsize{(0.5023)} & \scriptsize{(0.6776)} & \scriptsize{(0.3435)} & \scriptsize{(0.2218)} & \scriptsize{(0.1998)} \\ 
&\texttt{LLGMM}& 2.5181 & 1.1670 & 1.5578 & 0.9149 & 0.9383 & 0.7282 & 0.4452 & 0.4877 & 0.1764 & 0.3129 \\ 
&& \scriptsize{(5.6730)} & \scriptsize{(0.9750)} & \scriptsize{(3.3317)} & \scriptsize{(0.7419)} & \scriptsize{(1.5252)} & \scriptsize{(0.4698)} & \scriptsize{(0.7880)} & \scriptsize{(0.3133)} & \scriptsize{(0.3823)} & \scriptsize{(0.2244)} \\ 
\midrule

&\multicolumn{10}{c}{Case: T.3}\\
&\texttt{LLE} & 3.2804 & 1.3775 & 1.7902 & 1.0073 & 0.9352 & 0.7377 & 0.4639 & 0.5127 & 0.1775 & 0.3186 \\ 
&& \scriptsize{(4.9544)} & \scriptsize{(0.9220)} & \scriptsize{(2.4596)} & \scriptsize{(0.6888)} & \scriptsize{(1.2176)} & \scriptsize{(0.4858)} & \scriptsize{(0.6410)} & \scriptsize{(0.3446)} & \scriptsize{(0.2183)} & \scriptsize{(0.2009)} \\ 
&\texttt{LLWS}& 3.2564 & 1.3723 & 1.7889 & 1.0060 & 0.9413 & 0.7393 & 0.4611 & 0.5101 & 0.1791 & 0.3200 \\ 
&& \scriptsize{(4.9482)} & \scriptsize{(0.9206)} & \scriptsize{(2.4811)} & \scriptsize{(0.6920)} & \scriptsize{(1.2208)} & \scriptsize{(0.4857)} & \scriptsize{(0.6405)} & \scriptsize{(0.3420)} & \scriptsize{(0.2225)} & \scriptsize{(0.2025)} \\ 
&\texttt{LLGMM}& 2.9361 & 1.2996 & 1.6279 & 0.9844 & 0.9232 & 0.7368 & 0.4619 & 0.5101 & 0.1720 & 0.3109 \\ 
&& \scriptsize{(5.1293)} & \scriptsize{(0.9651)} & \scriptsize{(2.4444)} & \scriptsize{(0.6786)} & \scriptsize{(1.1947)} & \scriptsize{(0.4277)} & \scriptsize{(0.6291)} & \scriptsize{(0.2795)} & \scriptsize{(0.2670)} & \scriptsize{(0.1639)} \\ 
\midrule

&\multicolumn{10}{c}{Case: T.4}\\
&\texttt{LLE} & 3.0292 & 1.3058 & 1.5242 & 0.9301 & 0.8834 & 0.7045 & 0.4178 & 0.4810 & 0.1616 & 0.3016 \\ 
&& \scriptsize{(4.6230)} & \scriptsize{(0.9132)} & \scriptsize{(2.0874)} & \scriptsize{(0.6382)} & \scriptsize{(1.1701)} & \scriptsize{(0.4869)} & \scriptsize{(0.6464)} & \scriptsize{(0.3337)} & \scriptsize{(0.1967)} & \scriptsize{(0.1898)} \\ 
&\texttt{LLWS}& 2.9687 & 1.2889 & 1.5212 & 0.9246 & 0.8851 & 0.7044 & 0.4124 & 0.4769 & 0.1745 & 0.3084 \\ 
&& \scriptsize{(4.5720)} & \scriptsize{(0.9076)} & \scriptsize{(2.1073)} & \scriptsize{(0.6417)} & \scriptsize{(1.1668)} & \scriptsize{(0.4875)} & \scriptsize{(0.6462)} & \scriptsize{(0.3312)} & \scriptsize{(0.2883)} & \scriptsize{(0.2074)} \\ 
&\texttt{LLGMM}& 2.4915 & 1.1307 & 1.4612 & 0.8890 & 0.8843 & 0.7021 & 0.4109 & 0.4692 & 0.1675 & 0.3023 \\ 
&& \scriptsize{(3.6553)} & \scriptsize{(1.0011)} & \scriptsize{(2.9492)} & \scriptsize{(0.7331)} & \scriptsize{(1.5978)} & \scriptsize{(0.4878)} & \scriptsize{(0.7624)} & \scriptsize{(0.3092)} & \scriptsize{(0.3807)} & \scriptsize{(0.2279)} \\ 
\midrule

&\multicolumn{10}{c}{Case: T.5}\\
&\texttt{LLE} & 3.2795 & 1.3741 & 1.8026 & 1.0088 & 0.9343 & 0.7399 & 0.4655 & 0.5147 & 0.1765 & 0.3180 \\ 
&& \scriptsize{(4.8809)} & \scriptsize{(0.9269)} & \scriptsize{(2.4980)} & \scriptsize{(0.6933)} & \scriptsize{(1.1991)} & \scriptsize{(0.4807)} & \scriptsize{(0.6362)} & \scriptsize{(0.3441)} & \scriptsize{(0.2173)} & \scriptsize{(0.2000)} \\ 
&\texttt{LLWS}& 3.2583 & 1.3698 & 1.8043 & 1.0081 & 0.9409 & 0.7414 & 0.4621 & 0.5117 & 0.1783 & 0.3196 \\ 
&& \scriptsize{(4.8913)} & \scriptsize{(0.9288)} & \scriptsize{(2.5296)} & \scriptsize{(0.6977)} & \scriptsize{(1.2058)} & \scriptsize{(0.4814)} & \scriptsize{(0.6342)} & \scriptsize{(0.3409)} & \scriptsize{(0.2212)} & \scriptsize{(0.2016)} \\ 
&\texttt{LLGMM}& 2.8400 & 1.2891 & 1.6410 & 0.9832 & 0.9131 & 0.7406 & 0.4613 & 0.5125 & 0.1788 & 0.3171 \\ 
&& \scriptsize{(4.5039)} & \scriptsize{(0.9236)} & \scriptsize{(2.4036)} & \scriptsize{(0.6701)} & \scriptsize{(1.1644)} & \scriptsize{(0.4136)} & \scriptsize{(0.5899)} & \scriptsize{(0.2691)} & \scriptsize{(0.2418)} & \scriptsize{(0.1520)} \\ 
\bottomrule
\end{tabular}}
\end{table}
}

\section{Technical details}
\label{Chapter3-gmm-Section:technical}
In this section, we provide technical details of the proposed theorems in Section \ref{Chapter3-gmm-Section:theory}. 
We prove theorems \ref{Chapter3-gmm-Theorem:init} and \ref{Chapter3-gmm-Theorem:final} by proving the following lemmas. 
\subsection{Some useful lemmas}
\begin{lemma}
\label{Chapter3-gmm-Lemma:tight}
Under the conditions \ref{Chapter3-gmm-Cond:donsker}
$\frac{1}{\sqrt{n}}\sum\limits_{i=1}^{n}U_{i}(s_{0})\fM(\bX_{i})$ is tight. 
\end{lemma}
\begin{proof}
Consider the class of function $\sC = \{ U(s_{0})\fM(\bX_{i}): s_{0}\in [0,1]\}$. Therefore, due to the assumption \ref{Chapter3-gmm-Cond:donsker}, $\sC$ is a P-Donsker class. Therefore,  $\frac{1}{\sqrt{n}}\sum\limits_{i=1}^{n}U_{i}(s_{0})\fM(\bX_{i})$ is tight.
\end{proof}
\begin{lemma}
\label{Chapter3-gmm-Lemma:smapling-distribution}
Under the assumptions \ref{Chapter3-gmm-Cond:kernel}, \ref{Chapter3-gmm-Cond:density} and \ref{Chapter3-gmm-Cond:limit}, the following holds for any power $c \geq 0$. 
\begin{equation}
    \sup_{s\in [0,1]} \left|\int K_{h}(t-s)\left\{
    (t-s)/h\right\}^{c}d\Pi(t) - \Pi(t)\right| = O(1/(rh)^{-1/2}).
\end{equation}
The above bound can be replaced by $O(1/rh)$ for fixed design case. 
\end{lemma}
\begin{proof}
This can be proved by using the empirical process techniques by observing that the class $\left\{K((\cdot-s /h)) ((\cdot-s /h))^{c}: s\in [0,1] \right\}$ is a P-Donsker class \citep{zhu2012multivariate}. 
For the  balanced case, the results can be shown using Tayler's series expansion. 
\end{proof}

\begin{lemma}
\label{Chapter3-gmm-Lemma:Ibound}
Define $\bI(s_{0})
    = \frac{1}{nr}\sum\limits_{i=1}^{n}\sum\limits_{j=1}^{r}K_{h}(s_{j}-s_{0})\bQ_{ij}(s_{0})\bW_{ij}(s_{0})^{\tp}$. Under the conditions \ref{Chapter3-gmm-Cond:kernel}, \ref{Chapter3-gmm-Cond:density}, \ref{Chapter3-gmm-Cond:X} and \ref{Chapter3-gmm-Cond:limit}
    $\bI(s_{0}) = f(s_{0})\diag(1, \nu_{21}) \otimes \bOmega + O(h + \delta_{n1}(h))$  almost surely, where $\bOmega = \E\{\fM(\bX)\bX^{\tp}\}$. 
\end{lemma}
\begin{proof}
Observe the following.
\begingroup
\allowdisplaybreaks
\begin{align*}
    \bI(s_{0})
    &= \frac{1}{nr}\sum_{i=1}^{n}\sum_{j=1}^{r}K_{h}(s_{j}-s_{0})
        \bQ_{ij}(s_{0})\bW_{ij}(s_{0})^{\tp}\\
    &= \frac{1}{nr}\sum_{i=1}^{n}\sum_{j=1}^{r}K_{h}(s_{j}-s_{0})
        \left\{\bz_{h}(s_{j}-s_{0}) \otimes \fM(\bX_{i}) \right\}
        \left\{\bz_{h}(s_{j}-s_{0}) \otimes \bX_{i} \right\}^{\tp}\\
    &= \frac{1}{nR}\sum_{i=1}^{n}\sum_{j=1}^{r}
        K_{h}(s_{j}-s)
        \left\{\bz_{h}(s_{j}-s_{0})^{\otimes^{2}} \otimes \fM(\bX_{i})\bX_{i}^{\tp} \right\}\\
    &= \frac{1}{nr}\sum_{i=1}^{n}\sum_{j=1}^{r}K_{h}(s_{j}-s_{0})
        \begin{pmatrix}
            \fM(\bX_{i})\bX_{i}^{\tp} & \fM(\bX_{i})\bX_{i}^{\tp}(s_{j}-s_{0})/h\\
            \fM(\bX_{i})\bX_{i}^{\tp}(s_{j}-s_{0})/h & \fM(\bX_{i})\bX_{i}^{\tp}((s_{j}-s_{0})/h)^{2}\\
        \end{pmatrix}\\
    &:= \begin{pmatrix}
        {\bI}_{11}(s_{0}) & 
        {\bI}_{12}(s_{0})\\
        {\bI}_{21}(s_{0}) &
        {\bI}_{22}(s_{0})
    \end{pmatrix}.
    \numberthis
\end{align*}
\endgroup
Let us define $\bI_{a, b} = \frac{1}{nr}\sum\limits_{i=1}^{n}\sum\limits_{j=1}^{r} K_{h}(s_{j}-s_{0})(s_{j}-s_{0})^{a+b}\fM(\bX_{i})\bX_{i}^{\tp}$. Assume that $\nu_{41}$ is finite and due to condition \ref{Chapter3-gmm-Cond:density}, for general index $c$, we can derive the uniform bound of for all $s_{0}\in \sS$.
\begingroup
\allowdisplaybreaks
\begin{align*}
    \E\{\bI_{a,b}(s_{0})\} &= \E\left\{
        \frac{1}{nr}\sum_{i=1}^{n}\sum_{j = 1}^{r}
        K_{h}(s_{j} - s_{0})((s_{j}-s_{0})/h)^{c}\fM(\bX_{i})\bX_{i}^{\tp}
        \right\}\\
    &= \bOmega\E\left\{
        \frac{1}{r}\sum_{j=1}^{r}
        K_{h}(s_{j} - s_{0})((s_{j}-s_{0})/h)^{c}
        \right\}\\
    &= \bOmega\int K_{h}(u - s_{0})((u-s_{0})/h)^{c}f(u)du\\
    &= \bOmega\int K(u)u^{c}f(s_{0}+hu)du\\
    &= \bOmega\int K(u)u^{c}\left\{
                f(s_{0}) + huf'(s_{0}) + 0.5h^{2}u^{2}f''(s_{0}) + \cdots
            \right\} du\\
    &= \bOmega\begin{cases} 
      f(s_{0}) + O(h^{2}) & c = 0, \text{if $\nu_{21} < \infty$, $f''$ exists and finite}\\
      O(h) & c = 1, \text{if $\nu_{21} < \infty$, $f'$ exists and finite}\\
      f(s_{0})\nu_{21} + O(h^{2}) & c = 2, \text{if $\nu_{41} < \infty$, $f''$ exists and finite}\\
       O(h) & c = 3, \text{if $\nu_{41} < \infty$, $f'$ exists and finite.}
   \end{cases}
    \numberthis
\end{align*}
\endgroup
Moreover, under the condition \ref{Chapter3-gmm-Cond:X}, we have $\E\|\bX\|^{a}$ is finite for some $a > 2$ and can define, $b_{n} = h^{2} + h/r$ where $h \rightarrow 0$ such that $b_{n}^{-1}(\log n/n)^{1-2/a} = o(1)$. 
Thus, $\delta_{n1}(h) = \{b_{n}\log n/nh^{2}\}^{1/2}$.
Now to establish the uniform bound for $\bI(s_{0})$, by using Lemma 2 in \citet{li2010uniform} for each of $\bI_{a,b}(s_{0})$ for $a, b = 1, 2$, we have
\begin{equation}
\bI(s_{0}) = f(s_{0})(\diag(1, \nu_{21}))\otimes\bOmega + O(h + \delta_{n1}(h)) \qquad \text{almost surely.}
\end{equation}
\end{proof}
\begin{lemma}
\label{Chapter3-gmm-Lemma:Jbound}
Define, $\bJ(s_{0}) = \frac{1}{nr}\sum\limits_{i = 1}^{n}\sum\limits_{j = 1}^{r}
        K_{h}(s_{j}-s_{0})
        \bQ_{ij}(s_{0})\bX_{i}^{\tp}\bbeta_{0}(s_{j})$. 
Thus, under the conditions \ref{Chapter3-gmm-Cond:kernel}, \ref{Chapter3-gmm-Cond:density}, \ref{Chapter3-gmm-Cond:beta}, \ref{Chapter3-gmm-Cond:X} and \ref{Chapter3-gmm-Cond:limit}, 
$\bJ(s_{0}) - \bI(s_{0})\bgamma_{0}(s_{0}) = 0.5h^{2}\{f(s_{0})(\nu_{21}, 0)^{\tp}\otimes\bOmega\}\ddot{\bbeta}(s_{0}) + O(\delta_{n1}(h)+h)$ almost surely, where $\bgamma_{0}(s_{0}) = (\bbeta_{0}(s_{0})^{\tp},h\dot{\bbeta}_{0}(s_{0})^{\tp})^{\tp}$.
Moreover, $$\bT(s_{0}) = \frac{1}{nr}\sum_{i=1}^{n}\sum_{j=1}^{r}K_{h}(s_{j}-s_{0})\bQ_{ij}(s_{0})U_{ij} = O(\delta_{n1}(h))$$ almost surely.
\end{lemma}
\begin{proof}
Observe that, because of condition \ref{Chapter3-gmm-Cond:beta}, using Taylor's series expansion,  
\begin{align*}
    \bJ(s_{0}) &= \frac{1}{nr}\sum_{i=1}^{n}\sum_{j=1}^{r}K_{h}(s_{j}-s_{0})\bQ_{ij}(s_{0})\bX_{i}^{\tp}\bbeta_{0}(s_{0})\\
        &=\frac{1}{nr}\sum_{i = 1}^{n}\sum_{j = 1}^{r}
        K_{h}(s_{j}-s_{0})
        \bQ_{ij}(s_{0})\\
        &\times 
        \left\{
            \bX_{i}^{\tp}\bbeta_{0}(s_{0}) + (s_{j} - s_{0})\bX_{i}^{\tp}\dot{\bbeta}_{0}(s_{0}) 
            + 0.5(s_{j} - s_{0})^{2}\bX_{i}^{\tp}\ddot{\bbeta}_{0}(s_{0}) 
        \right\} + o(h^{2})\\
        &= \bI(s_{0})\bgamma_{0}(s_{0}) + 0.5h^{2}\bI_{21}(s_{0})\ddot{\bbeta}_{0}(s_{0}) + o(h^{2}).
    \numberthis
\end{align*}
Using similar arguments, due to Lemma 2 in \citep{li2010uniform}, under the conditions \ref{Chapter3-gmm-Cond:kernel}, \ref{Chapter3-gmm-Cond:density} and \ref{Chapter3-gmm-Cond:X}, with $\nu_{41}$ being finite, we have
\begin{align*}
    \bI_{21}(s_{0}) &= \frac{1}{nr}\sum_{i=1}^{n}\sum_{j=1}^{r}
            K_{h}(s_{j}-s_{0})
            \begin{pmatrix}
                ((s_{j}-s_{0})/h)^{2}\\
                ((s_{j}-s_{0})/h)^{3}
            \end{pmatrix}
            \fM(\bX_{i})\bX_{i}^{\tp}\\
            & = f(s_{0})(\nu_{21}, 0)^{\tp}\otimes\bOmega + O(\delta_{n1}(h) + h) \qquad \text{almost surely}
        \numberthis
\end{align*}
and
\begin{align*}
    \bT(s_{0}) = &\begin{pmatrix}
        \frac{1}{nr}\sum_{i = 1}^{n}\sum_{j = 1}^{r}
        K_{h}(s_{j}-s_{0})\fM(\bX_{i})U_{ij}\\
        \frac{1}{nr}\sum_{i = 1}^{n}\sum_{j = 1}^{r}
        K_{h}(s_{j}-s_{0})((s_{j}-s_{0})/h)\fM(\bX_{i})U_{ir}
        \end{pmatrix}\\
        &= O(\delta_{n1}(h)) \qquad \text{almost surely.}
    \numberthis
\end{align*}
\end{proof}

\begin{lemma}
\label{Chapter3-gmm-Lemma:Tdist}
Under conditions \ref{Chapter3-gmm-Cond:kernel},\ref{Chapter3-gmm-Cond:density}, \ref{Chapter3-gmm-Cond:X}, \ref{Chapter3-gmm-Cond:donsker}, \ref{Chapter3-gmm-Cond:limit},
$\sqrt{n}\bT(s_{0})(1+o_{a.s.}(1)) \xrightarrow{d} N(0, f^{2}(s_{0})(\sU \otimes  \bSigma_{\fM}^{*}(s_{0}, s_{0}))$ where $\bT(s_{0})$ is defined in Lemma \ref{Chapter3-gmm-Lemma:Jbound}, where the element of $(l, l')$ of the matrix $\sU$ is $\nu_{l-1}\nu_{l'-1}$ and $\bSigma_{\fM}^{*}(s_{0}, s_{0}) = \lim\limits_{n\rightarrow \infty}\frac{1}{n}\sum\limits_{i=1}^{n}\E\{\fM(\bX_{i})\fM(\bX_{i})^{\tp}\Sigma_{\bX_{i}}(s_{0}, s_{0})\}$.
\end{lemma}
\begin{proof}
Note that 
\begin{equation}
        \sqrt{n}\bT(s_{0}) = \frac{1}{\sqrt{n}r}\sum_{i=1}^{n}\sum_{j = 1}^{r}
        K_{h}(s_{j} - s_{0})
        \left[\bz_{h}(s_{j}-s_{0}) \otimes \fM(\bX_{i})\right]U_{ij}
\end{equation}
Therefore, the variance of the above quantity is
\begingroup
\allowdisplaybreaks
\begin{align*}
    &\Var\{\sqrt{n}\bT(s_{0})\}\\
    &= \frac{1}{n}\E\left\{
            \sum_{i=1}^{n}\sum_{j=1}^{r}\sum_{j'=1}^{r}
            K_{h}(s_{j}-s_{0})K_{h}(s_{j'}-s_{0})
            \left[
                \bz_{h}(s_{j}-s_{0})\bz_{h}(s_{j'}-s_{0})^{\tp} 
                \otimes\fM(\bX_{i})^{\otimes^{2}}
            \right]
            U_{ij}U_{ij'}
        \right\}\\
        &=\frac{1}{n}\sum_{i=1}^{n}\E\left\{
            \frac{1}{r^{2}}
            \sum_{j=1}^{r}\sum_{j' = 1}^{r}
            K_{h}(s_{j}-s_{0})K_{h}(s_{j'}-s_{0})
            \left[\bz_{h}(s_{j}-s_{0})\bz_{h}(s_{j'}-s_{0})
            \otimes\fM(\bX_{i})^{\otimes^{2}}\right]
            \Sigma_{\bX_{i}}(s_{j}, s_{j'})
        \right\}\\
        &=\E\left\{
            \frac{1}{r^{2}}
            \sum_{j=1}^{r}\sum_{j' = 1}^{r}
            K_{h}(s_{j}-s_{0})K_{h}(s_{j'}-s_{0})
            \bz_{h}(s_{j}-s_{0})\bz_{h}(s_{j}-s_{0})^{\tp}
            \otimes\bSigma^{*}_{\fM}(s_{j}, s_{j'})
        \right\}\\
        &=\E\{\bD_{1}(s_{0})\} + \E\{\bD_{2}(s_{0})\},
    \numberthis
\end{align*}
\endgroup
% \begingroup
% \allowdisplaybreaks
% \begin{align*}
%     &\Var\{\sqrt{n}\bT(s_{0})\}\\
%     &= \frac{1}{n}\E\left\{
%             \sum_{i=1}^{n}\sum_{j=1}^{r}\sum_{j'=1}^{r}
%             K_{h}(s_{j}-s_{0})K_{h}(s_{j'}-s_{0})
%             \left[
%                 \bz_{h}(s_{j}-s_{0})\bz_{h}(s_{j'}-s_{0})^{\tp} 
%                 \otimes\fM(\bX_{i})^{\otimes^{2}}
%             \right]
%             U_{ij}U_{ij'}
%         \right\}\\
%         &=\frac{1}{n}\sum_{i=1}^{n}\E\left\{
%             \frac{1}{r^{2}}
%             \sum_{j=1}^{r}\sum_{j' = 1}^{r}
%             K_{h}(s_{j}-s_{0})K_{h}(s_{j'}-s_{0})
%             \left[\bz_{h}(s_{j}-s_{0})\bz_{h}(s_{j'}-s_{0})
%             \otimes\fM(\bX_{i})^{\otimes^{2}}\right]
%             \Sigma(s_{j}, s_{j'})
%         \right\}\\
%         &=\E\left\{
%             \frac{1}{r^{2}}
%             \sum_{j=1}^{r}\sum_{j' = 1}^{r}
%             K_{h}(s_{j}-s_{0})K_{h}(s_{j'}-s_{0})
%             \bz_{h}(s_{j}-s_{0})\bz_{h}(s_{j}-s_{0})^{\tp}
%             \Sigma(s_{j}, s_{j'})
%         \right\}\otimes\bOmega\\
%         &=\left[\E\{\bD_{1}(s_{0})\} + \E\{\bD_{2}(s_{0})\}\right]\otimes\bOmega,
%     \numberthis
% \end{align*}
% \endgroup
where $\bSigma^{*}_{\fM}(s, s') = \lim\limits_{n\rightarrow \infty}\frac{1}{n}\sum\limits_{i=1}^{n}\E\{\fM(\bX_{i})^{\otimes^{2}}\Sigma_{\bx_{i}}(s, s')\}$,
$\bD_{1}(s_{0}) = \frac{1}{r^{2}}\sum_{j=1}^{n}K^{2}_{h}(s_{j}-s_{0})\bz_{h}(s_{j}-s_{0})^{\otimes^{2}}\otimes\bSigma_{\fM}^{*}(s_{j}, s_{j})$ and $\bD_{2}(s_{0}) = \frac{1}{r(r-1)}\underset{j\neq j'}{\sum_{j=1}^{n}\sum_{j'=1}^{r}}K_{h}(s_{j}-s_{0})K_{h}(s_{j'}-s_{0})\bz_{h}(s_{j}-s_{0})\bz_{h}(s_{j'}-s_{0})^{\tp}\otimes\bSigma_{\fM}^{*}(s_{j}, s_{j'})$. Note that 
\begingroup
\allowdisplaybreaks
\begin{align*}
\label{Chapter3-gmm-Eq:D21}
\E\{\bD_{1}(s_{0})\} &= 
        \E\left\{
            \frac{1}{r^{2}}\sum_{j=1}^{r}K^{2}(s_{j}-s_{0})
            \bz_{h}(s_{j}-s_{0})^{\otimes^{2}}
            \otimes\bSigma_{\fM}^{*}(s_{j}, s_{j})
        \right\}\\
        &= \frac{1}{r}\int K_{h}^{2}(t-s_{0})\bz_{h}(t-s_{0})^{\otimes^{2}}\otimes\bSigma_{\fM}^{*}(t, t) f(t)dt\\
        &= \frac{1}{hr}\int K^{2}(t)
        \begin{pmatrix}
            1 & t\\
            t & t^{2}
        \end{pmatrix}
        \otimes\bSigma_{\fM}^{*}(s_{0}+hu, s_{0}+hu)f(s_{0}+ht)dt\\
        &= \frac{1}{hr}\left\{ f(s_{0})\diag(\nu_{02}, \nu_{22})\otimes\bSigma_{\fM}^{*}(s_{0}, s_{0}) + O(h) \right\}.
\numberthis
\end{align*}
\endgroup
Now assume that $\Theta(s_{0}) = \E\{\bD_{2}(s_{0})\}$ with $(l, l')$-th entry $\theta_{l, l'}$ and $\sP(t) = \int_{\sS}K_{h}(t-s_{0})K_{h}(t'-s_{0})\bz_{h}(t-s_{0})\bz_{h}(t'-s_{0})\otimes\bSigma_{\fM}^{*}(t, t')f(t')dt'$ with $(l, l')$-the block element $\sP_{l, l'}$.
Therefore, using H\'ajek projection \citep{vaart1996weak}, we have 
\begin{equation}
    \bD_{2, l. l'}(s_{0})= \theta_{l, l'}(s_{0}) + \frac{2}{r}\sum_{j = 1}^{r}\left\{
        \sP_{l, l'}(s_{j}) - \theta_{l, l'}(s_{0})
    \right\} + \tilde{\epsilon}_{l, l'}(s_{0}),
\end{equation}
where $\frac{2}{r}\sum_{j = 1}^{r}\left\{
        \sP_{l, l'}(s_{j}) - \theta_{l, l'}(s_{0})
    \right\}$ is the projection on $\bD_{2, l, l'}(s_{0}) - \theta_{l, l'}(s_{0})$ onto the set of all statistics of the linear order form. 
Thus, it is easy to see $\Var\{\tilde{\epsilon}\} = O(1/(rh)^{2})$ 
\citep{zhu2012multivariate}. Since the Taylor series expansion for small $h \rightarrow 0$, we have $\theta_{l, l'}(s_{0}) = f(s_{0})^{2}\nu_{ l-1,}\nu_{l'-1,1}\bSigma_{\fM}^{*}(s_{0}, s_{0})$.
Therefore, in summary, 
we have $\Var\{\sqrt{n}\bT(s_{0})\} = f^{2}(s_{0})\sU\otimes\bSigma_{\fM}^{*}(s_{0}, s_{0})$, where the element $(l, l')$ of the matrix $\sU$ is $\nu_{l-1}\nu_{l'-1}$.

% Therefore, observe the following two cases, under the conditions \ref{Chapter3-gmm-Cond:density}, \ref{Chapter3-gmm-Cond:beta}

% \begin{equation}
% \label{Chapter3-gmm-Eq:D22}
%     \begin{split}
%         \E\left\{ \bD_{2}(s_{0}) \right\} &= \E\left\{
%             \frac{1}{r^{2}}
%             \underset{j \neq j' }{\sum_{j=1}^{r}\sum_{j'=1}^{r}}
%             K_{h}(s_{j}-s_{0})K_{h}(s_{j'}-s_{0})
%             \Sigma(s_{j}, s_{j'})
%         \right\}\\
%         &\sim  \int\int K_{h}(t-s_{0})K_{h}(t'-s_{0})
%             \Sigma(u, v)f(u)f(v)dudv\\
%         &= \int\int K(t)K(t')\Sigma(s_{0}+ht, s_{0}+ht')
%             f(s_{0}+ht)f(s_{0}+ht')dtdt'\\
%         &=  f^{2}(s_{0})\Sigma(s_{0}, s_{0}) + O(h^{2})
%     \end{split}
% \end{equation}
% Combining Equations (\ref{Chapter3-gmm-Eq:D21}) and (\ref{Chapter3-gmm-Eq:D22}), we obtain, 
% \begin{equation}
%     \Var\{\sqrt{n}\bT(s_{0})\} = f^{2}(s_{0})\Sigma(s_{0}, s_{0}) \otimes \bOmega + o(1)
% \end{equation}

To hold the above asymptotic results, we need to show that $ \sqrt{n}\bT(s_{0})$ be tight asymptotically. 
Therefore, consider the following, for suitable choice of $\underline{l}< \overline{l}$ after change of variables,
%\sqrt{n}\bI(s_{0})^{-1}\bT(s_{0})[1+o_{a.s.}(1)]\\
\begingroup
\allowdisplaybreaks
\begin{align*}
    &\sqrt{n}\bT(s_{0})
    = \frac{1}{\sqrt{n}r}\sum_{i=1}^{n}\sum_{r=1}^{r}K_{h}(s_{j}-s_{0})[\bz_{h}(s_{j}-s_{0}) \otimes \fM(\bX_{i})]U_{ij}\\
        &=\frac{1}{\sqrt{n}}\sum_{i=1}^{n}\left\{
            \frac{1}{r}\sum_{j=1}^{r}K_{h}(s_{j}-s_{0})\bz_{h}(s_{j}-s_{0})U_{ij}
            -\int_{0}^{1}K_{h}(t-s_{0})\bz_{h}(t-s_{0})U_{i}(t)f(t)dt
        \right\}\otimes\fM(\bX_{i})\\
        &\qquad + \frac{1}{\sqrt{n}}\sum_{i=1}^{n}U_{i}(s_{0})
        \int_{\underline{l}}^{\overline{l}}K(t)(1, t)^{\tp}f(s_{0}+ht)dt \otimes\fM(\bX_{i})\\
        &\qquad + \frac{1}{\sqrt{n}}\sum_{i=1}^{n}
        \int_{\underline{l}}^{\overline{l}}
        K(t)(1, t)^{\tp}\left\{
            U_{i}(s_{0} + ht) - U_{i}(s_{0})
        \right\}f(s_{0}+ht)dt \otimes\fM(\bX_{i})\\
        &:= \bT_{1}(s_{0}) + \bT_{2}(s_{0}) + \bT_{3}(s_{0}).
    \numberthis
\end{align*}
\endgroup
Note that,
\begingroup
\allowdisplaybreaks
\begin{align*}
    &\bT_{1}(s_{0})\\
    &= \frac{1}{r}\sum_{r=1}^{r}K_{h}(s_{j}-s_{0})\bz_{h}(s_{j}-s_{0})\left\{
        \frac{1}{\sqrt{n}}\sum_{i=1}^{n}U_{ij}\otimes\fM(\bX_{i}) - \frac{1}{\sqrt{n}}\sum_{i=1}^{n}U_{i}(t)\otimes\fM(\bX_{i})
        \right\}\\
        &\qquad + \left\{
            \frac{1}{r}\sum_{j=1}^{r}K_{h}(s_{j}-s_{0})\bz_{h}(s_{j}-s_{0}) - \int_{\underline{l}}^{\overline{l}}K_{h}(t-s_{0})\bz_{h}(t-s_{0})f(t)dt
        \right\}\\
        &\qquad\qquad\times
        \left\{
            \frac{1}{\sqrt{n}}\sum_{i=1}^{n}U_{i}(t)\otimes\fM(\bX_{i})
        \right\}\\
        & \qquad + \int_{\underline{l}}^{\overline{l}}K_{h}(t-s_{0})\bz_{h}(t-s_{0})\frac{1}{\sqrt{n}}\sum_{i=1}^{n}\left\{
            U_{i}(s_{0})-U_{i}(s)
        \right\}f(t)dt\otimes\fM(\bX_{i})\\
        &:= \bT_{11}(s_{0}) + \bT_{12}(s_{0}) + \bT_{13}(s_{0}).
    \numberthis
\end{align*}
\endgroup
Due to the Donsker Theorem, we have $\frac{1}{\sqrt{n}}\sum\limits_{i=1}^{n}\fM(\bX_{i})U_{i}(s)$ weekly converges to a centered Gaussian process and $\sup\limits_{s\in[0,1]}\left|\frac{1}{\sqrt{n}}\sum\limits_{i=1}^{n}\fM(\bX_{i})U_{i}(s)\right| = O_{p}(1)$ \citep{vaart1996weak}.
Therefore, 
\begingroup
\allowdisplaybreaks
\begin{align*}
    &|\bT_{11}(s_{0})| 
    \leq \frac{1}{r}\sum_{j=1}^{r}K_{h}(s_{j}-s_{0})\|\bz_{h}(s_{j}-s_{0})\|_{2}
        \left|\frac{1}{\sqrt{n}}\sum_{i=1}^{n}U_{ij}\otimes\fM(\bX_{i}) - 
        \frac{1}{\sqrt{n}}\sum_{i=1}^{n}U_{i}(s)\otimes\fM(\bX_{i})\right|\\
        & \leq \frac{1}{r}\sum_{j=1}^{r}K_{h}(s_{j}-s_{0})\|\bz_{h}(s_{j}-s_{0})\|_{2}\\
        & \qquad\times
        \sup_{|s-s_{0}|\leq h}
        \left|\frac{1}{\sqrt{n}}\sum_{i=1}^{n}U_{i}(s)\otimes\fM(\bX_{i}) - 
        \frac{1}{\sqrt{n}}\sum_{i=1}^{n}U_{i}(s_{0})\otimes\fM(\bX_{i})\right|\\
        &= o_{P}(1).
    \numberthis
\end{align*}
\endgroup
\begingroup
\allowdisplaybreaks
\begin{align*}
    &|\bT_{12}(s_{0})| \\
    & \leq \left|\frac{1}{r}\sum_{j=1}^{r}K_{h}(s_{j}-s_{0})\bz_{h}(s_{j}-s_{0})
        -\int_{\underline{l}}^{\overline{l}}K_{h}(t-s_{0})\bz_{h}(t-s_{0})f(t)dt\right|\\
    & \qquad \times \sup_{t\in [0,1]}\left\{
           \frac{1}{\sqrt{n}}\sum_{i=1}^{n}U_{i}(t)\otimes\fM(\bX_{i})
        \right\}\\
        &= O_{P}(1/\sqrt{rh})O_{P}(1) = o_{P}(1).
    \numberthis
\end{align*}
\endgroup
The above bound holds for Lemma \ref{Chapter3-gmm-Lemma:smapling-distribution} and Condition \ref{Chapter3-gmm-Cond:limit} so that $mh \rightarrow \infty$. 
\begingroup
\allowdisplaybreaks
\begin{align*}
    &|\bT_{13}(s_{0})| \\
    &\leq
        \sup_{|s-s_{0}|\leq h}
        \left|
        \frac{1}{\sqrt{n}}\sum_{i=1}^{n}U_{i}(s)\otimes \fM(\bX_{i}) - \frac{1}{\sqrt{n}}\sum_{i=1}^{n}U_{i}(s_{0})\otimes \fM(\bX_{i})  \right|\\
    & \times \int_{\underline{l}}^{\overline{l}}K_{h}(s_{j}-s_{0})\|\bz_{h}(s_{j}-s_{0})\|_{2}f(s)ds\\
        &= O_{P}(1).
    \numberthis
\end{align*}
\endgroup
By combining the above three bounds, due to conditions \ref{Chapter3-gmm-Cond:kernel},\ref{Chapter3-gmm-Cond:density}, \ref{Chapter3-gmm-Cond:X}, \ref{Chapter3-gmm-Cond:donsker}, \ref{Chapter3-gmm-Cond:limit}, 
we obtain $\bT_{1}(s_{0}) = o_{P}(1)$. 
Now, rewrite $\bT_{3}(s_{0})$ as 
\begingroup
\allowdisplaybreaks
\begin{align*}
    &\bT_{3}(s) = 
        \frac{1}{\sqrt{n}}\sum_{i=1}^{n}
        \int_{\underline{l}}^{\overline{l}}
        K(t)(1, t)^{\tp}\left\{
            U_{i}(s_{0} + ht) - U_{i}(s_{0})
        \right\}f(s_{0}+ht)dt \otimes\fM(\bX_{i})\\
        &= \int_{\underline{l}}^{\overline{l}}
        K(t)(1, t)^{\tp}\otimes
        \left\{U_{i}(s_{0} + ht) - U_{i}(s_{0})\right\}\fM(\bX_{i})
        f(s_{0}+ht)dt.
    \numberthis
\end{align*}
\endgroup
Since, $\frac{1}{\sqrt{n}}\sum\limits_{i=1}^{n}\fM(\bX_{i})U_{i}(s_{0})$ is asymptotically tight, for any $h \rightarrow 0$, we have the following \citep{vaart1996weak}.
\begin{equation}
        \sup_{s_{0}\in [0, 1]: |t| \leq 1}\frac{1}{\sqrt{n}}\sum_{i=1}^{n}
        \fM(\bX_{i})\left\{U_{i}(s_{0} + ht)-U_{i}(s_{0})\right\} = o_{P}(1).
\end{equation}
Now it is enough to show that $\bT_{2}(s_{0})$ is tight. First, observe that 
\begingroup
\allowdisplaybreaks
\begin{align*}
    &(1, 0)\int_{\underline{l}}^{\overline{l}}K(t)\diag(1, \nu_{21}^{-1})(1, t)^{\tp}f(s_{0}+ht)dt\\
        &=\int_{\underline{l}}^{\overline{l}}K(t)f(s_{0}+ht)dt\\
        &=\int_{\underline{l}}^{\overline{l}}K(t)\left\{f(s_{0})+htf'(s_{0}) + \cdots \right\}\\
        &= f(s_{0}) + o(h).
    \numberthis
\end{align*}
\endgroup
Therefore, $\bT_{2}(s_{0})(1+o_{P}(h)) = \frac{1}{\sqrt{n}}\sum\limits_{i=1}^{n}U_{i}(s_{0})\otimes\fM(\bX_{i})$. By assumption \ref{Chapter3-gmm-Cond:donsker}, $\bT_{2}(s_{0})$ is tight.
\end{proof}

\subsection{Proof of Theorem \ref{Chapter3-gmm-Theorem:init}}
Under the initial estimates, by considering $\fM(\bX) = \bX$, $\bOmega$ can be replaced by $\bOmega_{\bx}$ in Equation (\ref{Chapter3-gmm-Eq:Iinv}) and inverse of $\bOmega_{\bx}$ exits. Therefore, by using Lemma \ref{Chapter3-gmm-Lemma:Ibound}, it is easy to observe that, almost surely
\begin{equation}\label{Chapter3-gmm-Eq:Iinv}
    \bI(s_{0})^{-1} = f(s_{0})^{-1}(\diag(1, \nu_{21})^{-1})\otimes\bOmega_{\bx}^{-1} + O(h + \delta_{n1}(h)). 
\end{equation}
Similarly, for the numerator, we have the following.
\begin{align*}
    \label{Chapter3-gmm-Eq:Y-expansion}
    &\frac{1}{nr}\sum_{i = 1}^{n}\sum_{j = 1}^{r}
        K_{h}(s_{j}-s_{0})
        \left\{\bz_{h}(s_{j}-s_{0}) \otimes\bX_{i}\right\}Y_{ij}\\
        &= \frac{1}{nr}\sum_{i = 1}^{n}\sum_{j = 1}^{r}
        K_{h}(s_{j}-s_{0})
        \left\{\bz_{h}(s_{j}-s_{0}) \otimes\bX_{i}\right\}\left\{
            \bX_{i}^{\tp}\bbeta_{0}(s_{j}) + U_{ij}
        \right\}\\
        & = \bI(s_{0})\bgamma_{0}(s_{0})+ 0.5h^{2}\bI_{21}(s_{0})\ddot{\bbeta}_{0}(s_{0}) + \bT(s_{0}) + o(h^{2}).
    \numberthis
\end{align*}
Thus, using Equation (\ref{Chapter3-gmm-Eq:Iinv}) and (\ref{Chapter3-gmm-Eq:Y-expansion}), we can derive, 
\begingroup
\allowdisplaybreaks
\begin{align*}
    &\breve{\bbeta}(s_{0}) = [(1, 0) \otimes \bI_{p}]\bI(s_{0})^{-1}\left\{
            \bI(s_{0})\bgamma_{0}(s_{0})+ 0.5h^{2}\bI_{21}(s_{0})\ddot{\bbeta}_{0}(s_{0}) + \bT(s_{0}) + o(h^{2})
        \right\}
        \\
        &= \bbeta_{0}(s_{0}) + [(1,0)\otimes\bI_{p}]
        f(s_{0})^{-1}\left\{
            \diag(1, \nu_{21})^{-1}\otimes \bOmega_{\bx}^{-1}
        \right\}
        \left\{
            f(s_{0})(\nu_{21},0)\otimes\bOmega_{\bx} 
        \right\}0.5h^{2}\ddot{\bbeta}_{0}(s_{0})\\
        & \qquad + O(\delta_{n1}(h) + h)
        \\
        &= \bbeta_{0}(s_{0}) + 0.5h^{2}\nu_{21}\ddot{\bbeta}_{0}(s_{0}) + O(\delta_{n1}(h) + h)\\
        &= \bbeta_{0}(s_{0}) + O(\delta_{n1}(h) + h) \qquad \text{almost surely.}
    \numberthis
\end{align*}
\endgroup
Therefore, $\sup_{s_{0}\in\sS}\left|\breve{\bbeta}(s_{0}) - \bbeta_{0}(s_{0})\right| = O(\delta_{n1} + h)$ almost surely. Furthermore, observe that the bias of the initial estimator is 
\begin{equation}
    \E\{\breve{\bbeta}(s_{0})\} - \bbeta_{0}(s_{0}) = 0.5h^{2}\nu_{21}\ddot{\bbeta}_{0}(s_{0})\left\{ 1+ O_{P}(\delta_{n1}(h) + h)\right\}.
\end{equation}
Now, to calculate the variance, note that
\begingroup
\allowdisplaybreaks
\begin{align*}
    &\sqrt{n}\{\breve{\bbeta}(s_{0}) - \bbeta(s_{0}) - 0.5h^{2}\nu_{21}\ddot{\beta}_{0}(s_{0})\}(1+o_{a.s.}(1))\\
    &= [(1,0) \otimes \bI_{p}] 
    f(s_{0})
    \left\{\diag(1, \nu_{21})^{-1} \otimes\bOmega_{\bx}^{-1}\right\}\sqrt{n}\bT(s_{0}).
    \numberthis
\end{align*}
\endgroup
By Lemma \ref{Chapter3-gmm-Lemma:Tdist}, we have the variance of the above quantity $\bOmega_{\bx}^{-1}\bSigma_{\bx}^{*}(s_{0}, s_{0})\bOmega_{\bx}^{-1}$, where $\bSigma_{\bx}^{*}(s_{0}, s_{0}) = \lim\limits_{n\rightarrow\infty}\frac{1}{n}\sum\limits_{i=1}^{n}\E\{\bX_{i}\bX_{i}^{\tp}\Sigma_{\bx_{i}}(s_{0}, s_{0})\}$.

\subsection{Proof of Theorem \ref{Chapter3-gmm-Theorem:final}}
Define
$\bC_{\kappa_{0}}(s, s') = \sum\limits_{k=1}^{\kappa_{0}}\lambda_{k}\bphi_{k}(s)\bphi_{k}(s')^{\tp}$
and hence, we can define $\bC_{\kappa_{0}}^{-1}(s, s')$ with possible block matrix 
\begin{equation}
 \bC_{\kappa_{0}}^{-1}(s, s') = \sum_{k=1}^{\kappa_{0}}\lambda_{k}^{-1}\bphi_{k}(s)\bphi_{k}(s')^{\tp} = \begin{pmatrix}
    \bC_{\kappa_{0}, 1, 1}^{-1}(s, s') & 0\\
    0 & \bC_{\kappa_{0}, 2, 2}^{-1}(s, s')
\end{pmatrix}.   
\end{equation}
Also define, 
\begin{align*}
\widehat{\bgamma}_{\kappa_{0}}(s_{0})
    &= \left\{ 
        \sum_{k=1}^{\kappa_{0}}\frac{\lambda_{k}}{\lambda_{k}^{2}+\alpha}
        \sX_{k}(s_{0})\sX_{k}(s_{0})^{\tp}
    \right\}^{-1}
    \left\{
        \sum_{k=1}^{\kappa_{0}}\frac{\lambda_{k}}{\lambda_{k}^{2}+\alpha}
        \sX_{k}(s_{0})\sY_{k}(s_{0})
    \right\}.
    \numberthis
\end{align*}
where 
\begin{equation}
    \sX_{k}(s_{0}) = \frac{1}{nr}\sum_{j=1}^{n}\sum_{j=1}^{r}
    K_{h}(s_{j}-s_{0})
    \bW_{ij}(s_{0})\bQ_{ij}(s_{0})^{\tp}
    \bphi_{k}(s_{0}) + O(\delta(h))
\end{equation}
and 
\begin{equation}
    \sY_{k}(s_{0}) = \frac{1}{nr}\sum_{j=1}^{n}\sum_{j=1}^{r}
    K_{h}(s_{j}-s_{0})\bphi_{k}(s_{0})^{\tp}\bQ_{ij}(s_{0})Y_{ij} + O(\delta(h))
\end{equation}
almost everywhere. 
Therefore, we have the following.
\begingroup
\allowdisplaybreaks
\begin{align*}
    &\sum_{k=1}^{\kappa_{0}}\frac{\lambda_{k}}{\lambda_{k}^{2}+\alpha}
        \sX_{k}(s_{0})\sX_{k}(s_{0})^{\tp} + O(\delta(h))\\
        &=\left\{
            \frac{1}{nr}\sum_{i=1}^{n}\sum_{j=1}^{r}K_{h}(s_{j}-s_{0})
            \bW_{ij}(s_{0})\bQ_{ij}(s_{0})^{\tp}
        \right\}
        \sum_{k=1}^{\kappa_{0}}\lambda_{k}^{-1}\bphi_{k}(s_{0})
        \bphi_{k}(s_{0})^{\tp}\\
        & \qquad\times\left\{
            \frac{1}{nr}\sum_{i=1}^{n}\sum_{j=1}^{r}K_{h}(s_{j}-s_{0})
            \bW_{ij}(s_{0})\bQ_{ij}(s_{0})
        \right\}^{\tp}\\
        &=\bI(s_{0})^{\tp}\bC_{\kappa_{0}}^{-1}(s_{0}, s_{0})\bI(s_{0})\\
        &= f^{2}(s_{0})
        \left[
            \diag(1, \nu_{21})\otimes\bOmega^{\tp}
        \right]\bC_{\kappa_{0}}^{-1}(s_{0}, s_{0})
        \left[
            \diag(1, \nu_{21})\otimes\bOmega
        \right] + O(\delta(h))\\
        &=\sV(s_{0}) + O(\delta(h)),
    \numberthis
\end{align*}
\endgroup
where we define $\sV(s_{0}) = f^{2}(s_{0})\diag\left(\bOmega^{\tp}\bC_{\kappa_{0}, 1, 1,}^{-1}(s_{0}, s_{0})\bOmega, \nu_{21}^{2}\bOmega^{\tp}\bC_{\kappa_{0}, 2, 2,}^{-1}(s_{0}, s_{0})\bOmega\right)$
and 
\begingroup
\allowdisplaybreaks
\begin{align*}
&\sum_{k=1}^{\kappa_{0}}\frac{\lambda_{k}}{\lambda_{k}^{2}+\alpha}\sX_{k}(s_{0})\sY_{k}(s_{0})\\
    &=\left\{
        \frac{1}{nr}\sum_{i=1}^{n}\sum_{j=1}^{r}K_{h}(s_{j}-s_{0})
        \bW_{ij}(s_{0})\bQ_{ij}(s_{0})^{\tp}
    \right\}
    \sum_{k=1}^{\kappa_{0}}\lambda_{k}^{-1}\bphi_{k}(s_{0})
        \bphi_{k}(s_{0})^{\tp}\\
    &\qquad\times
    \left\{
        \frac{1}{nr}\sum_{i=1}^{n}\sum_{j = 1}^{r}K_{h}(s_{j}-s_{0})\bQ_{ij}(s_{0})Y_{ij}
    \right\}\\
    &=\bI(s_{0})^{\tp}\bC_{\kappa_{0}}^{-1}(s_{0}, s_{0})
    \left\{
        \bI(s_{0})\bgamma_{0}(s_{0}) + 0.5h^{2}\bI_{21}(s_{0})\ddot{\bbeta}(s_{0}) + \bT(s_{0}) + o(h^{2})
    \right\}.
    % &=\bI(s_{0})\bC_{\kappa_{0}}^{-1}(s_{0}, s_{0})\bI(s_{0})\bgamma_{0}(s_{0}) + 
    % 0.5h^{2}\bI(s_{0})\bC_{\kappa_{0}}^{-1}(s_{0}, s_{0})\bI_{21}(s_{0})\ddot{\bbeta}(s_{0}) +\bI(s_{0})\bC_{\kappa_{0}}^{-1}(s_{0}, s_{0})\bT(s_{0}) + o(h^{2})\\
    % &=\sV(s_{0})\bgamma_{0}(s_{0}) + 0.5h^{2}\sB(s_{0}) + \bI(s_{0})\bC_{\kappa_{0}}^{-1}(s_{0}, s_{0})\bT(s_{0}) +o(h^{2})
    \numberthis
\end{align*}
\endgroup
Hence, 
\begingroup
\allowdisplaybreaks
\begin{align*}
    &\widehat{\bbeta}(s_{0}) - \bbeta_{0}(s_{0})\\
    &=0.5h^{2}[(1, 0) \otimes \bI_{p}]\sV(s_{0})^{-1}\bI(s_{0})^{\tp}\bC_{\kappa_{0}}^{-1}(s_{0}, s_{0})\bI_{21}(s_{0})\ddot{\bbeta}(s_{0}) \\
    & \qquad + [(1, 0) \otimes \bI_{p}]\sV(s_{0})^{-1}\bI(s_{0})^{\tp}\bC_{\kappa_{0}}^{-1}(s_{0}, s_{0})\bT(s_{0}) + O(\delta(h))\\
    &= 0.5h^{2}f^{2}(s_{0})
        [
        (1, 0)\otimes \bI_{p}]\sV(s_{0})^{-1}
        [\diag(1, \nu_{21})\otimes\bOmega^{\tp}
        ]
        \bC_{\kappa_{0}}^{-1}(s_{0}, s_{0})
        [(\nu_{21}, 0)^{\tp} \otimes \bOmega]\ddot{\bbeta}(s_{0})\\
    & \qquad + f^{2}(s_{0})
         [(1, 0)\otimes \bI_{p}]
         \sV(s_{0})^{-1}
        [diag(1, \nu_{21})\otimes\bOmega^{\tp}]
        \bC_{\kappa_{0}}^{-1}(s_{0}, s_{0})\bT(s_{0}) + O(\delta(h))\\
    &= 0.5h^{2}\nu_{21}\ddot{\bbeta}(s_{0}) + \sC(s_{0}){\bT}(s_{0}) + O(\delta(h)),
    \numberthis
\end{align*}
\endgroup
where, $\sC(s_{0}) = f^{2}(s_{0})
         [(1, 0)\otimes \bI_{p}]
         \sV(s_{0})^{-1}
        [diag(1, \nu_{21})\otimes\bOmega^{\tp}]
        \bC_{\kappa_{0}}^{-1}(s_{0}, s_{0})$.
        Thus, in order to obtain the asymptotic variance, consider, using Lemmas \ref{Chapter3-gmm-Lemma:Ibound}, \ref{Chapter3-gmm-Lemma:Jbound} and \ref{Chapter3-gmm-Lemma:Tdist}, we have 
$\sqrt{n}\{ \widehat{\bbeta}(s_{0}) - \bbeta_{0}(s_{0}) - 0.5h^{2}\nu_{21}\ddot{\bbeta}(s_{0})\} \xrightarrow[]{d} N(0, \sA(s_{0}, s_{0}))$ where $\sA(s_{0}, s_{0}) = \sB(s_{0}, s_{0})^{-1}\bOmega^{\tp}\bC_{\kappa_{0}, 11}(s_{0}, s_{0})^{-1}
\bSigma_{\fM}^{*}(s_{0}, s_{0})
\bC_{\kappa_{0}, 11}(s_{0}, s_{0})^{-1}\bOmega\sB(s_{0}, s_{0})^{-1}$, with
$\sB(s_{0}, s_{0}) = \bOmega^{\tp}\bC_{\kappa_{0}, 11}^{-1}(s_{0}, s_{0})\bOmega$.

%$$\sA(s_{0}, s_{0}) = (\bOmega\bC_{\kappa_{0}, 11}^{-1}(s_{0}, s_{0})\bOmega^{\tp})^{-1}\bOmega\bC_{\kappa_{0}, 11}^{-1}(s_{0}, s_{0})\bSigma(s_{0}, s_{0})\bC_{\kappa_{0}, 11}^{-1}(s_{0}, s_{0})\bOmega(\bOmega\bC_{\kappa_{0}, 11}^{-1}(s_{0}, s_{0})\bOmega^{\tp})^{-1}.$$
% $\sA(s_{0}, s_{0})$ is the asymptotic variance of $\sqrt{n}\Tilde{\bT}(s_{0})$, where we derive, $\sA(s_{0}, s_{0}) = [(1,0)\otimes \bI_{p}]\sV^{-1}(s_{0})\tilde{\sA}(s_{0}, s_{0})\sV^{-1}(s_{0})[(1, 0)\otimes\bI_{p}]$ for $\tilde{\sA}(s_{0}, s_{0}) = [\diag(1, \nu_{21})\otimes\bOmega]\bC_{\kappa_{0}}^{-1}(s_{0}, s_{0})\diag(\Sigma(s_{0}, s_{0}), \nu_{11}^{2}\Sigma(s_{0}, s_{0}))\bC_{\kappa_{0}}^{-1}(s_{0}, s_{0})[\diag(1, \nu_{21})\otimes\bOmega]^{\tp}$. 
%By simple calculation, it can be shown that 
%$$\sA(s_{0}, s_{0}) = (\bOmega\bC_{\kappa_{0}, 11}^{-1}(s_{0}, s_{0})\bOmega^{\tp})^{-1}\bOmega\bC_{\kappa_{0}, 11}^{-1}(s_{0}, s_{0})\bSigma(s_{0}, s_{0})\bC_{\kappa_{0}, 11}^{-1}(s_{0}, s_{0})\bOmega(\bOmega\bC_{\kappa_{0}, 11}^{-1}(s_{0}, s_{0})\bOmega^{\tp})^{-1}.$$

{
\section{Discussion on the choice of IV}
Let $\fM^{*}(\bX) = \bX/\sigma^{2}(\bX)$ and $s_{a} = \E\{\sigma^{2a}(\bX)\}$ for $a = 1, 2$. 
%Now assume that $\Var\{\sigma(\bX)\}$ is low such that $\bX/\sigma(\bX)$ behaves like $\bX/\E\{\sigma^{2}(\bX)\}$. 
Thus, based on the $\fM^{*}(\bX)$, $\bOmega = \E\{\fM^{*}(\bX)\bX^{\tp}\} = \bOmega_{\bx}/s_{1}\{1+o(1)\}$, 
$\sB^{*}(s_{0}, s_{0}) = \bOmega_{\bx}\bC^{-1}_{\kappa_{0}, 11}(s_{0}, s_{0})\bOmega_{\bx}/s_{1}^{2}\{1+o(1)\}$ where $\bOmega_{\bx}=E(\bX\bX^{\tp})$. 
Then, the asymptotic variance under the choice of $\fM^{*}$ is
\begin{align*}
    &\sA^{*}(s_{0}, s_{0})\\
    &= s_{1}^{-2}s_{2}^{-1}\sB^{*-1}(s_{0}, s_{0})\bOmega_{\bx}^{\tp}\bC^{-1}_{\kappa_{0}, 11}(s_{0}, s_{0})\bSigma_{\bx}^{*}(s_{0}, s_{0})\bC^{-1}_{\kappa_{0}, 11}(s_{0}, s_{0})\bOmega_{\bx}\sB^{*-1}(s_{0}, s_{0})\\
    &=s_{2}^{-1}\bOmega_{\bx}^{-1}\bSigma_{\bx}^{*}(s_{0}, s_{0})\bOmega_{\bx}^{-1}.
\end{align*}
Now, define, 
\begin{align*}
    \fD_{i}(s_{0}) &= \bOmega^{\tp}\bC_{\kappa_{0}, 11}(s_{0}, s_{0})^{-1}\left\{r^{-1}\sum_{j=1}^{r}K_{h}(s_{j}-s_{0})\fM(\bX)U_{ij} \right\}\;\mbox{and}\\
    \fD_{i}^{*}(s_{0}) &= r^{-1}\sum_{j=1}^{r}K_{h}(s_{j}-s_{0})\bX U_{ij}/\sigma^{2}(\bX).
\end{align*}
Therefore, by some calculation, it is not difficult to show the following: 
\begin{align*}
\E\{\fD_{i}(s_{0})\fD_{i}(s_{0})^{\tp}\} &= \bOmega^{\tp}\bC_{\kappa_{0}, 11}(s_{0}, s_{0})^{-1}\bSigma_{\fM}^{*}(s_{0}, s_{0})\bC_{\kappa_{0}, 11}(s_{0}, s_{0})^{-1}\bOmega \{1+ o(1)\};\\
\E\{\fD_{i}(s_{0})\fD_{i}^{*}(s_{0})^{\tp}\} &= \bOmega^{\tp}\bC_{\kappa_{0}, 11}(s_{0}, s_{0})^{-1}\bOmega/s_{1} \{1+ o(1)\};\\
\E\{\fD_{i}^{*}(s_{0})\fD_{i}^{*}(s_{0})^{\tp}\} &= \bOmega_{\bx}/s_{2} \{1+ o(1)\}.
\end{align*}
Therefore, 
\begingroup
\allowdisplaybreaks
\begin{align*}
    &\sA(s_{0}, s_{0}) - \sA^{*}(s_{0}, s_{0})\\
    &= \sB(s_{0}, s_{0})^{-1}\bOmega^{\tp}\bC_{\kappa_{0}, 11}(s_{0}, s_{0})^{-1}\bSigma^{*}_{\fM}(s_{0}, s_{0})\bC_{\kappa_{0}, 11}(s_{0}, s_{0})^{-1}\bOmega\sB(s_{0}, s_{0})^{-1}\\
    & \qquad - s_{2}^{-1}\bOmega_{\bx}^{-1}\bSigma_{\bx}^{*}(s_{0}, s_{0})\bOmega_{\bx}^{-1}\\
    & = [\E\{\fD(s_{0}, s_{0})\fD^{*}(s_{0}, s_{0})^{\tp}\}]^{-1}\\
    & \qquad \times \left(
        \E\{\fD(s_{0}, s_{0})\fD(s_{0}, s_{0})^{\tp}\} \right.\\
        &\qquad\left.- \E\{\fD(s_{0}, s_{0})\fD^{*}(s_{0}, s_{0})^{\tp}\}[\E\{\fD^{*}(s_{0}, s_{0})\fD^{*}(s_{0}, s_{0})^{\tp}\}]^{-1}\E\{\fD^{*}(s_{0}, s_{0})\fD(s_{0}, s_{0})^{\tp}\}
    \right)\\
    & \qquad [\E\{\fD^{*}(s_{0}, s_{0})\fD(s_{0}, s_{0})^{\tp}\}]^{-1}\\
    &= \E\{\fR(s_{0}, s_{0})\fR^{\tp}(s_{0}, s_{0})\}\geq 0,
\end{align*}
\endgroup
where 
\begin{align*}
    \fR(s_{0}, s_{0}) 
    &= [\E\{\fD(s_{0}, s_{0})\fD^{*}(s_{0}, s_{0})^{\tp}\}]^{-1}
    \Big\{\fD(s_{0}, s_{0})\\
    &- \E\{\fD(s_{0}, s_{0})\fD^{*}(s_{0}, s_{0})^{\tp}\}[\E\{\fD^{*}(s_{0}, s_{0})\fD^{*}(s_{0}, s_{0})^{\tp}\}]^{-1}\fD^{*}(s_{0}, s_{0})\Big\}.
\end{align*}
Thus, the chosen IV estimator is optimal among the class of all local linear GMM estimators of the varying coefficient model. 
}

%% use bibfile 
\bibliographystyle{chicago}      % Chicago style, author-year citations
\bibliography{main}   % name your BibTeX data base

\bibhang=1.7pc
\bibsep=2pt
\fontsize{9}{14pt plus.8pt minus .6pt}\selectfont
\renewcommand\bibname{\large \bf References}
%\begin{thebibliography}{11}
\expandafter\ifx\csname
natexlab\endcsname\relax\def\natexlab#1{#1}\fi
\expandafter\ifx\csname url\endcsname\relax
  \def\url#1{\texttt{#1}}\fi
\expandafter\ifx\csname urlprefix\endcsname\relax\def\urlprefix{URL}\fi

%%%%%%%%%%%%%%%%%%%%%%%%%%%%%%%%%%%%%%%%%%%%%%%%%%%%%%%%%%%%%%%%%%%%%%%%%%%%%%%%%%%%%%%%%%%%%%%%%%%%%%%%%%%%%%%%%%%%%%%%%%%%
\vskip .65cm
\noindent
Department of Biostatistics, Johns Hopkins University, 
615 North Wolfe Street,
Baltimore, MD 21205, USA\\
% \vskip 2pt
\noindent
E-mail: (pnyogi1@jhmi.edu)
\vskip 2pt
\noindent
\vskip 2pt
\noindent
Department of Mathematics, Statistics and Computer Science, University of Illinois at Chicago, 
851 S. Morgan Street, Chicago, Illinois 60607, USA \\
% \vskip 2pt
\noindent
E-mail: (pszhong@uic.edu)
\vskip 2pt
\noindent
\vskip 2pt
\noindent
Center for MR Research and Departments of Radiology, Neurosurgery, and Bioengineering, University of Illinois at Chicago,
1801 West Taylor St., Chicago, Illinois 60612, USA\\
% \vskip 2pt
\noindent
E-mail: (xjzhou@uic.edu)
\vskip .3cm
%\centerline{(Received ???? 20??; accepted ???? 20??)}\par
\end{document}